\documentclass[a4paper,11pt]{article}
\pdfoutput=1
\usepackage{jcappub}
\usepackage{colortbl}
\usepackage{tablefootnote}
\usepackage{tabularx,booktabs}
\newcolumntype{Y}{>{\centering\arraybackslash}X}

\usepackage[utf8]{inputenc}
\usepackage[font=small,labelfont=bf,tableposition=top]{caption}
\usepackage{float}
\usepackage{adjustbox}
\usepackage{arydshln}

\usepackage{amsmath,amssymb,amsbsy,amstext, amsthm, simplewick, amsfonts, bm}
\usepackage{graphicx}
\usepackage{siunitx}
\usepackage{upgreek}
\usepackage{framed}
\usepackage{wrapfig}
\usepackage{multirow}
\usepackage{bbm}
\usepackage[svgnames,dvipsnames,x11names]{xcolor}
\usepackage{subcaption}
\usepackage{hyperref}
\usepackage{fontawesome}
\usepackage{sidecap}

\usepackage[top=2.5cm,bottom=2.5cm,right=2.5cm,left=2.5cm]{geometry}
\def\be{\begin{equation}}
\def\ee{\end{equation}}
\def\bea{\begin{eqnarray}}
\def\eea{\end{eqnarray}}
\def\ba{\begin{align}}

\def\bi{\begin{itemize}}
\def\ei{\end{itemize}}

\def\bx{{\bf x}}
\def\bk{{\bf k}}
\def\bq{{\bf q}}

\def\bd{{\bf d}}
\def\nbar{\bar{n}}

\def\cG{{\mathcal G}}

\def\cQ{{\mathcal Q}}
\def\cL{{\mathcal L}}

\def\cP{{\mathcal P}}
\def\cS{{\mathcal S}}
\def\cO{{\mathcal O}}

\def\calP{\mathcal{P}}
\def\calS{\mathcal{S}}
\def\gpch{{h^{-1}\rm{Gpc}}}
\def\mpch{{h^{-1}\rm{Mpc}}}
\def\hmpc{{{\rm Mpc}^{-1} h}}

\newcommand{\ft}[1]{\mathcal{FT}\left\{{#1}\right\}}

\newcommand{\av}[1]{\left\langle{#1}\right\rangle} 

\title{Cosmological Information in Skew Spectra of \\ [.45cm] Biased Tracers in Redshift Space \vspace{0.15in} \\ }

\author[a,b]{\fontsize{15.84}{23}\selectfont Jiamin Hou,}
\author[c]{Azadeh Moradinezhad Dizgah,}
\author[d]{ChangHoon Hahn,}
\author[e,f]{Elena Massara}
\affiliation[a]{Department of Astronomy, University of Florida, 211 Bryant Space Science Center, Gainesville, FL 32611, USA}
\affiliation[b]{Max-Planck-Institut f\"ur Extraterrestrische Physik, Postfach 1312, Giessenbachstrasse 1, 85748 Garching bei M\"unchen, Germany}
\affiliation[c]{ D\'epartement de Physique Th\'eorique,
Universit\'e de Gen\`eve, 24 quai Ernest Ansermet, \\ 1211 Gen\`eva 4, Switzerland}
\affiliation[d]{Department of Astrophysical Science, Princeton University, Peyton Hall, Princeton NJ 08544, USA}
\affiliation[e]{Waterloo Centre for Astrophysics, University of Waterloo, 200 University Ave W, Waterloo, ON N2L 3G1, Canada}
\affiliation[f]{Department of Physics and Astronomy, University of Waterloo, Waterloo, ON N2L 3G1, Canada}
\emailAdd{jiamin.hou@ufl.edu, Azadeh.MoradinezhadDizgah@unige.ch}

\abstract{Extracting the non-Gaussian information encoded in the higher-order clustering statistics of the large-scale structure is key to fully realizing the potential of upcoming galaxy surveys. We investigate the information content of the redshift-space {\it weighted skew spectra} of biased tracers as efficient estimators for 3-point clustering statistics. The skew spectra are constructed by correlating the observed galaxy field with an appropriately-weighted square of it. We perform numerical Fisher forecasts using two synthetic datasets; the halo catalogs from the Quijote N-body simulations and the galaxy catalogs from the Molino suite. The latter serves to understand the effect of marginalization over a more complex matter-tracer biasing relation. Compared to the power spectrum multipoles, we show that the skew spectra substantially improve the constraints on six parameters of the $\nu\Lambda$CDM model, $\{\Omega_m, \Omega_b, h, n_s, \sigma_8, M_\nu\}$. Imposing a small-scale cutoff of $k_{\rm max}=0.25\,\hmpc$, the improvements from skew spectra  alone range from 23\% to 62\%  for the Quijote halos and from 32\% to 71\% for the Molino galaxies. Compared to the previous analysis of the bispectrum monopole on the same data and using the same range of scales, the skew spectra of Quijote halos provide competitive constraints. Conversely, the skew spectra outperform the bispectrum monopole for all cosmological parameters for the Molino catalogs. This may result from additional anisotropic information, particularly enhanced in the Molino sample, that is captured by the skew spectra but not by the bispectrum monopole.  Our stability analysis of the numerical derivatives shows comparable convergence rates for the power spectrum and the skew spectra, indicating potential underestimation of parameter uncertainties by at most 30\%.}

\begin{document}
\maketitle

\clearpage
\section{Introduction}
The non-Gaussian nature of cosmic large-scale structure (LSS) implies that statistics beyond the power spectrum carry significant information about the universe's origin, evolution, and constituents. The volume and precision of the data from upcoming galaxy surveys, such as, PFS\footnote{\url{https://pfs.ipmu.jp/}} \cite{PFSTeam:2012fqu}, DESI\footnote{\url{https://www.desi.lbl.gov}} \cite{DESI:2016fyo}, SPHEREx\footnote{\url{https://www.jpl.nasa.gov/missions/spherex}} \cite{Dore:2014cca}, Euclid\footnote{\url{https://www.euclid-ec.org}} \cite{Amendola:2016saw}, and Roman Space Telescope\footnote{\url{https://roman.gsfc.nasa.gov/}} \cite{Wang:2021oec} will allow for higher signal-to-noise measurement of higher-order statistics such as the bispectrum. However, in addition to challenges in accounting for observational effects, modeling the signal and the covariance estimation of the galaxy bispectrum is rather challenging. Therefore, in addition to a growing number of recent works on extracting information from the LSS bispectrum\footnote{See \cite{Philcox:2021hbm,Gualdi:2022kwz,Hou:2022wfj,Philcox:2022hkh} for recent works on galaxy 4-point statistics, \cite{Valogiannis:2021chp,Eickenberg:2022qvy, Valogiannis:2022xwu} for statistics beyond $n$-point correlation functions using wavelets, and \cite{White:2016yhs,Massara:2020pli,Massara:2022zrf,Bonnaire:2021sie,Paillas:2022wob} for summary statistics capturing the environmental dependence of clustering.} (e.g. \cite{Gil-Marin:2016wya,Hahn:2019zob, Hahn:2020lou,Gualdi:2021yvq,MoradinezhadDizgah:2020whw,Eggemeier:2021cam,Philcox:2021kcw,Ivanov:2021kcd,Cabass:2022wjy,Cabass:2022ymb,Philcox:2022frc,DAmico:2022gki,DAmico:2022osl,Coulton:2022rir}), several alternative and more efficient estimators that compress the information in the bispectrum via some weighted averaging, have been proposed in the literature \cite{Fergusson:2010ia,Schmittfull:2012hq,Schmittfull:2014tca,Chiang:2015pwa,Eggemeier:2015ifa,Eggemeier:2016asq,Wolstenhulme:2014cla,Byun:2017fkz}.

Among the proposed statistics, the weighted galaxy skew spectra \cite{Schmittfull:2014tca,MoradinezhadDizgah:2019xun,Dai:2020adm,Schmittfull:2020hoi,Chakraborty:2022aok} are simple and computationally efficient proxy statistics for the galaxy bispectrum, constructed from the cross-correlation of the observed galaxy density field with an appropriately weighted square of it\footnote{Optimally weighted skew spectra in harmonic space, which saturate the Cramer-Rao bound of the bispectrum in the weakly non-Gaussian limit, were first introduced in the context of constraining primordial non-Gaussianity from the CMB data \cite{Komatsu:2003iq,Munshi:2009ik} (see \cite{Munshi:2020ofi,Munshi:2021uwn,Chakraborty:2022aok} for some recent works on the harmonic-space skew and kurto spectra).}. Since the forms of the skew spectra are derived from the maximum likelihood estimator of the amplitude of the bispectrum \cite{Schmittfull:2014tca}, in the limit of weak non-Gaussianity, they carry the same Fisher information as the full bispectrum for parameters that appear as overall amplitudes of the separable contributions of tree-level bispectrum. Therefore they represent a lossless compression of the bispectrum for constraining the galaxy biases, the logarithmic growth rate, and the amplitude of the primordial power spectrum and bispectrum\footnote{Analogous estimators for trispectrum, often referred to as kurto spectra, involve correlations of two quadratic fields or that of a cubic and linear field \cite{Abidi:2018eyd,Munshi:2021uwn}.} \cite{MoradinezhadDizgah:2019xun}. In real space, and considering Gaussian initial conditions (GICs), the list of quadratic fields include a squared galaxy density, a tidal term, and a shift term \cite{Schmittfull:2014tca}. While the first two depend on nonlinear biases $b_2$ and $b_{\mathcal G_2}$, the third only depends on linear bias $b_1$. Therefore, the corresponding three skew spectra optimally capture the bispectrum information on the three galaxy biases. In the presence of primordial non-Gaussianity, additional skew spectra should be constructed \cite{Schmittfull:2014tca}. For instance, the correlation of galaxy density filtered by the inverse of the matter transfer function with a weighted square of it captures the imprint of primordial bispectrum on matter bispectrum. The weights, in this case, are determined by the shape of the primordial bispectrum \cite{,MoradinezhadDizgah:2019xun}. In redshift space (and for GICs), to capture the full information of the bispectrum on amplitude-like parameters, one needs a total of fourteen skew spectra, the forms of which are determined by the tree-level bispectrum in redshift-space \cite{Schmittfull:2020hoi}. In this case, each of the spectra has a different dependence on the logarithmic growth rate, the amplitude of fluctuations, and galaxy biases. 

There are several advantages to using the skew spectra as opposed to the bispectrum; first, they are simple to interpret since they are functions of a single wavenumber and not triangle shapes. Second, the computational cost of measuring them from the data is comparable to that of the power spectrum. While accounting for all bispectrum triangles requires ${\mathcal O}(N^2)$ operations\footnote{See \cite{Scoccimarro:2015bla,Slepian:2015qwa} for fast estimators of Fourier- and configuration-space 3-point statistics.}, capturing the full information of the bispectrum using the skew spectra requires ${\mathcal O}(N \log N)$ operations, where $N = (k_{\rm max}/\Delta k)^3$ is the number of 3D Fourier-space grid points at which the fields are evaluated given a small-scale cutoff of $k_{\rm max}$. Third, estimating the covariance matrices of the skew spectra from mocks requires a significantly smaller number of mocks\footnote{For a given data vector estimating the covariance from the mocks requires a significantly larger number of realizations than the length of the data vector. Several proposals are made in the literature to speed up and decrease the number of required simulations for the estimation of bispectrum covariance \cite{Joachimi:2016xhk,Hall:2018umb,Chartier:2021frd,Philcox:2020zyp}.}. Lastly, in comparison to the bispectrum \cite{Pardede:2022udo}, accounting for the survey window function is expected to be considerably simpler for the skew spectra \cite{Hou:2022aaa}.

Previously, the perturbative models of the weighted skew spectra in real and redshift space have been tested against N-body simulations \cite{Schmittfull:2014tca, Schmittfull:2020hoi}, and their constraining power for amplitude-like model parameters in real space has been investigated \cite{Schmittfull:2014tca,MoradinezhadDizgah:2019xun}. However, the full information content of the galaxy skew spectra for constraining cosmological parameters has not been explored before\footnote{See also Ref. \cite{Dai:2020adm} for forecasts for a small subset of skew spectra, corresponding to the $\langle \delta^2 \delta \rangle$ correlation.}. In this paper, we investigate how much the complete set of redshift-space galaxy skew spectra, introduced in Ref. \cite{Schmittfull:2020hoi}, can improve the cosmological information from the power spectrum multipoles and how the improvement compares with that of the bispectrum. For this purpose, we use simulated datasets that are constructed to perform numerical Fisher forecasts. Since the forecasts do not rely on utilizing theoretical models of the observables (the derivatives and covariance matrices are directly measured from simulations), we can explore the information encoded on the smaller-scale fluctuations, which normally are excluded since the perturbative models of the observables fail as one approaches the non-linear scales. 

We first focus on the Fisher forecasts using the halo catalogs from the Quijote suite of simulations \cite{Villaescusa-Navarro:2019bje} \href{https://github.com/franciscovillaescusa/Quijote-simulations}{\faGithub}\footnote{\url{ https://github.com/franciscovillaescusa/Quijote-simulations}}. This analysis can be considered an idealized case, where the halo biases are perfectly known, and only cosmological parameters are varied. To make the analysis more realistic, similar to some of the previous forecasts with this dataset~\cite{Naidoo:2021dxz,Coulton:2022rir}, we vary a single nuisance parameter, the minimum halo mass cut $M_{\rm min}$ as a proxy for unknown halo bias. The halo mass cut captures the difference in the halo number densities and is marginalized over. Next, we go beyond a simple single-parameter proxy of the halo bias
and investigate the impact of marginalization over parameters of a more complex tracer-to-matter biasing relation. For this analysis, we use the Molino galaxy mock catalogs \cite{Hahn:2020lou} \href{https://changhoonhahn.github.io/molino}{\faGlobe}\footnote{\url{ https://changhoonhahn.github.io/molino}}, which are built on Quijote halo catalogs by applying a five-parameter Halo Occupation Distribution (HOD) prescription. Therefore, they allow us to simultaneously vary cosmological and HOD parameters and study the extent to which the information content of the skew spectra is affected by marginalization. For both datasets, we test the robustness of the numerical forecasts in computing the derivatives~\cite{Coulton:2022rir} as well as the noise due to the estimation of the covariance matrices from a finite number of simulations. We find that on both datasets, the skew spectra provide significant improvements over the power spectrum results. We, thus, advocate the skew spectra as powerful, efficient, and easy-to-interpret summary statistics for extracting non-Gaussian information of the LSS. 

The rest of the paper is organized as follows. In \S \ref{sec:th_skewspecs}, we review the theoretical construction of the redshift-space skew spectra of biased tracers and describe the measurement pipeline that we apply to the simulated data. We then briefly describe the synthetic datasets from the Quijote and Molino simulations in \S \ref{sec:sims}, and outline the numerical Fisher forecast methodology in \S \ref{sec:fisher}. We present our results in \S \ref{sec:res}, together with the measurements of the skew spectra, their signal-to-noise ratio, and the forecasted parameter constraints. We draw our conclusions in \S \ref{sec:conc}. We present additional results in a series of appendices pertaining to the information in individual skew spectra \S\ref{app:indSkew}, tests of numerical stability of Fisher forecasts \S\ref{app:convergence}, the effect of shot noise subtraction \S\ref{app:shot}, and lastly, the results for Molino galaxies in light of the peculiar characteristics of the sample \S\ref{app:Molino}.

\section{Weighted Skew Spectra in Redshift Space}\label{sec:th_skewspecs}
This section reviews the basic ingredients for constructing the galaxy skew spectra and the perturbative model of its leading contributions, focusing on redshift-space estimators. We also describe the numerical pipeline we use to measure the skew spectra from simulated halo/galaxy catalogs. We refer the interested reader to Ref.~\cite{Schmittfull:2020hoi} for more details on modeling and the comparison of the theory predictions against the simulated dark matter overdensity field. 

\subsection{Skew Spectra as Maximum Likelihood Estimators}
For a theoretical model of the galaxy bispectrum, $A_b^{\rm th} B_g^{\rm th}(\bk_1,\bk_2,\bk_3)$, the maximum likelihood estimator of its amplitude, $A_b^{\rm th}$, in the weak non-Gaussian limit is given by the projection of the inverse-variance weighted cubic estimator onto the theoretical template \cite{Fergusson:2010ia},
\begin{align}\label{eq:maxlike}
{\hat A}_b^{\rm th} &= \iint\frac{d^3 k\, d^3q}{N_{\rm th}(2\pi)^3} \  \frac{B_g^{\rm th}(\bq,-\bk,\bk-\bq)}{P_g^{\rm obs}(\bk)P_g^{\rm obs}(\bk-\bq)P_g^{\rm obs}(\bk)} \notag \\ 
& \hspace{1in} \times \left[\delta_g^{\rm obs}(\bq)\delta_g^{\rm obs}(\bk-\bq)\delta_g^{\rm obs}(-\bq) - 3 \av{\delta_g^{\rm obs}(\bq)\delta_g^{\rm obs}(\bk-\bq)} \delta_g^{\rm obs}(-\bk)\right].
\end{align}
Here, $N_{\rm th}$ is the normalization which depends on the assumed theory bispectrum, $\delta_g^{\rm obs}$ represents a noisy measurement of the galaxy and $P_g^{\rm obs}$ is the measured power spectrum including the shot noise. The term linear in $\delta_g^{\rm obs}$ is only relevant for the statistically inhomogeneous field and is dropped in the rest of our discussions. 

For a theoretical bispectrum that consists of a sum of product-separable terms, 
$B_g^{\rm th}(\bk_1,\bk_2,\bk_3)\\ =\sum_i f_i(\bk_1) g_i(\bk_2)h_i(\bk_3)$ (as is the case for the leading-order contributions due to gravitational evolution and most types of primordial bispectrum), the above estimator for each separable contribution can be recast in terms of cross-correlations between appropriately filtered quadratic and linear fields,
\begin{align}\label{eq:skew_opt}
{\hat A}_b^{\rm th} &= \int \frac{d^3 k}{N_{\rm th}} \left[\frac{f\delta_g^{\rm obs}}{P_g^{\rm obs}} \star \frac{g\delta_g^{\rm obs}}{P_g^{\rm obs}}\right](\bk) \left[\frac{h \delta_g^{\rm obs}}{P_g^{\rm obs}}\right](-\bk) \notag \\
&= \frac{1}{N_{\rm th}}\int k^2 dk \ {\hat \cP}_{\frac{f\delta_g}{P_g}\star \frac{g\delta_g}{P_g}, \frac{h \delta_g}{P_g}}(k).
\end{align}
Here, $\star$ denotes the convolution of two filtered galaxy fields, and we defined $\hat \cP$ as the angle-averaged skew spectrum between the filtered quadratic and linear fields. Instead of summing over all Fourier modes to obtain the overall amplitude of the bispectrum, one can retain the shape information and consider a set of skew spectra to extract the information on various contributions to the bispectrum optimally. The estimator in Eq.~\eqref{eq:maxlike} includes the information from all triangles and is unbiased. Therefore, the weighted skew spectra defined in Eq.~\eqref{eq:skew_opt} provide a lossless compression of the (tree-level) bispectrum information for parameters appearing as overall amplitudes of separable contributions to the model bispectrum. In the rest of this paper, as in Ref.~\cite{Schmittfull:2020hoi}, we drop the inverse-variance weighting of the skew spectra estimators and denote the redshift-space skew spectra as $\hat \cP_{\cS_n {\tilde \delta_g}}(k)$, with $\cS_n$ and $\tilde \delta_g$ being the filtered quadratic and linear fields, respectively. We note that when analyzing real data with both shot noise and observational effects, the inverse-variance weighting of the observed density field should be accounted for.

Since the skew spectra involve integration over a wide range of Fourier modes, some of the information from small scales are imprinted on larger scales. The input field should be smoothed to remove the information from small scales (in particular if using perturbative theoretical models of the skew spectra based on tree-level bispectrum). For a tophat filter, smoothing the fields is equivalent to imposing a cutoff $k_{\rm max}$ to ensure that wavevectors above some threshold are not included in the analysis. As in Ref. \cite{Schmittfull:2014tca, Schmittfull:2020hoi}, to avoid edge effects in our measurements, we apply a Gaussian smoothing $W_R(k) = \exp({- k^2R^2/2})$ to the observed density field when computing the quadratic fields. 

\subsection{Clustering Component}
To define the redshift-space skew spectra, we consider the tree-level galaxy bispectrum. Neglecting the derivative bias operators and stochastic contributions, up to second order in perturbation theory, the over-density of a biased tracer (such as halos or galaxies denoted generically as $\delta_g$) is given in terms of the matter overdensity and tidal field as \cite{Desjacques:2016bnm}, 
\be
\delta_g(\bx) = b_1 \delta(\bx) + \frac{1}{2} b_2 \delta^2(\bx) + b_{\cG_2} \cG_2(\bx).
\label{eqn:deltag_b1b2bg2}
\ee
Here, $\delta$ refers to matter overdensity and $\cG_2$ is the Galileon operator given in terms of the Newtonian gravitational potential $\Phi$ as 
\be
\cG_2(\Phi) \equiv \left(\partial_i \partial_j \Phi \right)^2 - \left(\partial^2 \Phi\right)^2.
\label{eq:Galileon_G2}
\ee
Neglecting the effective field theory counterterms (including fingers-of-god contributions), the tree-level gravitationally induced halo/galaxy bispectrum in redshift space is given by
\be\label{eq:rsd_bis}
B_g(\bk_1,\bk_2,\bk_3) = 2 Z_1(\bk_1) Z_1(\bk_2) Z_2(\bk_1,\bk_2) P_{\rm lin}(k_1) P_{\rm lin}(k_2) + 2 \ {\rm perms.},
\ee
where $P_{\rm lin}$ is the linear matter power spectrum. The functions $Z_1$ and $Z_2$ are the redshift-space perturbation theory kernels and are given by
\begin{align}
    Z_1(\bk_1) &= b_1 + f \frac{k^2_{1\parallel}}{k_1^2}, \notag \\
   Z_2(\bk_1,\bk_2) &=  b_1F_2(\bk_1,\bk_2) + \frac{b_2}{2} + b_{{\mathcal G}_2} S^2(\bk_1,\bk_2) 
  + f \frac{k_{\parallel}^2}{k^2}  G_2(\bk_1,\bk_2)
  \nonumber \\
&\quad
+ b_1 f \frac{k_{\parallel}}{2}  \left(\frac{k_{1\parallel}}{k_1^2}+\frac{k_{2\parallel}}{k_2^2}\right)
+f^2 \frac{k_{\parallel}}{2}
\left(
\frac{k_{1\parallel}}{k_1^2}\frac{k_{2\parallel}^2}{k_2^2} + \frac{k_{1\parallel}^2}{k_1^2}\frac{k_{2\parallel}}{k_2^2}
\right).
\end{align}
Here, $f$ is the logarithmic growth rate, and $k_{\parallel}$ is parallel to the line-of-sight (LoS) component of wavevector $\bf k$. The $F_2$ and $G_2$ functions are the standard perturbation theory kernels for matter density and velocity, while $S^2$ is the Fourier transform of the Galileon operator,
\be
S^2(\bk_1,\bk_2) \equiv \left(\frac{\bk_1\cdot\bk_2}{k_1 k_2}\right)^2 - 1.
\ee
The tree-level bispectrum in Eq.~\eqref{eq:rsd_bis} can be cast as a sum of separable contributions of the general form $B_n(\bk_1,\bk_2,\bk_3) = P_{\rm lin}(\bk_1)P_{\rm lin}(\bk_2) D_n(\bk_1,\bk_2) h(\bk_3)$. Therefore, the full information of the bispectrum on galaxy biases, growth rate, and the amplitude of the primordial power spectrum and bispectrum can be fully captured by a set of 14 skew spectra $\cP_{\cS_n \tilde \delta}$, where $\cS_n$ are smoothed fields quadratic in galaxy overdensity (see the definition below). The general form of the expectation value of the skew spectra estimators can be written as 
\be
\cP_{\cS_n \tilde \delta}(k) = \int  \frac{d\Omega_k}{4\pi}  \av {\cS_n(\bk) \tilde \delta_g(-\bk)}, 
\label{eqn:quad_field}
\ee
where the quadratic fields $\cS_n$ of two smoothed fields $\delta_g^R(\bk) = W_R(k) \delta_g(\bk)$ are defined as
\be
\cS_n(\bk) = \int \frac{d^3 q}{(2\pi)^3} D_n(\bq,\bk-\bq) W_R(\bq) W_R(\bk-\bq) \delta_g(\bq)\delta_g(\bk-\bq),
\label{eq:ss_definition}
\ee
and the filtered field is given by $\tilde \delta_g(\bk) = h(\bk) \delta_g(\bk)$. In redshift space, the kernels $D_n$ can have a dependence on the direction with respect to the LoS. To make the notation simpler, in the rest of the paper, we drop the tilde of the filtered density field. 

Each quadratic operator $\mathcal S_n$ picks up a different combination of bias parameters and growth rate $f$. The explicit forms of the skew spectra were obtained in Ref. \cite{Schmittfull:2020hoi}. Here, we duplicate them for completeness,
\begin{align}\label{eq:skew_list}
    b_1^3: & \qquad \mathcal S_1 = F_2[\delta,\delta],\\
    b_1^2b_2: & \qquad \mathcal S_2 = \delta^2, \\
    b_1^2 b_{\mathcal G_2}: & \qquad \mathcal S_{3} = S^2[\delta,\delta],\\
    b_1^3f: & \qquad \mathcal S_4 = \hat z_i\hat z_j\,\partial_i\left(\delta\frac{\partial_j}{\nabla^2}\delta\right), \label{eq:S4}\\
    b_1^2f: & \qquad \mathcal S_5 = 2F_2[\delta^\parallel,\delta] + G_2^\parallel[\delta,\delta], \\
    b_1b_2 f: & \qquad  \mathcal S_6 = \delta\delta^\parallel, \\
     b_1 b_{\mathcal G_2} f: & \qquad \mathcal S_7 =  S^2[\delta,\delta^{\parallel}], \\
     b_1^2f^2: & \qquad \mathcal S_8 = \hat z_i\hat z_j\partial_i\left(\delta\frac{\partial_j}{\nabla^2}\delta^\parallel
    +2\delta^\parallel \frac{\partial_j}{\nabla^2}\delta
    \right),
    \\
    b_1f^2: & \qquad \mathcal S_9 = F_2[\delta^\parallel,\delta^\parallel] + 2 G_2^\parallel[\delta^\parallel,\delta],\\
    b_2f^2: & \qquad \mathcal S_{10} = \big(\delta^\parallel\big)^2,\\
    b_{\mathcal G_2} f^2: & \qquad  \mathcal S_{11} = S^2(\delta^{\parallel}, \delta^{\parallel}),\\
    b_1f^3: & \qquad \mathcal S_{12} = \hat z_i\hat z_j\partial_i\left(\delta^{\parallel\parallel}\frac{\partial_j}{\nabla^2}\delta
    +2\delta^\parallel \frac{\partial_j}{\nabla^2}\delta^\parallel
    \right),\\
    f^3: & \qquad \mathcal S_{13} = G_2^\parallel[\delta^\parallel,\delta^\parallel],\\
    f^4: & \qquad \mathcal S_{14} = \hat z_i\hat z_j\partial_i\left(\delta^{\parallel\parallel}\frac{\partial_j}{\nabla^2}\delta^\parallel \right).
\end{align}
In the above expressions, all products are in pixel space. We defined the redshift-space operators, 
\begin{align}
    \mathcal O^\parallel &= 
    \hat z_i\hat z_j\frac{\partial_i\partial_j}{\nabla^2}\mathcal O, \\
    \mathcal O^{\parallel\parallel} &=
    \hat z_i\hat z_j\hat z_m\hat z_n\frac{\partial_i\partial_j\partial_m\partial_n}{\nabla^4}\mathcal O,
\end{align}
and the operators $\mathcal O [a,b]$ that act on arbitrary fields $a$ and $b$, 
\be
  \label{eq:F2Def}
  {\mathcal O}[a,b](\bk) \equiv \int\frac{d^3 q}{(2\pi)^3}  \,\frac{1}{2}\Big[a(\bq)b(\bk-\bq)+b(\bq)a(\bk-\bq)\Big] \, {\mathcal O}(\bq,\bk-\bq). 
\ee

\vspace{0.1in}
\subsection{Shot Noise Component}\label{subsec:shot}
In addition to the clustering component described above, the skew spectra receive a shot noise contribution due to the discrete nature of halos and galaxies. Assuming the Poisson shot noise of the bispectrum, it is straightforward to derive the expressions for the skew spectra shot noise. Extending the calculation of real-space shot noise of the skew spectra \cite{Schmittfull:2014tca} to redshift space, for each of the 14 skew spectra, the (Poisson) shot noise is given by
\be 
\cP^{\rm shot}_{{\mathcal S_n}\delta}(k) = \frac{1}{2} \int d\mu_k \left\{ \left[\frac{1}{\nbar^2} + \frac{P_g(k,\mu_k)}{\bar{n}}\right] J_{D_n}(k,\mu_k) + \frac{2}{\bar{n}}\tilde{J}_{D_n}(k,\mu_k)\right\},
\label{eq:shot_th}
\ee
where $\mu_k = k_{\parallel}/k$ and
\begin{align}
    J_{D_n}(k,\mu_k) &= \int \frac{d^3q}{(2\pi)^3}W_R(\bq)W_R(\bk-\bq) D_n(\bq, \bk-\bq), \notag \\
    \tilde{J}_{D_n}(k,\mu_k) &= \int \frac{d^3q}{(2\pi)^3} W_R(\bq)W_R(\bk-\bq) D_n(\bq, \bk-\bq) P_g(q,\mu_q).
\end{align}
Deviations from the Poisson shot noise can be captured by including additional nuisance parameters in the $1/{\bar n}$ and $1/{\bar n}^2$ terms, in analogy with the bispectrum \cite{Schmittfull:2014tca,MoradinezhadDizgah:2019xun}.  

For the theoretical prediction of the shot noise, which we compare with the measured shot noise of the skew spectrum estimator, we assume the shot noise to be Poissonian, and use the linear Kaiser model of the redshift-space halo power spectrum,
\be
P_g(k,\mu_k) = (b_1 + f\mu_k^2)^2 P_{\rm lin}(k).
\ee

\subsection{Measurement Pipeline}\label{subsec:pipeline}
To measure the clustering components of the skew-spectra, we use the Python package \textsc{skewspec} \cite{Schmittfull:2020hoi} \href{https://github.com/mschmittfull/skewspec}{\faGithub}\footnote{\url{https://github.com/mschmittfull/skewspec}}, which employs \textsc{nbodykit} \cite{Hand:2017pqn} \href{https://github.com/bccp/nbodykit}{\faGithub}\footnote{\url{https://github.com/bccp/nbodykit}} to compute the quadratic fields $\cS_n$ and their cross-spectrum with the input halo/galaxy density fields. While the quadratic fields can be computed efficiently as the product of fields in configuration space at the same location $\bx$, applying the filters in Fourier space is more straightforward than taking derivatives of the real-space data. Therefore, to compute the quadratic fields, we Fourier transform two copies of the input density, apply the filters corresponding to each $\cS_n$ by multiplying each copy by appropriate factors in $k$, Fourier transform back to real space, and multiply the two fields there. In order to reduce the statistical noise on the measurements of the skew spectra, we extend the \textsc{skewspec} package to include the three LoS directions and average over the measurements along the three directions. 

Furthermore, we implement routines for measuring the shot noise of biased tracers. In performing the numerical Fisher forecasts, while for numerical derivatives, the shot noise cancels at the leading order, it can non-trivially enter the covariance and lead to artificial response when mocks have different number densities (for instance, when applying different halo mass-cut as discussed in figure~\ref{fig:2dcontour_ss_smooth10_shotonxoff_nuisMmin}). Therefore, we account for the contribution of the shot noise of the estimator to the covariance by subtracting the shot noise components from the skew spectra estimators. Below we describe in more detail how the shot noise estimator is constructed.

We define a shot-noise estimator based on the theoretical prediction in Eq.~\eqref{eq:shot_th}. 
To refrain from performing convolution in Fourier space, we define a unity field in Fourier space, $\mathbb{I}(\bk)$, on which the relevant filters are applied. With this, the estimator of the shot noise of the skew spectra, which we apply to simulated halo/galaxy catalogs, is given by
\begin{align}\label{eqn:shot_pipeline}
\av{{\hat \cP}^{\rm shot}_{{\mathcal S_n}\delta} (k)}&= \frac{1}{L^3} \int \frac{d\Omega_k}{4\pi} {\tilde \cO}^{\bf m''}_{n''}(-\bk) \Biggl\{
\left(\frac{1}{{\bar n}^2}+\frac{P_g(\bk)}{\bar{n}}\right)\ft{\cO^{\bf m m'}_{nn'}(\bx)} +\ft{\frac{2}{\bar n}\cQ^{\bf mm'}_{nn'}(\bx)} \Biggr\}, 
\end{align}
where $\mathcal{FT}$ refers to Fourier transform, we use the tilde symbol to denote quantities in Fourier space. Here, $\tilde{\cO}^{\bf m}_n(\bk)$ is the filtered unity field in Fourier space, $\cO^{\bf m m'}_{nn'}(\bx)$ is the filtered quadratic unity field in real space, and $\cQ^{\bf mm'}_{nn'}(\bx)$ is a filtered cubic field built from quadratic density field of the biased tracer and a unity field in real space. Here indices ${\bf m, m',m''}$ are 3-vectors, e.g. ${\bf m}=(m_x,m_y,m_z)$, and $n',n'$ and $n''$ are real numbers. These parameters specify the filtering of the unity and halo density fields by
\begin{align}
    {\tilde \cO}_n^{\bf m}(\bk) &\equiv k^n \bk^{\bf m} \mathbb{I}(\bk), \notag \\
     {\tilde \cQ}_n^{\bf m}(\bk) &\equiv k^n \bk^{\bf m} \cQ(\bk),
\end{align}
where the filtered 3D power spectrum $\cQ(\bk)\equiv W_R(\bk)P_g(\bk)$. We define the element-wise exponentiation of a 3D vector as
\be
    \bk^{\bf m} \equiv (k_x)^{m_x} (k_y)^{m_y} (k_z)^{m_z},
\ee
where e.g. $(k_x)^{m_x}$ denotes the $x$-component of $\bk$ raised to the power $m_x$. The powers $k^n$ are the square of the magnitude of the wavevectors. The quadratic fields are defined as 
\begin{align}
    {\tilde \cO}_{nn'}^{\bf m m'}(\bk) &\equiv \int \frac{d^3q}{2\pi^3} {\tilde \cO}_n^{\bf m}(\bq) {\tilde \cO}_{n'}^{\bf m'}(\bk-\bq) \notag \\
    \tilde{\cQ}_{nn'}^{\bf m m'}(\bk) &\equiv \int \frac{d^3q}{2\pi^3} {\tilde \cQ}_n^{\bf m}(\bq) {\tilde \cO}_{n'}^{\bf m'}(\bk-\bq).
\end{align}
These convolutions in Fourier space can be evaluated by taking the Fourier transform of the real-space products of the fields,
\begin{align}
    \cO_{nn'}^{\bf m m'}(\bx) &= \cO_{n}^{\bf m}(\bx) \cO_{n'}^{\bf m'}(\bx) \notag \\
    \cQ_{nn'}^{\bf m m'}(\bx) &= \cQ_{n}^{\bf m}(\bx) \cO_{n'}^{\bf m'}(\bx)
\end{align}

In the current implementation, measuring all 14 skew-spectra on a single CPU core takes about four minutes; the most time-consuming calculations are those involving the tidal Galileon operators, including $\mathcal{P}_{\cS_3\delta}$, $\mathcal{P}_{\cS_7\delta}$, and $\mathcal{P}_{\cS_{11}\delta}$ since they involve a nested loop over the three LoS directions. The shot noise measurements require almost the same amount of time. Although the quadratic unity fields only need to be computed once for the first two contributions in Eq.~\eqref{eqn:shot_pipeline}, the last term requires the operator to be applied to a 3D power spectrum-like field and a unity field, whereas the power spectrum varies from mocks to mocks. 

\section{Synthetic Datasets}\label{sec:sims}
In this section, we review the characteristics of the two sets of simulations that we use in our Forecasts, the Quijote halo catalogs~\cite{Villaescusa-Navarro:2019bje}, and the Molino galaxy catalogs~\cite{Hahn:2020lou}. We refer to the corresponding references for more details. 

\subsection{N-body Quijote Halo Catalogs}
The Quijote simulation suite \cite{Villaescusa-Navarro:2019bje} consists of 44,000 N-body simulations run with several different cosmological models to perform numerical Fisher forecasts and as training data for machine learning applications. The simulations have a box size of $L_{\rm box}=1 \ \gpch$ and are run with Gadget-III TreePM+SPH code~\cite{Springel:2005mi}, evolving the particles from initial condition at redshift $z=127$ to $z=0$. In this work, we use $15,000$ realizations of Quijote Suite with fiducial cosmology at ${\bm \theta}=\{\Omega_{\rm m}, \Omega_{\rm b}, h, n_s, \sigma_8\} = \{0.3175, 0.049, 0.6711, 0.9624, 0.834\}$ with total neutrino mass being zero. For each parameter we use a pair of 500 realizations at a step lower $\Delta\theta_i^-=\theta_i^--\theta_i$ and higher $\Delta\theta_i^+=\theta_i^+-\theta_i$ compared to the fiducial set. Additionally, we use a sets of simulations including massive neutrinos $\{M_{\nu}^+, M_{\nu}^{++}, M_{\nu}^{+++}\}$. The simulations with massless neutrinos have $512^3$ dark matter particles, and they were generated using initial conditions with second-order perturbation theory (2LPT) to compute particle displacements and peculiar velocities for a given input matter power spectrum. The simulations with massive neutrinos use the Zel'dovich approximation for generating the initial condition, where the $512^3$ neutrino particles and $512^3$ dark matter particles are both treated as collisionless and pressureless fluids. The Friend-of-Friend (FoF) halo finding algorithm \cite{Davis:1985rj} is used to identify halos, assuming a linking length of $b=0.2$. The step size for cosmological parameters are given by $\{\Delta\Omega_{\rm m}, \Delta\Omega_{\rm b}, \Delta h, \Delta n_s, \Delta \sigma_8\}=\{0.02, 0.004, 0.04, 0.04, 0.03\}$, with $\Delta\theta_i \equiv \Delta\theta_i^+-\Delta\theta_i^-$.

We apply a halo mass cut of $M_{\rm min}>3.2 \times 10^{13} M_{\odot}$, which corresponds to a mean halo density of $\bar{n}=1.55 \times 10^{-4}\, h^3 {\rm Mpc}^{-3}$. As in Ref. \cite{Coulton:2022rir}, for computing the derivatives, we take the average of the measurement along the three LoS directions, while for the covariance matrices, we only use the projection along a single LoS to avoid biasing the covariance estimation by the correlations between different LoS directions. 

\subsection{HOD-based Molino Galaxy Catalogs}

The Molino suite of galaxy catalogs~\cite{Hahn:2020lou} consists of 75,000 mocks that are constructed upon the application of a standard 5-parameter HOD model to Quijote halo catalogs. In general, a HOD model provides a prescription for building the relationship between the underlying dark matter and the biased tracers; the halos are populated by galaxies such that the mean number of galaxies is given by the sum of central and satellite galaxies,
\be
\langle N_{\rm gal} \rangle = \langle N_{\rm cent} \rangle + \langle N_{\rm sat} \rangle.
\ee
In the standard five-parameter HOD model \cite{Zheng:2007zg}, used in Molino catalogs, the mean occupation function of the central galaxies can be described by a step-like function with a soft cutoff profile to account for the scatter between galaxy luminosity and host halo mass. The mean satellite galaxy occupation is often modeled as a power law form at high halo masses (with the slope close to unity) and drops steeper than the power law at lower masses;
\be
 \langle N_{\rm cent} \rangle = \frac{1}{2}\left[1+{\rm erf}\left(\frac{\log M_h - \log M_{\rm min}}{\sigma_{\log M}} \right)\right], \qquad \langle N_{\rm sat} \rangle = \langle N_{\rm cent} \rangle\left(\frac{M_h-M_0}{M_1}\right)^\alpha.
\ee
The two free parameters of the central distribution are the minimum mass scale of halos that can host central galaxies above the luminosity threshold $M_{\rm min}$, and the width of the cutoff profile $\sigma_{\log M}$. The three free parameters of the satellite galaxies are the mass scale $M_0$ at which the mean occupation of satellites drops faster than a power law, $M_1$ characterizes the amplitude of the satellite mean occupation function, and $\alpha$ is the asymptotic slope at high halo mass. Central galaxies are placed at the center of the halo, while the spatial distribution of satellite galaxies inside halos is assumed to follow the NFW profile~\cite{Navarro:1996gj}. 

The fiducial cosmology of Molino catalogs assumes the following values of the HOD parameters, $\left\{ \log M_{\rm min}, \sigma_{\log M}, \log M_0, \alpha, \log M_1 \right\} = \left\{13.65, 0.2, 14.0, 1.1, 14.0 \right\}$. To compute the numerical derivatives with respect to the HOD parameters, the Molino suite additionally includes mocks with a variation of a single parameter above and below the fiducial values of HOD parameters applied to 500 Quijote boxes in the fiducial cosmology. The step size of these variations is $\left\{ \Delta \log M_{\rm min}, \Delta \sigma_{\log M}, \Delta \log M_0, \Delta \alpha, \Delta \log M_1 \right\} = \left\{ 0.05, 0.02, 0.2, 0.2, 0.2 \right\}$. To obtain more accurate measurements of the derivatives, each simulation box is populated with galaxies with five different initial seeds. The redshift-space measurements are performed along the three LoS directions. Therefore, the derivatives used in the Fisher forecasts are the average over 15 measurements. As in Quijote  measurements, the covariance matrices are obtained from the simulations with the fiducial cosmology along a single LoS. 

\section{Forecasting Methodology}\label{sec:fisher}

The accuracy with which a given dataset can constrain cosmological and nuisance parameters are commonly estimated using the Fisher information matrix formalism. In general the Fisher matrix is defined as \cite{Tegmark:1996bz,Tegmark:1997rp}
\be
    F_{\alpha \beta} = -\left\langle \frac{\partial^2 \mathcal{L(\bd|{\bm \theta})}}{\partial\theta_\alpha\partial\theta_\beta}\right\rangle,
\ee
where $\cL$ is the likelihood of the data ${\bd} $ given the parameters ${\bm \theta}$. Assuming that the likelihood $\cL$ is Gaussian, the forecasted marginalized uncertainty on the $\alpha$-th parameter is given by $\sigma^2(\theta_\alpha) = (F^{-1})_{\alpha \alpha}$. As a consequence of the Cramér-Rao inequality, for an unbiased estimator, this uncertainty is always larger than the unmarginalized uncertainty $1/\sqrt{F_{\alpha \alpha}}$. For multivariate Gaussian probability distribution with nonzero mean and covariance that is independent of parameters of interest, the Fisher matrix is further simplified and can be written in terms of the change of the mean of the data with respect to the model parameters and the data covariance. 

In our analysis, we consider the data vector consisting of the 14 redshift-space skew spectra in the selected $k$-bins, $\bd \equiv \left\{ {\bm\cP}_{\cS_1 \delta}, ...,  {\bm\cP}_{\cS_{14} \delta} \right\}$. When exploring the constraining power of power spectrum and skew spectra combined, the data vector additionally consists of power spectrum multipoles, $\bd \equiv \left\{ {\bm\cP}_{\cS_1 \delta}, ...,  {\bm\cP}_{\cS_{14}\delta}, {\bf P}_0, {\bf P}_2, {\bf P}_4 \right\}$, where ${\bf P}_n $ are the data vectors power spectrum multipoles of different $k$-bins, respectively. The shot noise is subtracted in the estimators for both skew spectra and power spectrum multipoles\footnote{As will be described later, in our analysis of the Molino galaxies, we do not subtract the shot noise contributions of the skew spectra.}. The Fisher information matrix for the above data vectors is then given by
\be\label{eq:fisher}
F_{\alpha\beta} = \sum_{i,j = 1}^{N_{\rm d}} \frac{\partial d_i}{\partial \theta_\alpha} \mathbb{C}^{-1}_{ij}  \frac{\partial d_j}{\partial\theta_\beta}\, ,
\ee
where $N_{\rm d}$ is the total size of the data vector in consideration. Binning the skew spectra and the power spectrum multipoles in $N_{\rm b}$ bins of width $\Delta k$, we have $N_{\rm d} = 14 N_{\rm b}$ for skew spectra only and $N_{\rm d} = 14 N_{\rm b} + 3 N_{\rm b}$ for the joint data vector of skew spectra and power spectrum multipoles. The covariance matrix in Eq.~\eqref{eq:fisher} is the full $N_{\rm d} \times N_{\rm d}$ covariance matrix for the data vector $\bf d$ defined by
\be
\mathbb{C}_{ij} \equiv 
\frac{1}{N_{\rm s}-1}\sum_n^{N_{\rm s}} \left(d_i^{(n)}-\bar{d}_i\right)\left(d_j^{(n)}-\bar{d}_j\right),
\ee
with $N_{\rm s}$ being the number of mocks. We correct the inverse of the covariance by the Hartlap factor~\cite{Hartlap:2006kj} given that the covariance is estimated from a limited number of mocks. We use the Quijote and the Molino suites to compute the derivatives and the covariance matrix numerically.

For cosmological parameters, $\{\Omega_{\rm m}, \Omega_{\rm b}, h, n_s, \sigma_{8}\}$, and the HOD parameters, we evaluate the derivatives from the averaged measurements of the skew spectra and power spectrum over realization at $\theta^+$ and $\theta^-$,
\be
 \frac{\partial d_i}{\partial \theta_i} = \frac{d_i(\theta^+) - d_i(\theta^-)}{\theta^+ - \theta^-}.
\ee
For $M_\nu$, since the fiducial value is zero and negative masses do not have any physical meaning, we use simulations with $\{M_{\nu}^{+}, M_{\nu}^{++}, M_{\nu}^{+++}\}=\{0.1\, {\rm eV}, 0.2\, {\rm eV}, 0.4\,{\rm eV}\}$, and estimate the derivatives using
\be
    \frac{\partial d_i}{\partial M_{\nu}} = \frac{-21 d_i(\theta_{\rm fid}^{\rm ZA})+32 d_i(M_{\nu}^+)-12 d_i(M_{\nu}^{++})+d_i(M_{\nu}^{+++})}{12\,\delta M_{\nu}},
\ee
which is accurate up to the second order in $\mathcal{O}(\delta M_{\nu}^2)$\footnote{See Ref.~\cite{Massara:2022zrf} for more detailed discussion of the different estimators for neutrino derivatives.}. For the Quijote halos, we add the minimum halo mass-cut $M_{\rm min}$ as a nuisance parameter to account for an unknown effective bias parameter. We take $M^+_{\rm min} = 3.3 \times 10^{13}\, h^{-1}M_\odot$ and $M^-_{\rm min} = 3.1 \times 10^{13}\, h^{-1}M_\odot$. In Appendix~\ref{app:convergence} we study the stability of the numerical derivatives and the covariance matrix estimation from a finite number of mocks.

We set the $k$-bin width to be the fundamental frequency of the simulation box with side length $L$, {\it i.e.}, $k_f=2\pi/L$ and the minimum $k$-bin edge to be half of the fundamental mode.
We chose the maximum $k$-value for our base analysis to be $k_{\rm max} = 0.25\, \hmpc$. Therefore  we have $N_{\rm b} = 39$ for each of the observables in the data vector. Our base analysis assumes the smoothing scale of $R=10\, \mpch$, which is motivated by half of the average inter-particle separation distance $\av{\Delta r}\sim 18\,\mpch$ for the Quijote halo catalogs. To test the dependence of the constraints on the smallest scales included in the forecasts, we also present the results for the smoothing scales of $R= 20\, \mpch$, and a range of other values of $k_{\rm max}$.

\section{Results}\label{sec:res}

In this section, we first present the measurements of the skew spectra and their covariance matrix. Next, we investigate the impact of the non-Gaussian off-diagonal elements of the covariance matrix due to mode coupling on the signal-to-noise ratio (SNR) of individual skew spectra. Finally, we present the forecasted parameter constraints for both Quijote halo and Molino galaxy samples from skew spectra alone and in combination with power spectrum multipoles. For the covariance estimation, we measure 15,000 skew spectra at fiducial cosmology. This is the same for both Quijote and Molino. For the derivative estimation, we measure 500 pairs of simulations left- and right-ward of the reference cosmology for 6 cosmological parameters. In addition, for Quijote, we also carry out measurement using two different halo mass cuts; while for Molino, instead of the halo mass cuts we measure the skew spectra with the 5 HOD parameters. In summary, at each smoothing scale, we measure a total of $24,000 + 15,000 = 39,000$ skew spectra for Quijote and $180,000 + 15,000 = 195,000$ for Molino.

\subsection{Measured Skew Spectra and Their Covariance Matrix}
\begin{figure}[t]
    \centering
    \includegraphics[width=0.86\textwidth]{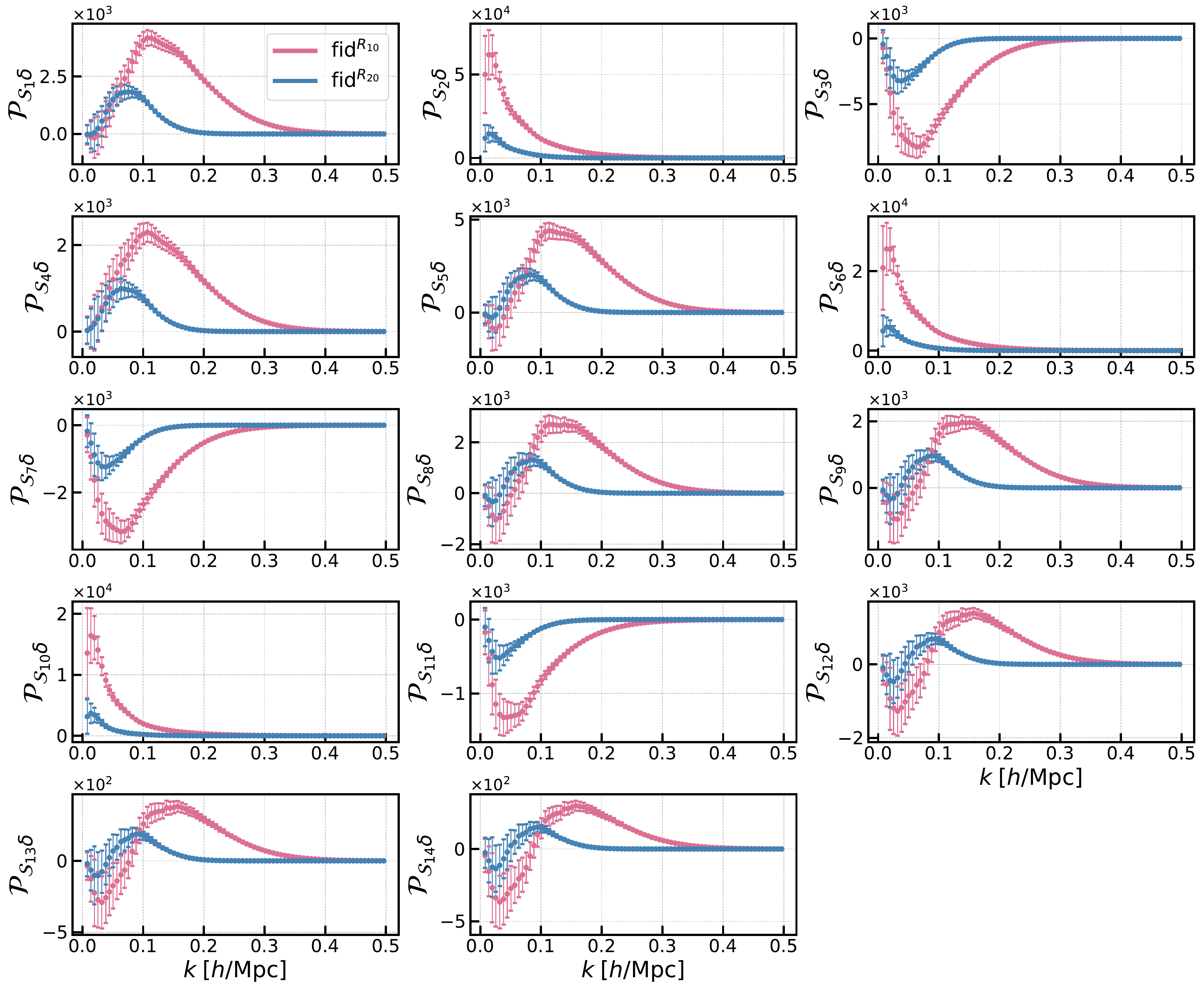}
    \caption{Measured skew spectra with fiducial cosmology on Quijote halo catalogs setting smoothing scales of  $R=10\, \mpch$  (red) and $R=20\, \mpch$ (blue). The error bars correspond to the variance of the spectra, averaged over 15,000 simulations at reference cosmology.}
    \label{fig:quijote_ss_smooth10x20_shoton_fid} 
\end{figure}

In figure \ref{fig:quijote_ss_smooth10x20_shoton_fid} we show the measurements of the 14 halo skew spectra at fiducial cosmology of Quijote simulations, applying smoothing scales of $R=10\, \mpch$ (red) and $R=20\, \mpch$ (blue). The error bars correspond to the variance of the spectra, averaged over 15,000 simulations at reference cosmology. As expected, for the larger smoothing scale, all skew spectra drop faster towards higher-$k$ modes, and almost all vanish when approaching $k_{\rm max}\sim 0.3\, \hmpc$. When using a smaller smoothing scale, the overall amplitudes of the skew spectra increase and the peak positions shift towards smaller scales. The shape of skew spectra can be classified into three categories: (i) Constant type with density square $\delta^2$: $\mathcal{P}_{\mathcal{S}_2\delta}$, $\mathcal{P}_{\mathcal{S}_6\delta}$, $\mathcal{P}_{\mathcal{S}_{10}\delta}$, which are characterized by a peak on very large scales and a relatively sharp fall off towards smaller scales.
(ii) Displacement type with operator $\partial_i\partial_j/\nabla^2$: $\mathcal{P}_{\mathcal{S}_1\delta}$, $\mathcal{P}_{\mathcal{S}_{4-5}\delta}$, $\mathcal{P}_{\mathcal{S}_{8-9}\delta}$, $\mathcal{P}_{\mathcal{S}_{12-14}\delta}$, which are characterized by a positive ``bump". For the operators that have dependence on the LoS, the resulting skew spectra have a ``zero-crossing" feature on large scales.
(iii) Tidal type with operator $S^2$: $\mathcal{P}_{\mathcal{S}_3\delta}$, $\mathcal{P}_{\mathcal{S}_7\delta}$, $\mathcal{P}_{\mathcal{S}_{11}\delta}$, these skew spectra all have a negative ``bump" feature and are anti-correlated with the rest of skew spectra on small scales.
The three categories are intrinsically related to the real-space skew spectra (see figure 1 in~\cite{Schmittfull:2014tca}). Despite the similarity within the three groups, there are high correlations among these 14 skew spectra, and each of them contributes almost equally to the total information content, and we can not simply truncate the data vector by selecting only a subset of them. However, the data vector of skew spectra can be potentially optimally compressed by finding combinations of different spectra that maximize the Fisher information. We defer further investigation of this compression to future works.

\begin{figure}[t]
     \centering
    \includegraphics[width=.47\textwidth]{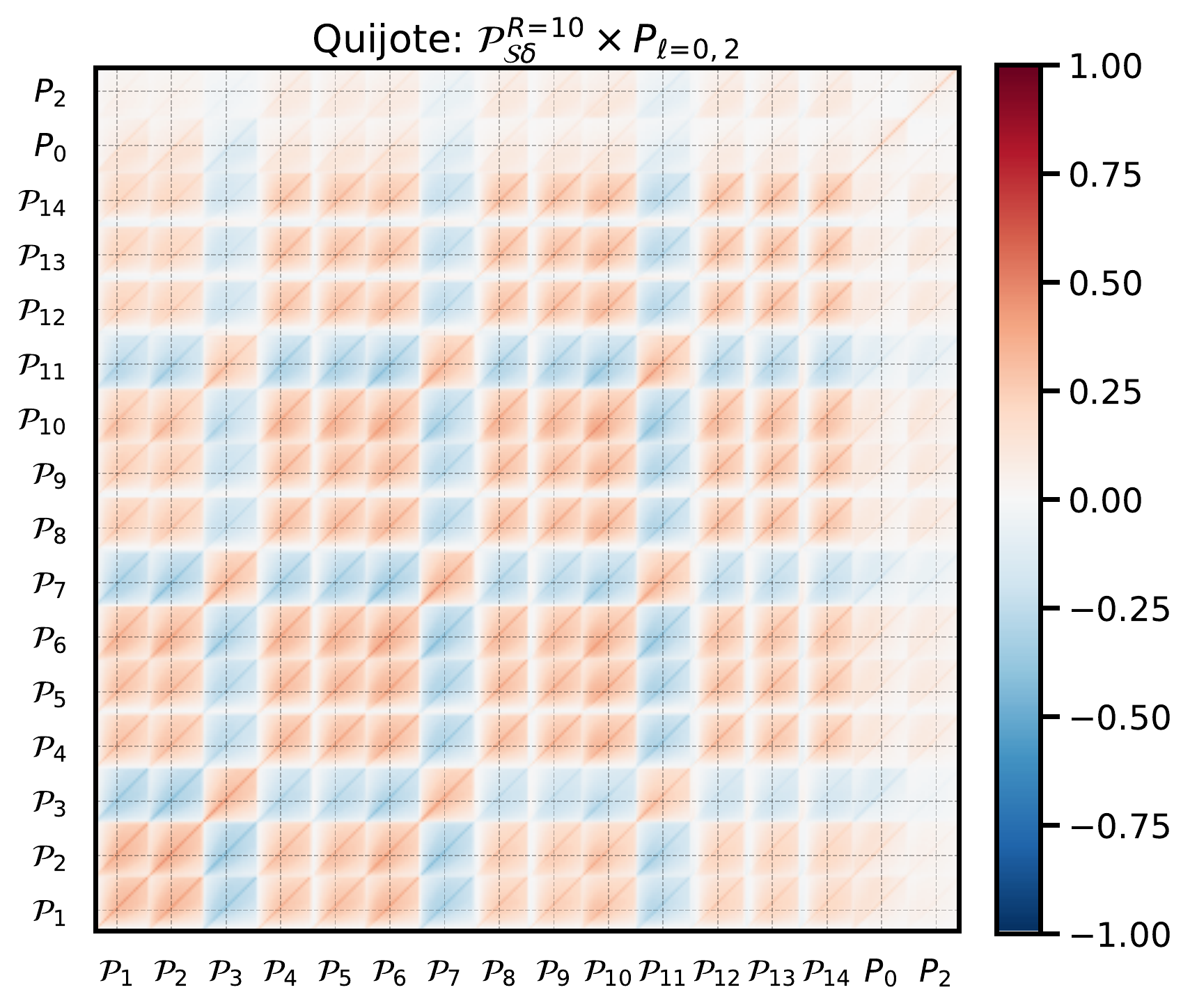}
    \includegraphics[width=.47\textwidth]{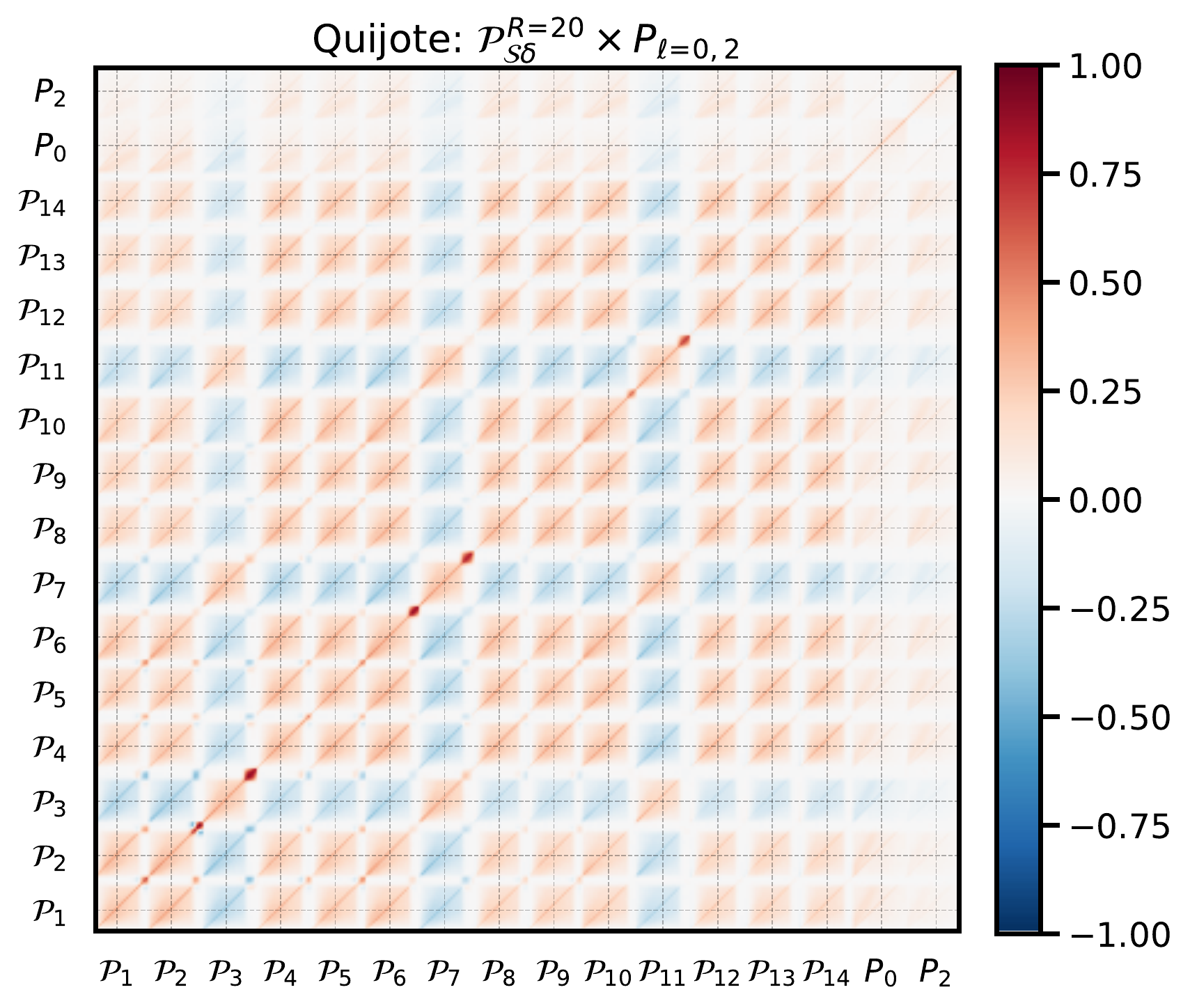}
    \caption{Correlation matrix of the combined data vector $\mathbf{d}\equiv \{\mathcal{P}_{\cS_1\delta}, \ldots \mathcal{P}_{\cS_{14}\delta}, P_0, P_2\}$ for halo skew spectra and power spectrum multipoles setting $k_{\rm max}=0.5\, \hmpc$ measured from Quijote simulations. On the left, we show the combined data vector assuming a smoothing scale of $R=10\, \mpch$ for the skew spectra, while on the right, we used $R=20\, \mpch$. Each block denotes the covariance between the $i$-th component in the combined data vector with the $j$-th component. Within each block, the structure shows the covariance for different wave numbers $k$.}
    \label{fig:quijote_covariance_smooth10x20_shoton_fid}
\end{figure}
Having a large number of mocks of the Quijote simulation suite, we estimate the full covariance matrix of the redshift-space skew spectra and the power spectrum multipoles for the first time. Figure \ref{fig:quijote_covariance_smooth10x20_shoton_fid} displays the joint correlation matrix of the lowest two power spectrum multipoles ($\ell=0,2$) and the skew spectra with $R=10\,\mpch$ (left panel) and $R=20\,\mpch$ (right panel). Each block denotes the (cross-)covariance structure between two skew spectra $\mathcal{P}_{\cS_i\delta}$ with $\mathcal{P}_{\cS_j\delta}$ for $i, j=\{1,\ldots,14\}$, between power spectrum multipoles $P_\ell$ for $\ell =\{ 0,2\}$, and between $P_\ell$ and $\mathcal{P}_{\cS_i\delta}$. Within each block, the structure shows the covariance for different wave numbers, i.e., the mode coupling contributions. The white bands correspond to scales where the covariance vanishes since smoothing the field washes out small-scale fluctuations. The covariance for the larger smoothing scale goes to zero at $k \sim\!  0.3\, \hmpc$, the scale at which the skew spectra themselves approach zero as shown in figure \ref{fig:quijote_ss_smooth10x20_shoton_fid}. The dark blobs appearing in the white bands on the right plot are numerical artifacts from the fact that to compute the correlation matrix, we divided the covariance matrix elements by diagonal elements, which on those scales approach zero. Overall, the correlations of the skew spectra and power spectrum multipoles are smaller than the cross-correlations between different skew spectra. Unlike the power spectrum, the skew spectra can have positive or negative signs, as shown in figure \ref{fig:quijote_ss_smooth10x20_shoton_fid}. Therefore, different spectra can be correlated (red blocks) or anti-correlated (blue blocks), depending on the sign of the skew spectra. We note that for Quijote halos, we only include the power spectrum monopole and quadrupole for this paper. This is because the hexadecapole for Quijote halos has negligible signal-to-noise. As we will discuss later in this section, this is not the case for Molino galaxies. Therefore, in Molino analysis, we include the hexadecapole. 
\begin{figure}[t]
    \centering
\includegraphics[width=0.55\textwidth]{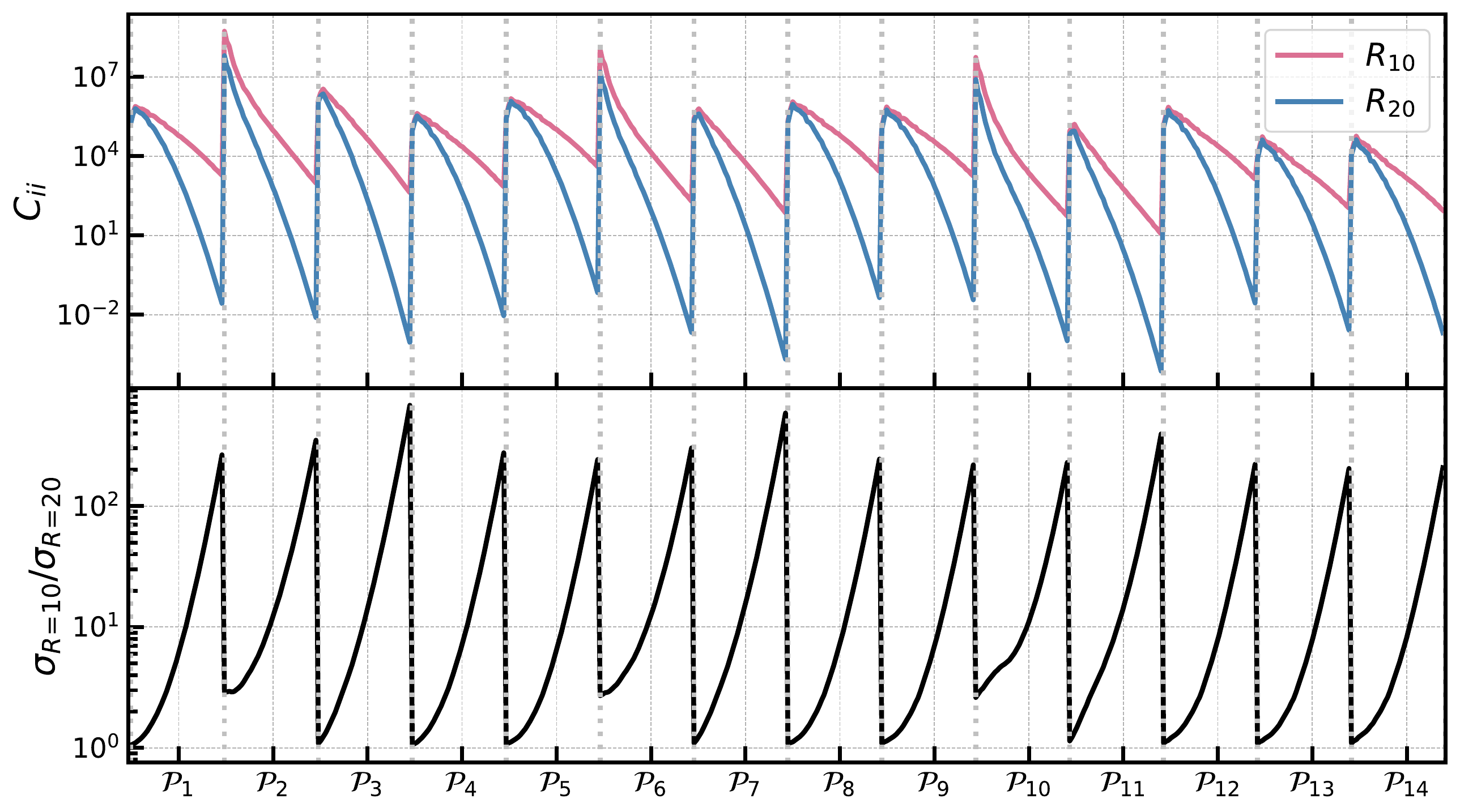}
    \caption{{\it Upper panel}: Diagonal of the 14 skew spectra covariance matrix for smoothing scale $R=10\, \mpch$ (red) and $=20\, \mpch$ (blue) with $k_{\rm max}=0.25\, \hmpc$. {\it Lower panel}: square root ratio of the diagonal elements of 14 skew spectra covariance matrix for the two smoothing scales.}
\label{fig:quijote_diag_cov_smooth10x20_shoton_fid} 
\vspace{.1in}
\end{figure}
To better compare the relative size of the covariance matrices for the two smoothing scales, in the top panel of figure \ref{fig:quijote_diag_cov_smooth10x20_shoton_fid}, we show the diagonal elements of the covariance matrix of the skew spectra for $R=10\, \mpch$ in red and $R=20\, \mpch$ in blue. The square root of the ratio of the two is shown in the bottom panel. The points lying between each of the two vertical dotted lines correspond to a given skew spectrum with wavenumbers increasing from left to right. Overall, the skew spectra with a smaller smoothing scale have a larger variance, with a slower drop towards the smaller scales. This is due to the fact that applying a large smoothing scale washes away the fluctuations on small scales. While this trend is seen for all skew spectra, the scale dependencies of the variances of different spectra differ from one another, reflecting the difference in their shapes shown in figure \ref{fig:quijote_ss_smooth10x20_shoton_fid}. 

\subsection{Impact of the Non-Gaussian Covariance}
Before presenting the parameter constraints, we investigate the importance of the off-diagonal contributions\footnote{The non-Gaussian contributions to diagonal elements are accounted for in the measured covariances.} to the covariance matrix, focusing on the signal-to-noise ratio (SNR) of individual skew spectra. While the impact of off-diagonal elements for the halo/galaxy power spectrum and bispectrum has been previously studied ({\it e.g.}~\cite{Sato:2013mq,Kayo:2012nm,Chan:2016ehg,Sugiyama:2018yzo}), here we present it for the first time for the skew spectra. It is worth noting that the off-diagonal elements of the skew spectra covariance matrix receive both Gaussian and non-Gaussian contributions. This was explicitly shown in the theoretical prediction of the covariance matrix of real-space skew spectra using tree-level perturbation theory \cite{Schmittfull:2014tca, MoradinezhadDizgah:2019xun}. Even if retaining only the diagonal Gaussian contributions of the bispectrum covariance, the skew spectra (even on large scales) have non-vanishing off-diagonal elements. 

\begin{figure}
    \centering
\includegraphics[width=0.85\textwidth]{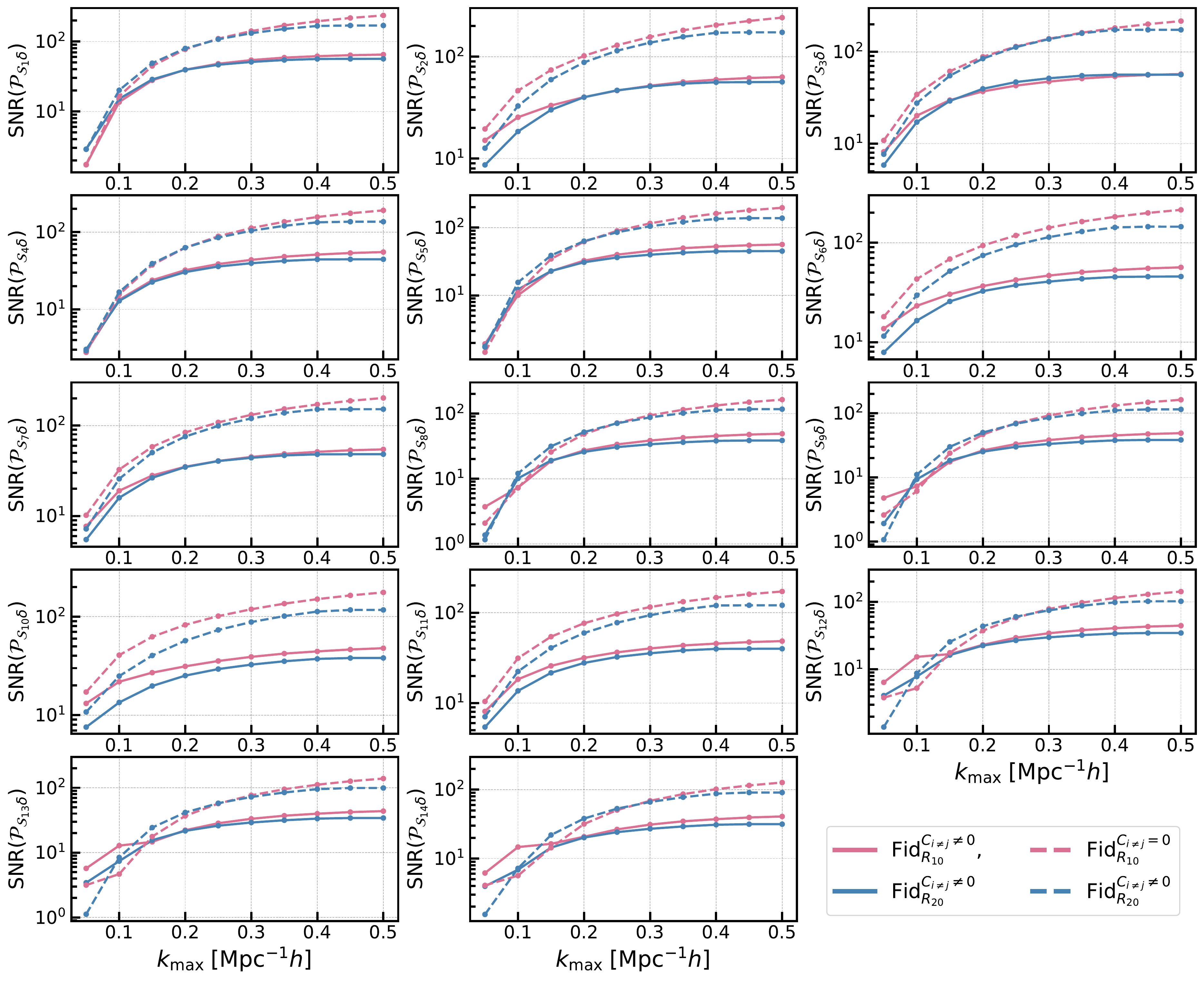}
    \caption{The signal-to-noise ratio of individual skew spectra given by ${\rm SNR}=(\mathcal{P}_{\mathcal{S}_i\delta} C^{-1}_{ij} \mathcal{P}_{\mathcal{S}_j\delta})^{1/2}$ for $R=10\, \mpch$ (red) $R=20\, \mpch$ (blue). The solid lines are computed using the full covariance matrix, while only the diagonal elements are included for the dashed lines.}
    \label{fig:quijote_snr_smooth10x20_shoton_Mnu}
\end{figure}

In figure \ref{fig:quijote_snr_smooth10x20_shoton_Mnu}, we show the SNR for each of the skew spectra as a function of small-scale cutoff using $R=10\, \mpch$ (red) and $R=20\, \mpch$ (blue). The dashed lines show the SNR assuming the covariance matrix to be diagonal, while the solid lines include the off-diagonal elements. As expected, the diagonal covariance approximation artificially enhances the SNR. When accounting for the mode coupling in the covariance, the SNR grows slower\footnote{This saturated trend of the SNR of the skew spectra resembles one of the power spectra in figure 19 of Ref.~\cite{Chan:2016ehg}.} and nearly levels off at $k_{\rm max} \sim 0.4\, \hmpc$. Going from $k_{\rm max}=0.3 \, \hmpc$ to $k_{\rm max}=0.4 \, \hmpc$, there is an increase by 25\% in signal-to-noise ratio when using the full covariance, while an increase by 54\% under the diagonal covariance approximation.
This trend is the same for both values of $R$.

The impact of the off-diagonal covariance element on the SNR can be clearly understood with an analytic approximation, as was done for the power spectrum covariance in Ref.~\cite{Carron:2014hja}. We assume that the covariance matrix $\mathbb{C}_{[i]}$ of the $i$-th skew spectrum can be decomposed into a diagonal and an off-diagonal components, $\mathbb{C}_{[i]}\equiv\mathbb{D}+ \sigma_{\rm min}^2 {\calP_{\calS_{i}\delta}}{\calP_{\calS_{i}\delta}}^T$, with $\mathbb{D}$ being the diagonal part for which $\sigma_{\rm min}^2$ is zero. In the case of power spectrum, $\sigma^2_{\rm min}$ has the interpretation of minimum achievable variance \cite{Carron:2014hja}, while for the skew spectra, the definition can be extended to the cumulants of $\av{\delta^3(x_1)\delta^3(x_2)}_c \propto \sigma^8\xi(x_1-x_2)$ \cite{Bernardeau:1995ty}. Using the Sherman-Morrison-Woodbury formula~\cite{Sherman:1950, Woodbury:1950}, the inverse covariance can be expanded in terms of the inverse of the diagonal part plus a correction,
\begin{eqnarray}
    \mathbb{C}_{[i]}^{-1} \equiv \left(\mathbb{D}+ \sigma_{\rm min}^2{\calP_{\calS_{i}\delta}}{\calP_{\calS_{i}\delta}}^T\right)^{-1} = \mathbb{D}^{-1} - \sigma_{\rm min}^2\frac{\mathbb{D}^{-1} {\calP_{\calS_{i}\delta}}{\calP_{\calS_{i}\delta}}^T\mathbb{D}^{-1}}{1+ \sigma_{\rm min}^2{\calP_{\calS_{i}\delta}}^T \mathbb{D}^{-1} {\calP_{\calS_{i}\delta}}}.
    \label{eqn:SM_expansion}
\end{eqnarray}
Therefore, the SNR for the full covariance, $({\rm S}/{\rm N})^2$, is related to SNR in the limit of diagonal covariance, $({\rm S}/{\rm N})^2_G$, as  \cite{Carron:2014hja}
\begin{eqnarray}
    ({\rm S}/{\rm N})^2 = \frac{({\rm S}/{\rm N})^2_G}{1+\sigma_{\rm min}^2({\rm S}/{\rm N})^2_G}.
\end{eqnarray}
It is thus clear that including the non-diagonal part of the covariance denoted by the term associated with $\sigma_{\rm min}^2$ always decreases the SNR.  

Our results for the SNR align with the Fisher forecasts presented in Ref.~\cite{MoradinezhadDizgah:2019xun}, which showed that neglecting the off-diagonal elements significantly affects the parameter constraints, underestimating the expected uncertainties. Therefore, our forecasts presented below use the full covariance matrix shown in figure \ref{fig:quijote_covariance_smooth10x20_shoton_fid}.

\subsection{Forecasted Parameter Constraints}
\label{subsec:forecast_param}

\vspace{0.05in}
\underline{Quijote Halo Catalogs}
\vspace{0.05in}

\begin{figure}[t]
    \centering
    \includegraphics[width=0.8\textwidth]{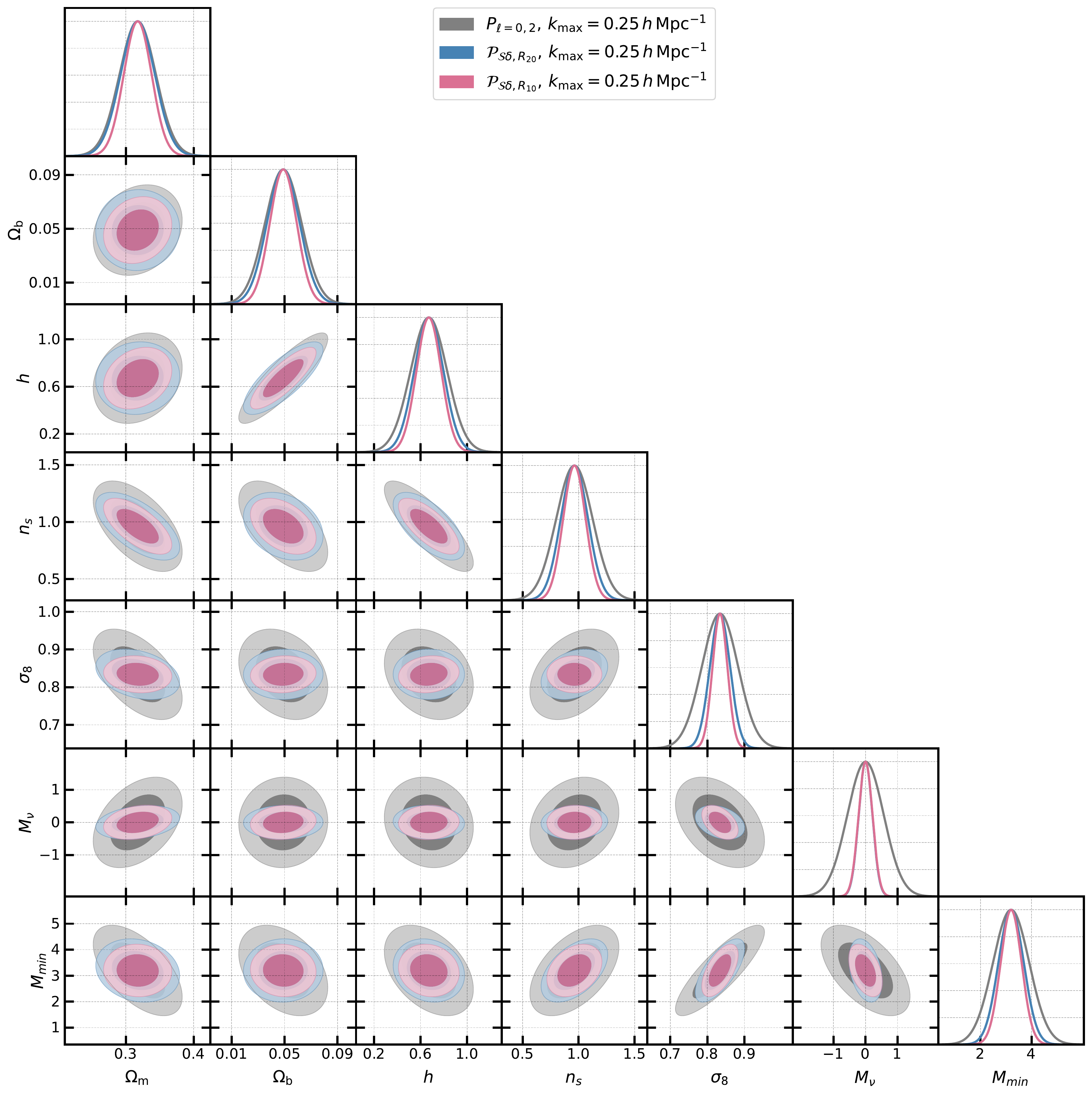}
    \caption{Marginalized 1- and 2-$\sigma$ parameter constraints from Quijote halo power spectrum multipoles (grey), and skew spectra with two smoothing scales of $R=10\,\mpch$ (blue) and $R=20\, \mpch$ (red). The small-scale cutoff is set to $k_{\rm max} = 0.25 \ \hmpc$ in all cases.}
    \label{fig:2dcontour_pk_ss_smooth10x20}\vspace{0.2in}
\end{figure}
\noindent To compare the information content of the skew spectra with that of the power spectrum multipoles ($\ell = 0,2$), in figure \ref{fig:2dcontour_pk_ss_smooth10x20}, we show the 2D marginalized constraints on cosmological parameters and minimum halo mass (as a proxy for the unknown halo bias) from the full skew spectra data vector. The gray contours correspond to the power spectrum multipoles, while the blue and red ones are from the skew spectra with $R=20\, \mpch$ and $R=10\, \mpch$, respectively. The small-scale cutoff is set to $k_{\rm max} = 0.25\, \hmpc$ in all three cases. Compared to the power spectrum multipoles, the skew spectra improve the constraints on all parameters by $\sim (23-62)\%$, with the improvements of $M_\nu$ and $\sigma_8$ being the two most prominent. The choice of the smoothing scale has nearly no impact on the constraint on the total mass of neutrinos. On the contrary, the constraint on $\sigma_8$ improves by $\sim 35\%$ when reducing the smoothing scale from $R=20\ \mpch$ to $R=10\ \mpch$. For other parameters, the improvement is at most $\sim\!20\%$. The insensitivity of the neutrino constraints to the choice of smoothing scale can be better understood by inspecting the shape of the normalized derivatives of skew and power spectra as a function of wavenumber. As shown in figure~\ref{fig:quijote_response_smooth10x20_Mnu} of Appendix~\ref{app:responses}, the responses (which are the inverse-variance normalized derivatives) for both smoothing scales are of similar amplitudes. For individual skew spectra, the off-diagonal mode-coupling contributions of the covariance counteract the small difference between the response functions for the two choices of smoothing scale such that the parameter constraints, shown in figure \ref{fig:1sigma_kmax_ind_smoothing_mnu}, are nearly unaffected by the value of smoothing scale. When combining all skew spectra, the constraints become even more comparable, presumably due to additional covariance between different skew spectra.

\begin{figure}[t]
    \centering
    \includegraphics[width=0.9\textwidth]{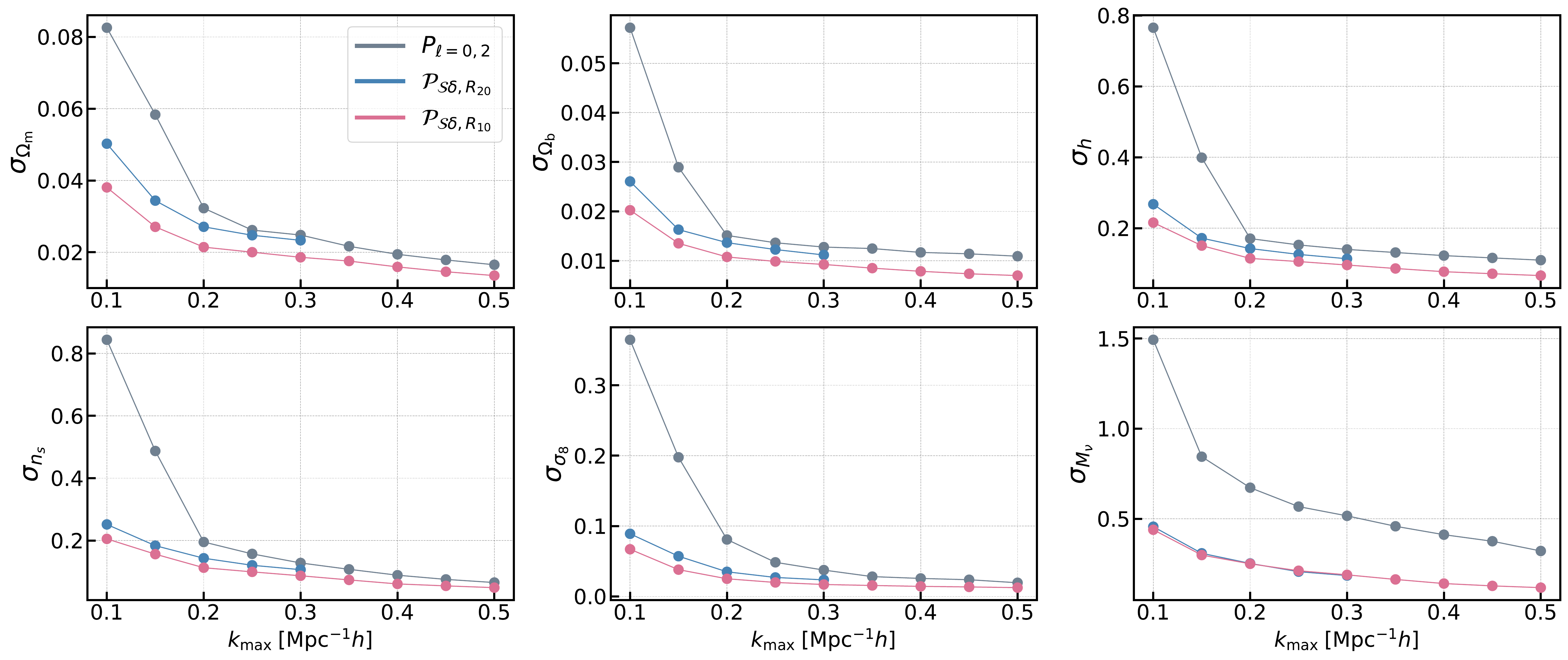}
    \caption{Marginalized 1-$\sigma$ parameter constraint from power spectrum multipoles (grey), skew spectra with smoothing scales of $R=10\,\mpch$ (blue) and $R=20\, \mpch$ (red), as a function of small-scale cutoff $k_{\rm max}$.}
    \label{fig:quijote_1sigma_ss_smooth10x20_shotoff_pk02_nuisMmin}
 \end{figure}
In the results above, we have applied a conservative small-scale cutoff, regardless of the smoothing scale. We illustrate the dependence of the constraints on the choice of $k_{\rm max}$ in figure~\ref{fig:quijote_1sigma_ss_smooth10x20_shotoff_pk02_nuisMmin}. The magenta and blue lines show the results from skew spectra with $R=10\, \mpch$ and $R=20\, \mpch$, while the grey lines are from the power spectrum multipoles. For $R=20\, \hmpc$, we only plot the errors up to $k_{\rm max} = 0.3\, \hmpc$ since beyond this scale the covariance is nearly vanishing and the Fisher matrix is not reliable. On the largest scale (the lowest values of $k_{\rm max}$), the skew spectra provide remarkably better constraints on all cosmological parameters compared to the power spectrum multipoles. This is not unexpected since being a correlation between composite quadratic fields and the original halo field, the non-Gaussian information from small-scale fluctuations gets imprinted on the large-scale skew spectra. Therefore, even when imposing the small-scale cutoff of $k_{\rm max}$, modes smaller than this cutoff still contribute to the observed skew spectra due to convolution in the quadratic field. For $M_\nu$, the skew spectra consistently provide tighter constraints than the power spectrum for all choices of $k_{\rm max}$ (ranging from a factor of 2.5 to 3 better). For $\Omega_b$ and $h$, the ratio between the parameter constraints from skew and power spectra stays nearly constant for $k_{\rm max} \gtrsim 0.2 \, \hmpc$, while for $\Omega_m, n_s$ and $\sigma_8$ the skew spectra and power spectrum constraints approach one another increasing the $k_{\rm max}$. These conclusions may differ when considering a more complex bias model. Breaking degeneracies between cosmological and nuisance bias parameters can lead to even more superior constraints from the skew spectra. This is because they capture the information in higher-order statistics not imprinted on the power spectrum. We investigate this by considering the Molino galaxy sample and varying the HOD parameters. The discussion will follow later in this section. 

\begin{figure}[t]
    \centering
\includegraphics[width=0.8\textwidth]{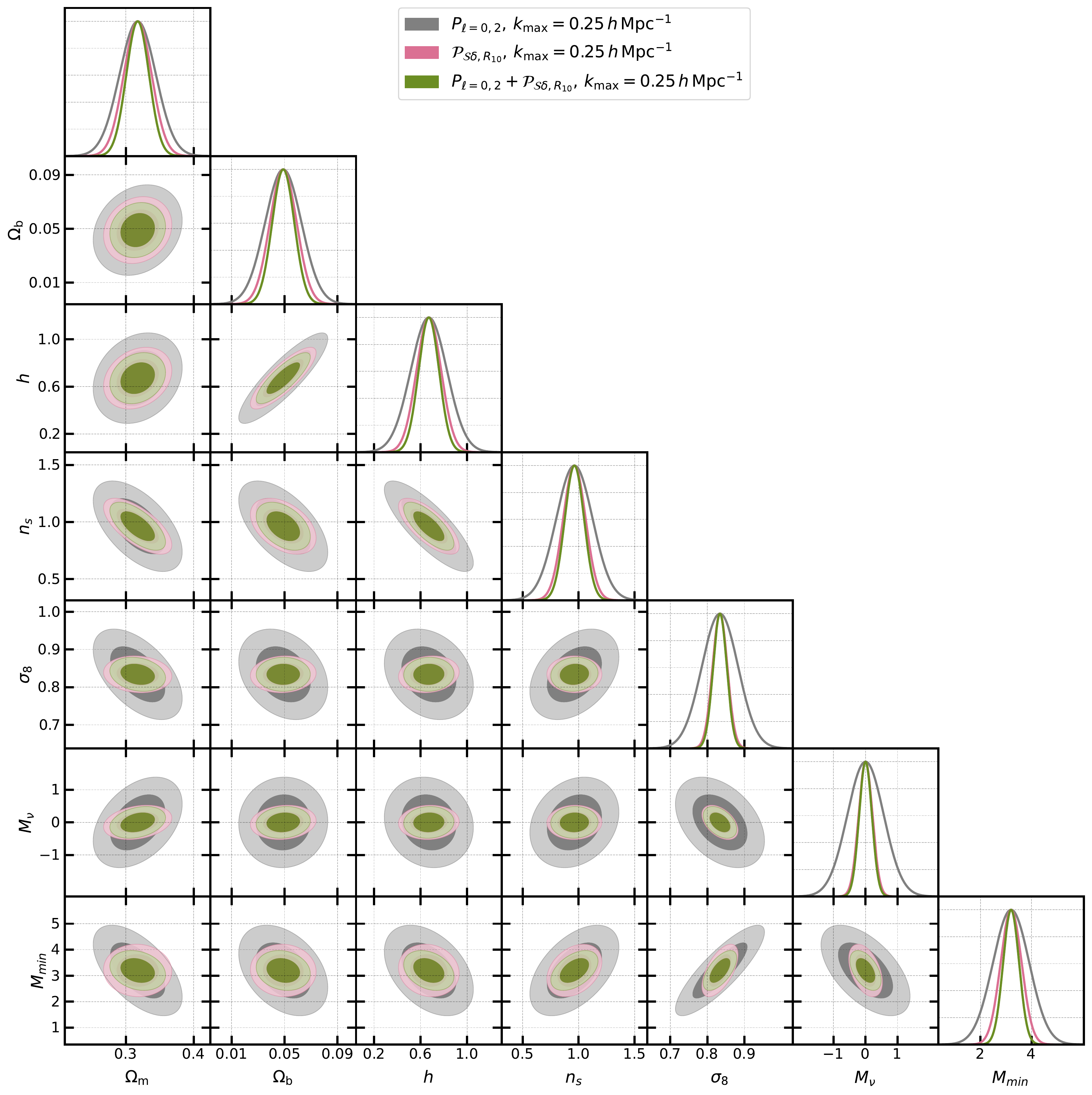}
    \caption{Marginalized 1- and 2-$\sigma$ constraints from Quijote halo power spectrum multipoles (grey), skew spectra with smoothing scales $R=10\, \mpch$ (red), and the two combined (green). The smoothing scale is set to $k_{\rm max} = 0.25 \ \hmpc$ in all cases.}
    \label{fig:2dcontour_pk_ss_joint_smooth10}
\end{figure}
Next, we demonstrate how much the joint analysis of the skew spectra and power spectrum multipoles improves upon their individual constraints. In figure \ref{fig:2dcontour_pk_ss_joint_smooth10}, we compare the 2D marginalized 1- and 2-$\sigma$ constraints from the power spectrum multipoles (grey), the skew spectra with $R= 10\, \mpch$ (red), and the two combined (green). We set $k_{\rm max}=0.25\, \hmpc$ in all three cases. The constraints on cosmological parameters from the joint analysis are dominated by the skew spectra, and the combination with the power spectrum multipoles only improves the constraints slightly, ranging from $(9-18)\%$, the smallest and largest gains are on $M_\nu$ and $\Omega_{\rm b}$, respectively. Most parameters exhibit identical degeneracy directions in power and skew spectra. Therefore, the improvement in the constraints is not driven by the breaking of degeneracies when the two statistics are combined. As we will show later in this section, for Molino galaxies, the conclusions differ as a result of varying a larger number of nuisance parameters characterizing a more complex matter-tracer biasing relation. In table~\ref{tab:quijote}, we report the marginalized 1-$\sigma$ uncertainties on cosmological parameters from skew and power spectra individually and combined, marginalizing over the minimum halo mass cut $M_{\rm min}$. 

In the previous analysis of the halo power spectrum (monopole and quadrupole) and bispectrum (monopole) of the Quijote dataset presented in Ref.~\cite{Hahn:2019zob}, in addition to $M_{\rm min}$, an additional nuisance parameter characterizing the overall amplitude of the two summary statistics was varied. To compare our results with theirs, we perform an additional forecast (setting $k_{\rm max}=0.2\ \hmpc$ as in their forecasts), allowing the overall amplitudes of the power spectrum multipoles and the skew spectra to vary. In that case, our power spectrum constraints are comparable to theirs\footnote{We use a slightly different k binning than Ref.~\cite{Hahn:2019zob}, but the overall effect in terms of parameter constraints is no more than 10\%}, and on average, the constraints from skew spectra are very competitive to the one from bispectrum.
We note, however, that the comparison between constraints from skew spectra and bispectrum monopole has scale dependence. The constraints from skew spectra reach a plateau, resulting from the imposed smoothing scale. On the contrary, the bispectrum constraints continue improving with increased $k_{\rm max}$. For example, at $k_{\rm max}=0.3\,\hmpc$, we find the constraints from skew spectra on most parameters are weaker. 

\begin{table}[t]
\centering
\begin{tabular}{c|c|cccccc} 
\toprule
Quijote & \begin{tabular}[c]{@{}c@{}}$k_{\rm max}$\\$[{\rm Mpc}^{-1} h]$\end{tabular} & $\Omega_{\rm m}$ & $\Omega_{\rm b}$ & $h$ & $n_s$ & $\sigma_8$ & \begin{tabular}[c]{@{}c@{}}$M_{\nu}$\\ $[{\rm eV}]$\end{tabular} \\ 
\hline
\multirow{3}{*}{$\mathcal{P}^{R_{20}}_{\mathcal{S}\delta}$} & $0.15$ & $0.035$ & $0.016$ & \begin{tabular}[c]{@{}c@{}}$0.171$\\\end{tabular} & $0.185$ & $0.058$ & $0.315$ \\
& $0.20$ & $0.027$ & $0.014$ & $0.141$ & $0.144$ & $0.036$ & $0.258$ \\ 
& $0.25$ & $0.025$ & $0.012$ & $0.123$ & $0.122$ & $0.027$ & $0.212$ \\ 
\hline
\multirow{4}{*}{$\mathcal{P}^{R_{10}}_{\mathcal{S}\delta}$} & $0.15$ & $0.027$ & $0.014$ & $0.151$ & $0.156$ & $0.038$ & $0.299$ \\
& $0.20$ & $0.021$ & $0.011$ & $0.115$ & $0.113$ & $0.025$ & $0.251$ \\
& $0.25$ & $0.020$ & $0.010$ & $0.106$ & $0.100$ & $0.020$ & $0.212$ \\
 & $0.50$ & $0.014$ & $0.007$ & $0.066$ & $0.049$ & $0.013$ & $0.118$ \\ 
\hline
\multirow{4}{*}{$P_{\ell=0,2}$} & $0.15$ & $0.058$ & $0.029$ & $0.399$ & $0.487$ & $0.198$ & $0.845$ \\
 & $0.20$ & $0.032$ & $0.015$ & $0.171$ & $0.195$ & $0.081$ & $0.673$ \\
 & $0.25$ & $0.026$ & $0.014$ & $0.153$ & $0.157$ & $0.049$ & $0.568$ \\
 & $0.50$ & $0.017$ & $0.011$ & $0.110$ & $0.065$ & $0.019$ & $0.322$ \\ 
\hline
$\mathcal{P}^{R_{10}}_{\mathcal{S}\delta}+P_{\ell=0,2}$ & $0.25$ & $0.017$ & $0.008$ & $0.088$ & $0.086$ & $0.018$ & $0.196$ \\
\bottomrule
\end{tabular} \vspace{0.2in}
\caption{Marginalized 1-$\sigma$ constraints on cosmological parameters from halo skew and power spectra measured on Quijote simulations. We show the results from skew spectra with $R=20\,\mpch$ and $R=10\,\mpch$, contrasted with those from the power spectrum monopole and quadrupole. The last row shows the constraints from the combination of the skew spectra with $R=10\  \mpch$ and the power spectrum multipoles. }
\label{tab:quijote}\vspace{-0.2in}
\end{table}

An important question to address in order to establish the skew spectra as (nearly) optimal proxy statistics for the bispectrum is how much of the information of the bispectrum is captured by the skew spectra. While at $k_{\rm max} = 0.5\, \hmpc$, the comparison with bispectrum results of Ref.~\cite{Hahn:2019zob} clearly indicates information loss (even compared to just bispectrum monopole), we note that, as pointed out in Ref.~\cite{Coulton:2022rir}, the results of numerical Fisher forecasts of the bispectrum should be interpreted with some caution. This is primarily due to the fact that a lack of convergence of the numerical derivatives with respect to the number of mocks used in the measurements can result in artificially tight parameter constraints~\cite{Coulton:2022rir}. Therefore, a more conclusive statement of the information content of the skew spectra in comparison with the bispectrum is not possible at the moment. We perform extensive stability tests for convergence of numerical derivatives of skew spectra in Appendix~\ref{app:convergence}, quantifying the amount by which the parameter uncertainties may be underestimated due to lack of full convergence and estimating the required number of simulations to achieve full convergence. We conclude that the convergence rate of the skew spectra is comparable with that of the power spectrum multipoles. Therefore, the relative gain in constraints from the skew spectra compared to the power spectrum should not be affected by the convergence of the numerically estimated derivatives and covariance.

\begin{figure}[t]
    \centering
\includegraphics[width=\textwidth]{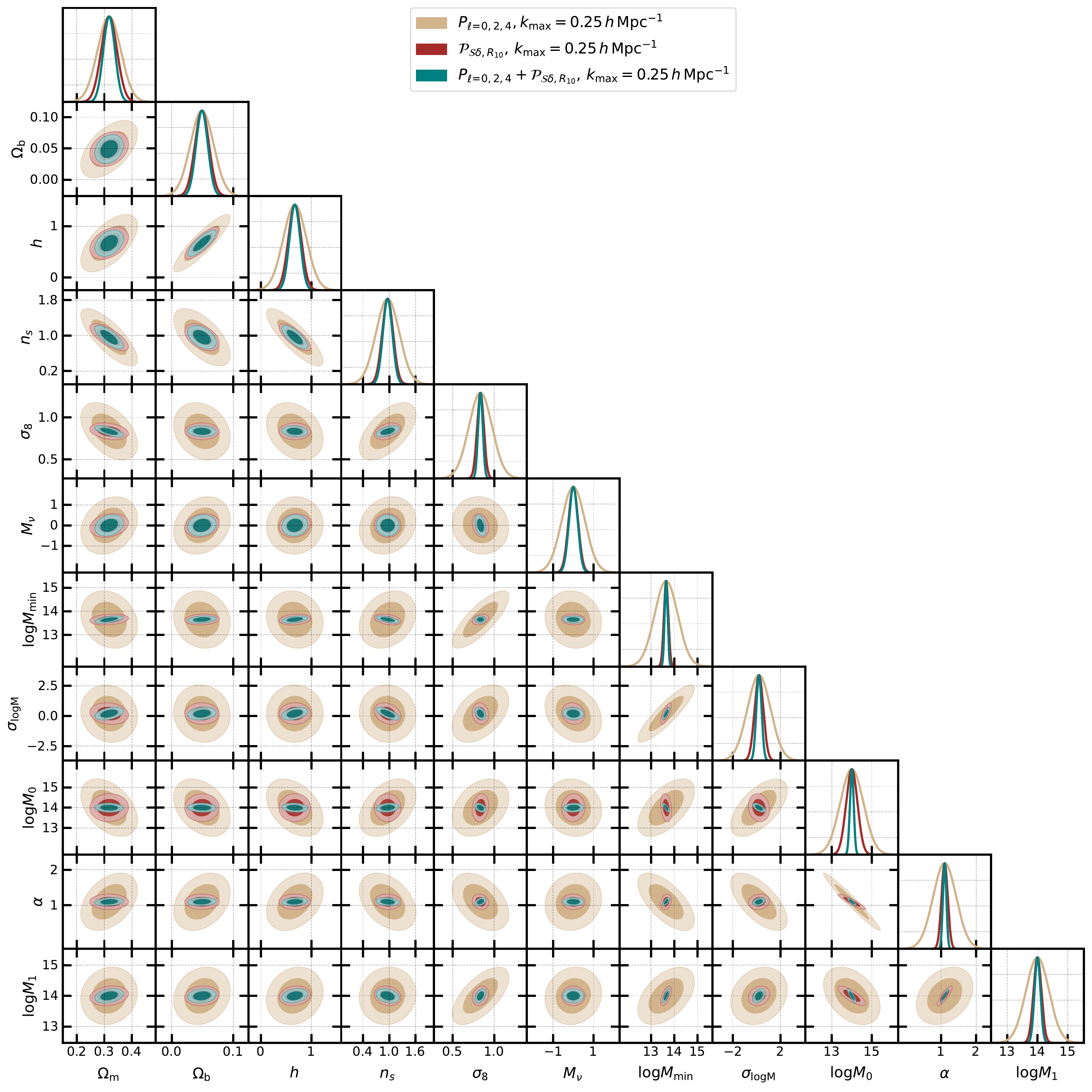}
    \caption{Marginalized 1- and 2-$\sigma$ constraints for Molino galaxy mocks. The plots show constraints obtained from power spectrum multipoles (tan), skew spectra with smoothing scale of $R=10\, \mpch$ (red), and the two combined (teal).}
    \label{fig:molino_2dcontour_pk4shot1xssshot1_k0d25}
\end{figure}

\vspace{0.2in}
\noindent \underline{Molino Galaxy catalogs}
\vspace{0.05in}  

\noindent In order to understand the impact of marginalization over parameters of a more complex tracer-matter relation, we next study the galaxy catalog from the HOD-based Molino suite. Before presenting the results, let us make two remarks on the differences between our analysis of Quijote and Molino datasets. First, in contrast to the Quijote analysis, we include the three lowest multipoles of the power spectrum. This is because the Molino galaxy sample has a large hexadecapole and a fair comparison with skew spectra (which capture the anisotropic clustering information) should include the power spectrum hexadecapole. In Appendix~\ref{app:Molino}, we test the information content of the hexadecapole for this sample and provide further discussion of how the adopted calibration of the HOD model is at the root of the observed large hexadecapole. Second, we compare the results of the skew spectra without subtraction of the shot noise to the shot-noise-subtracted power spectrum. This choice was made solely for computational convenience in measuring the skew spectra. We found that for the power spectrum of Molino galaxies, the shot noise plays a marginal role in terms of constraints on cosmological parameters. Given that we found that the cosmological constraints are largely insensitive to subtraction of the shot noise for Quijote catalogs (presented in Appendix \ref{app:shot}), we expect that our results for Molino should also not be affected by the shot noise subtraction.

Figure~\ref{fig:molino_2dcontour_pk4shot1xssshot1_k0d25} shows marginalized 1- and 2-$\sigma$ constraints for Molino galaxy mocks obtained from power spectrum multipoles (orange), skew spectra with smoothing scale of $R=10\, \mpch$ (red), and the two combined (teal) at $k_{\rm max}=0.25\,\hmpc$. We find that marginalization over the HOD parameters, which capture a complex biasing relation between dark matter and tracers, does not affect the degeneracy directions among the cosmological parameters (in comparison with Quijote results in figure~\ref{fig:2dcontour_pk_ss_joint_smooth10}). Although there is no guarantee that parameter degeneracies should always remain the same when opening up parameter space, we do not observe a strong (anti)correlation between the HOD parameters and any of the cosmological parameters in Molino. This is likely to be the reason why in Molino analysis, the degeneracies among cosmological parameters are still determined by the underlying physical processes. Nonetheless, as we discuss in Appendix~\ref{app:Molino}, further investigation of the consequence of the galaxy assignment scheme using a HOD parameterization that is tuned to realistic samples remains interesting. 
\begin{figure}[t]
    \centering
    \includegraphics[width=0.9\textwidth]{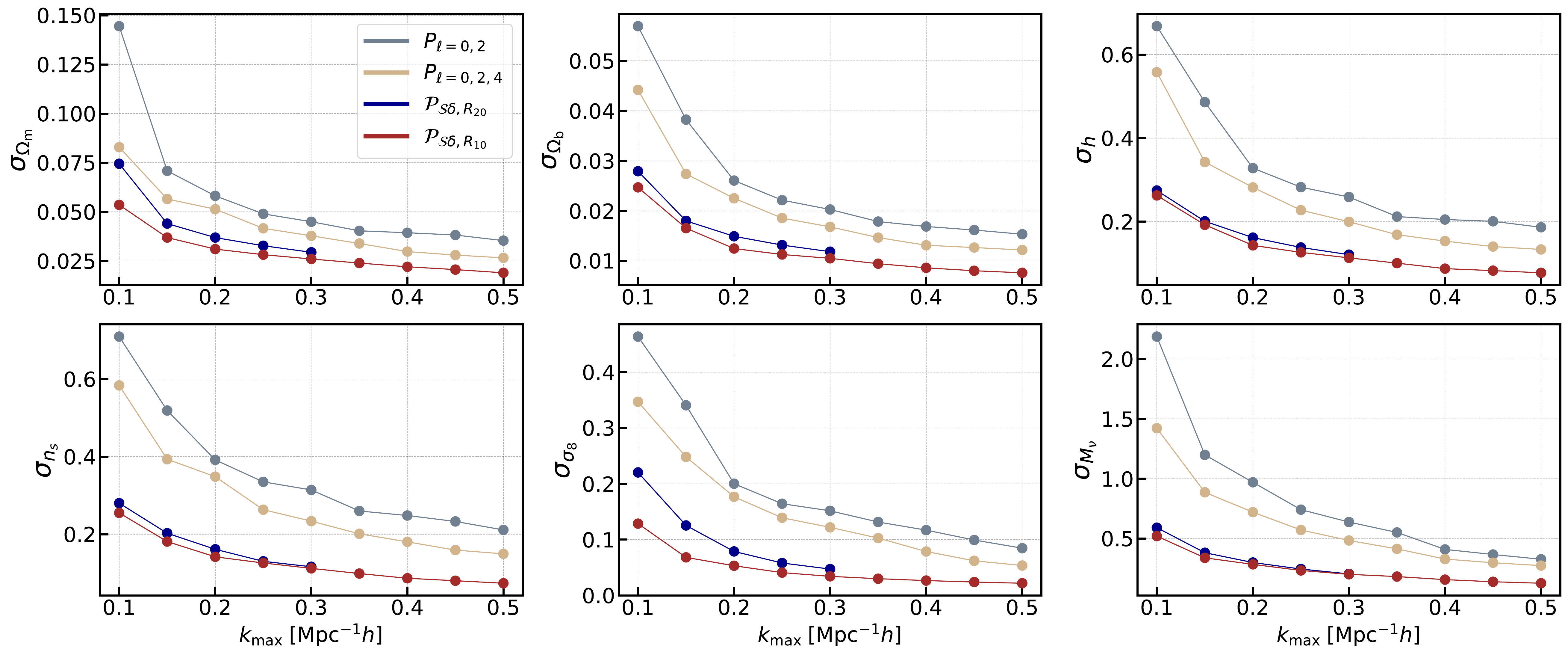}
    \caption{Marginalized 1-$\sigma$ constraints as a function of $k_{\rm max}$ from Molino galaxies. We show the constraints from the lowest two (three) multipoles in gray (yellow), from skew spectra with $R=20\, \mpch$ in blue, and from skew spectra with $R=10\, \mpch$ in red.}
    \label{fig:molino_1sigma_ss_smooth10x20_shot1_pk024_shot0}
\end{figure}

Compared to the power spectrum up to hexadecapole, skew spectra improve the constraints by $(32-71)\%$. Combining the power spectrum and skew spectra improves the constraints from skew spectra alone by an additional $(10-34)\%$, with the least improvement again in $M_\nu$ and the largest in $\sigma_8$. Overall, the relative gain between skew spectra and power spectrum in cosmological parameters is more significant on Molino than Quijote, by at most $50\%$ in $n_s$ (for $M_\nu$ the constraints are slightly degraded). In addition, we also see that skew spectra, alone and combined with power spectra, tightly constrain the HOD parameters. In particular, the degeneracy in the ${\rm log}M_{\rm min}-{\rm log}M_{0}$ plane and $\sigma_{\rm log M}-{\rm log}M_{0}$ is broken by the skew spectra.

We found that the skew spectra can achieve very competitive results compared to the bispectrum monopole at linear to mildly non-linear scales ($k<0.3 \ \hmpc$). In particular, we found that for Molino the skew spectra not only outperform the bispectrum monopole but also have a larger relative gain compared to the power spectrum than using the bispectrum monopole. In Table \ref{tab:molino}, we summarize the 1-$\sigma$ constraints from skew spectra, power spectrum multipoles ($\ell=0,2,4$), and the combination of the two measured on Molino galaxies. Going from Quijote to Molino, the gain in these higher-order statistics estimators to power spectrum is more obvious for skew spectra than bispectrum monopole at linear to mildly non-linear scales. There could be various reasons driving this difference in the relative gain. For instance, the power excess (rising slope of satellite power spectrum towards high-$k$) in the satellite galaxies implies that the satellite velocities can induce large anisotropy in redshift space. This information can be captured by the LoS-dependent skew spectra while is averaged over in bispectrum monopole. As we discuss in more detail in Appendix~\ref{app:Molino}, the conclusions can be different for a realistic galaxy sample.

We demonstrate the dependence of the constraints on the small-scale cutoff in figure~\ref{fig:molino_1sigma_ss_smooth10x20_shot1_pk024_shot0}, which shows the marginalized 1-$\sigma$ constraints as a function of $k_{\rm max}$. To highlight the information content of hexadecapole of Molino galaxies, we show the constraints for the lowest two and three multipoles of the power spectrum in grey and in yellow, respectively. Additionally, we show the constraints from the skew spectra with smoothing scales of $R=20\,\mpch$ in blue and $R=10\,\mpch$ in red. Again, for $R=20\,\mpch$, we show only up to $k_{\rm max}=0.3\,\hmpc$ because this smoothing washes out the small-scale information. The 2D marginalized constraints are shown in figure~\ref{fig:molino_2dcontour_pk_shot0_ss_shot1_k0d25} of Appendix \ref{app:Molino}. For the Molino sample, we also find that the skew spectra provide more significant improvement in cosmological constraints at large scales in comparison to the power spectrum. The improvements gradually plateau when approaching smaller scales. Furthermore, the hexadecapole adds non-negligible information to the constraints (due to the large satellite contribution in the Molino sample). 

The Molino galaxy mocks provide valuable insight into the impact of marginalization over (nuisance) parameters characterizing a more complex biasing relation. However, for realistic spectroscopic galaxy samples, the conclusions about the relative gain in parameter constraints from skew spectra could be different. This is due to the fact that apart from the underestimation of the parameter uncertainties intrinsic to all Fisher forecasts, as described above and in Appendix D, the characteristics of the Molino sample do not resemble the known populations of galaxies probed by spectroscopic galaxy surveys. Therefore, establishing the impact of marginalization over HOD nuisance parameters requires performing a similar study as ours applied to a HOD parameterization calibrated to match realistic clustering signals.

\begin{table}[t]
\centering
\begin{tabular}{c|c|cccccc} 
\toprule
Molino & \begin{tabular}[c]{@{}c@{}}$k_{\rm max}$\\$[{\rm Mpc}^{-1} h]$\end{tabular} & $\Omega_{\rm m}$ & $\Omega_{\rm b}$ & $h$ & $n_s$ & $\sigma_8$ & \begin{tabular}[c]{@{}c@{}}$M_{\nu}$\\ $[{\rm eV}]$\end{tabular} \\ 
\hline
\multirow{3}{*}{$\mathcal{P}^{R_{20}}_{\mathcal{S}\delta}$} & $0.15$ & $0.044$ & $0.018$ & $0.201$ & $0.201$ & $0.127$ & $0.384$ \\
& $0.20$ & $0.036$ & $0.015$ & $0.162$ & $0.161$ & $0.078$ & $0.301$ \\ 
& $0.25$ & $0.032$ & $0.013$ & $0.139$ & $0.132$ & $0.058$ & $0.244$ \\ 
\hline
\multirow{4}{*}{$\mathcal{P}^{R_{10}}_{\mathcal{S}\delta}$} & $0.15$ & $0.037$ & $0.017$ & $0.192$ & $0.181$ & $0.068$ & $0.339$ \\
& $0.20$ & $0.031$ & $0.012$ & $0.143$ & $0.142$ & $0.053$ & $0.285$ \\
 & $0.25$ & $0.028$ & $0.011$ & $0.126$ & $0.126$ & $0.041$ & $0.234$ \\
 & $0.50$ & $0.019$ & $0.008$ & $0.077$ & $0.074$ & $0.022$ & $0.127$ \\ 
\hline
\multirow{3}{*}{$P_\ell=0,2,4$} & $0.15$ & $0.056$ & $0.027$ & $0.339$ & $0.392$ & $0.247$ & $0.868$ \\
 & $0.20$ & $0.051$ & $0.022$ & $0.283$ & $0.347$ & $0.179$ & $0.708$ \\
 & $0.25$ & $0.042$ & $0.019$ & $0.227$ & $0.261$ & $0.140$ & $0.565$ \\
 & $0.50$ & $0.028$ & $0.012$ & $0.133$ & $0.148$ & $0.053$ & $0.275$ \\ 
\hline
$\mathcal{P}^{R_{10}}_{\mathcal{S}\delta}+P_\ell=0,2,4$ & $0.25$ & $0.021$ & $0.009$ & $0.106$ & $0.109$ & $0.027$ & $0.216$ \\
\bottomrule
\end{tabular}\vspace{0.2in}
\caption{Same as table~\ref{tab:quijote}, but for Molino galaxy catalogs.}
\label{tab:molino}
\vspace{-0.2in}
\end{table}

\section{Conclusion}\label{sec:conc}

In this paper, we have focused on quantifying the cosmological information of skew spectra of biased tracers in redshift space. The skew spectra are cross-correlations between the observed galaxy field and several appropriately weighted squares of it. They have been proposed as efficient proxy statistics to extract the non-Gaussian information of the LSS encoded in the bispectrum of biased tracers. By construction, the skew spectra correspond to the maximum-likelihood estimators for the parameters that appear as overall amplitudes in the tree-level bispectrum model. Therefore, they optimally constrain (perturbative) galaxy bias parameters, growth rate, and amplitudes of the primordial power spectrum and bispectrum. Their information content for other cosmological parameters, which is the focus of this paper, has not been explored before. 

The main advantages of the skew spectra compared to the bispectrum lies in their efficiency and interpretability, which result from their low dimensionality; In contrast to the redshift-space bispectrum, which is a function of five variables, the skew spectra are just a function of a single variable, similar to the power spectrum multipoles. Furthermore, being pseudo-power spectra, it is expected that for the skew spectra, accounting for observational effects (in particular the survey window function) is simpler than the bispectrum. Lastly, the extensive previous works on the analysis of galaxy clustering power spectrum, which now has reached considerable maturity, should be largely applicable to the analysis of the skew spectra. Motivated by these potential advantages, we set out to investigate cosmological constraints from redshift-space skew spectra.    

We used two sets of synthetic data, the Quijote halo and the Molino galaxy catalogs, to perform numerical Fisher forecasts for six-parameter $\nu\Lambda$CDM model, varying $\{\Omega_{\rm m},\Omega_{\rm b}, h, n_s, \sigma_8, M_\nu\}$. After presenting the measured skew spectra and their covariance matrix, we illustrated the impact of off-diagonal elements of the covariance on the SNR from individual skew spectra. We then showed the forecasted parameter constraints from the redshift-space skew spectra, contrasted with those from the power spectrum multipoles. We investigated several ingredients and assumptions of the Fisher forecasts, including the impact of subtraction of the shot noise and the choice of the smoothing scale, the information content of individual skew spectra, and the stability of the forecasts w.r.t. a number of mocks used in the measurements. 

For Quijote halos and using scales up to $k_{\rm max}=0.25\,\hmpc$, we found that the skew spectra (with $R=10\ \mpch$) provide constraints on cosmological parameters that are tighter than those from the power spectrum multipoles $(\ell=0,2)$ by $(23-62)\%$. Combining the skew and power spectra, the constraints are further improved (by up to 18\%), compared to skew spectra alone. We did not find the shot noise to have a major impact on the forecasted constraints of cosmological parameters. The constraints from skew spectra are competitive with those from the bispectrum monopole at $k_{\rm max}=0.2\,\hmpc$ and become less constraining at $k\sim\!0.3\,\hmpc$ due to the smoothing of the density field. Using Molino galaxy catalogs, we investigated the effect of marginalization over parameters of a more complex biasing relation between tracers and DM. For this dataset, upon marginalization over HOD parameters, we found an improvement of $(32-71)\%$ in cosmological constraints from skew spectra compared to power spectrum multipoles ($\ell = 0, 2, 4$). Given the excessive power of satellite galaxies at small scales, in this analysis, we included the power spectrum hexadecapole. Interestingly enough, despite the strong random-motion-associated redshift space distortion on the small scales, the shapes of the 14 skew spectra and degeneracy directions between cosmological parameters on the Molino sample are unaffected.  

We examined how the choice of smoothing scale affects the constraints and found that using a smoothing scale of $R=10\,\mpch$ tightens the constraints on cosmological parameters, on average by $\sim 20\%$, compared to the analysis with $R=20\,\mpch$. While the perturbative theoretical model of the skew spectra (based on tree-level bispectrum prediction) was shown to be more reliable with $R=20\,\mpch$ \cite{Schmittfull:2020hoi}, in this paper, we presented both scales but considered the results for the smaller smoothing scale as our baseline analysis. On the one hand, the theoretical model can be improved by going beyond the tree-level bispectrum, and on the other hand, non-standard analysis methods, which do not rely on the availability of theoretical predictions, such as likelihood-free inference~\cite{Hahn2022simbig1, Hahn2022simbig2}, can potentially enable us to extract the information from these smaller scales.

Apart from the simplifying assumptions intrinsic to Fisher forecasts, in interpreting simulation-based numerical forecasts, testing the stability of the results with respect to variation of the number of mocks used in the estimation of derivatives and covariance matrix is essential. We estimated the additional variance that enters the sampling covariance, which is estimated from a limited number of realizations, and found that for 15,000 mocks, the noise in the covariance leads to an additional variance of 3.8\% for the skew spectra, compared to a sub-percent level for the power spectrum. Testing the convergence with regards to the noise in the measurement of the derivatives, we showed that the derivatives estimated from the pairs of mocks and averaged over the 3 LoS, using 500 of such pairs, could lead to an underestimation of the constraints by $(20-30)\%$ on neutrino masses. This is the case for skew spectra or power spectra and both on Quijote or Molino samples. For this reason, instead of quoting the absolute number, we largely focused on the relative improvement between the two statistics. 

While the results presented in this work are certainly very encouraging, the information content of the skew spectra is far from established for realistic observational data, and this work is just the first step in this regard. There are several directions in which this work can be extended, some of which we will pursue in upcoming works. First, the fact that the 14 skew spectra can be classified into three categories implies that there may be more efficient ways to group the information. Currently, sorting the skew spectra according to their power in growth rate and bias is motivated by being a maximum-likelihood estimator for the ``amplitude-like" parameters. It remains an interesting question to compress the data vector in a more efficient way for generalized applications. Second, In order to perform theory-based likelihood analysis of the observational data, more accurate perturbative theoretical models of the skew spectra, in particular when assuming a smaller smoothing scale, are necessary. Third, in order to be applied to survey data, the construction of survey estimators, including the observational effects, and accounting for the survey window in theoretical predictions, is required.

\section*{Acknowledgments}
We are grateful to William Coulton, Yangyang Li, and Zvonomir Vlah for very helpful discussions. We thank Paco Francisco Villaescusa-Navarro for guidance on the use of Quijote data, and Oliver Philcox, Marcel Schmittfull, and Zack Slepian for their helpful feedback on the draft of this manuscript. JH has received funding from the European Union’s Horizon 2020 research and innovation program under the Marie Sk\l{}odowska-Curie grant agreement No 101025187. AMD acknowledges funding from Tomalla Foundation for Research in Gravity and Boninchi Foundation. The authors acknowledge the University of Florida Research Computing for providing computational resources and support that have contributed to the research results reported in this publication.

\appendix

\section{Information in Individual Skew Spectra}
\label{app:indSkew}

All the results presented in the main body of this paper were obtained considering a data vector consisting of all 14 skew spectra defined in Eq. \eqref{eq:skew_list}. In this section, we dissect the information content of the full data vector by considering individual skew spectra. We study the response functions of each one to changes in cosmological parameters, as well as the marginalized parameter constraints from individual skew spectra.

\subsection{Responses to Change of Parameters} \label{app:responses}

\begin{figure}[htbp!]
\centering
\includegraphics[width=0.75\textwidth]{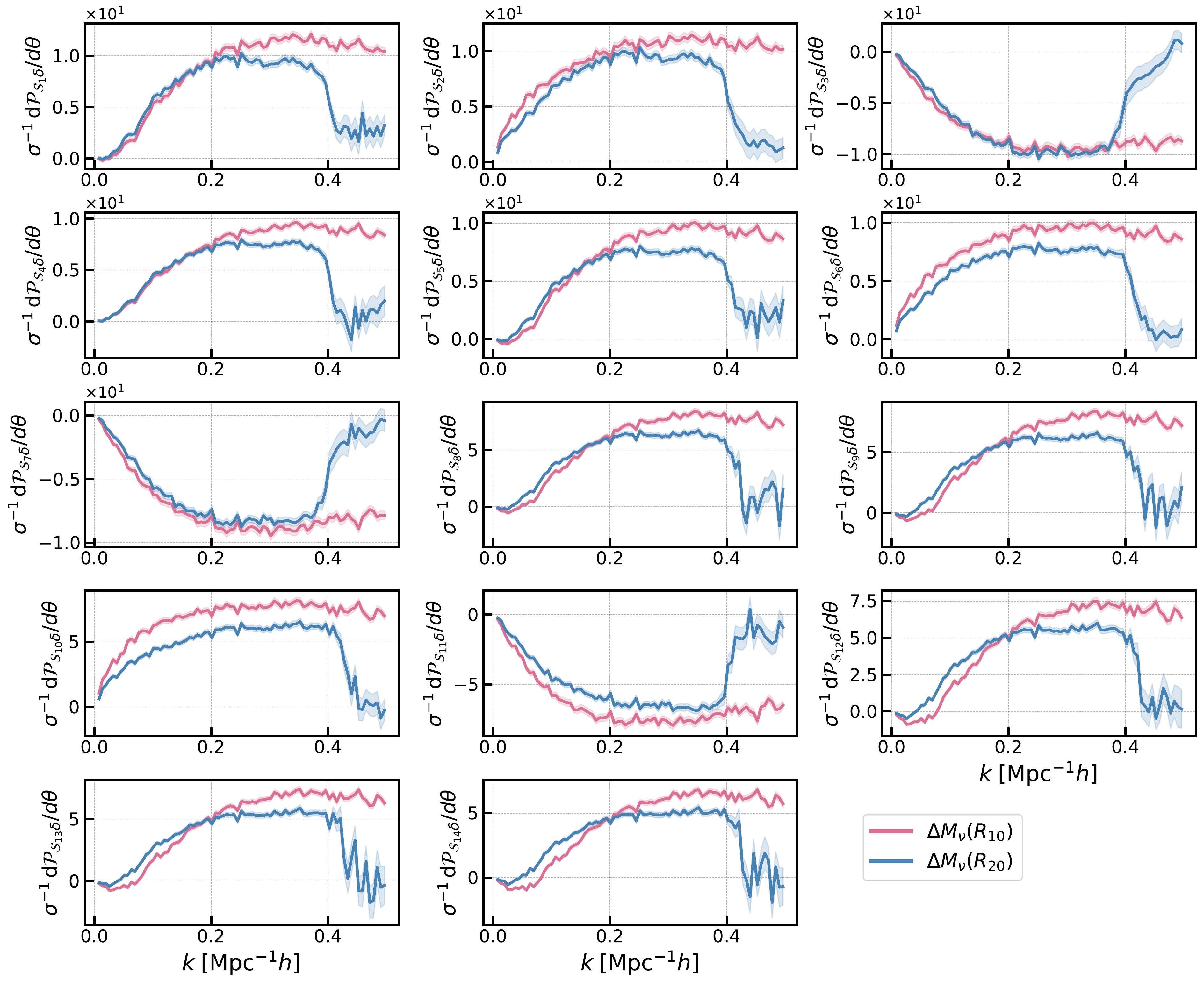}
    \caption{Response of the 14 skew spectra Quijote halos to variation of the total mass of neutrinos assuming smoothing scales $R=10 \mpch$ (red) and $=20 \mpch$ (blue). The responses are normalized by the standard deviation of the 15,000 mocks at fiducial cosmology, $\sigma$. The shaded regions are inferred from the standard error of the mean of the 500 mock pairs.}
\label{fig:quijote_response_smooth10x20_Mnu}
\end{figure}

We first study the responses (derivatives) of individual skew spectra of Quijote halos to the variation of neutrino mass around the reference value of $M_{\nu}=0.4\, {\rm eV}$. This provides guidance for understanding the results shown in figure \ref{fig:quijote_1sigma_ss_smooth10x20_shotoff_pk02_nuisMmin} as to why the constraint on $M_\nu$ is insensitive to the choice of smoothing scale. The measured responses of individual skew spectra are shown in figure \ref{fig:quijote_response_smooth10x20_Mnu}, applying smoothing scales of $R = 10 \ \mpch$ in magenta and $R = 20 \ \mpch$ in blue. Here, the responses are normalized by the standard deviation of the 15,000 mocks at fiducial cosmology, $\sigma$, so that the 14 responses have similar amplitudes. The overall amplitudes of the scaled responses for both smoothing scales are comparable, with a slightly higher amplitude for those with $R=10\mpch$ for almost all skew spectra. The cutoff at $k \simeq 0.4 \ \hmpc$ is due to the fact that smoothing with $R=20\,\mpch$ completely washes out the fluctuations on those scales, i.e., the skew spectra themselves are vanishing. The cutoff isn't exactly at $k_{\rm max} = 2\pi/R$ since we use a Gaussian smoothing instead of a top hat in Fourier space. While the responses of some of the skew spectra for the smaller smoothing scales are clearly larger, as a result of the high correlation between the 14 skew spectra, the non-vanishing off-diagonal elements have a non-trivial impact on the Fisher information. Therefore, we can not directly map the amplitude of the response for a given parameter to its constraint.

\begin{figure}[htbp!]
\centering
\includegraphics[width=0.75\textwidth]{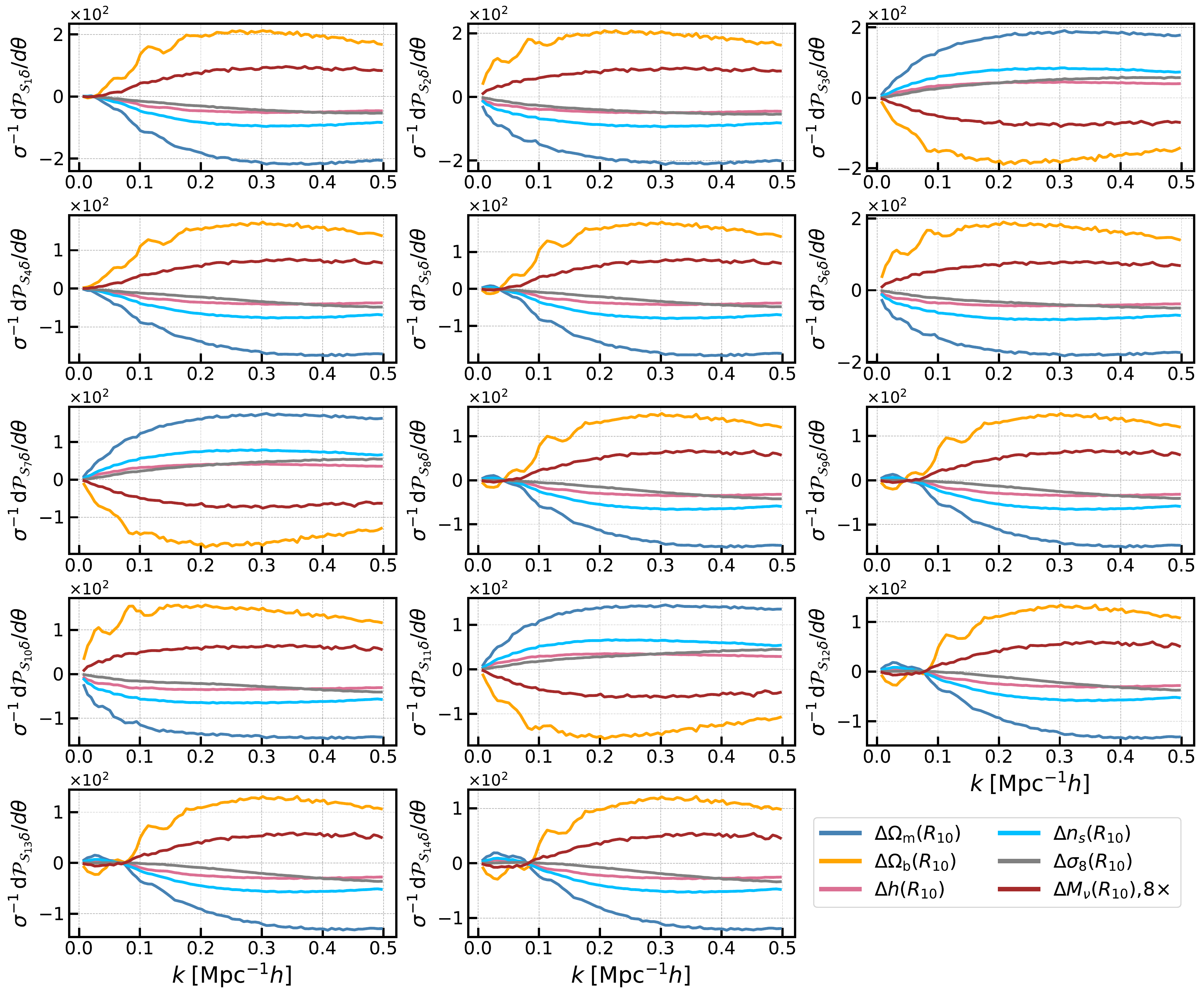}
\caption{Responses of the 14 skew spectra of Quijote halos to cosmological parameters. The responses are normalized by the standard deviation, $\sigma$, of the 15,000 mocks at fiducial cosmology. The smoothing scale is set to $R=10 \,\mpch$. The response to variation in the total mass of neutrinos is further rescaled by a factor of 8 for visibility.}
\label{fig:quijote_normed_response_smooth10_shotoff_6cosmo}\vspace{-0.2in}
\end{figure}

Figure \ref{fig:quijote_normed_response_smooth10_shotoff_6cosmo} shows the responses of the 14 skew spectra to cosmological parameters measured on the Quijote halo catalogs, assuming the smoothing scale $R=10\, \mpch$. Again, the responses are normalized by the standard deviation of the 15,000 mocks at fiducial cosmology. We further multiply the response to variation of total neutrino mass by a factor of 8 as it is much smaller compared to the other parameter. Except for the change in sign, the 14 responses have similar behavior in that they are most sensitive to $\Omega_{\rm m}$ and $\Omega_{\rm b}$, followed by $n_s$, while $h$ and $\sigma_8$ are of comparable size. The shapes of the skew spectra are also reflected in the response; for example, for $\mathcal{P}_{\mathcal{S}_9\delta}$, $\mathcal{P}_{\mathcal{S}_{12-14}\delta}$ we also see the zero-crossing feature at $k\!\sim\! 0.1\,\hmpc$ in the responses. Additionally, they all saturate at $k \sim 0.3 \ \mpch$. 

\begin{figure}[htbp!]
\centering
\includegraphics[width=0.8\textwidth]{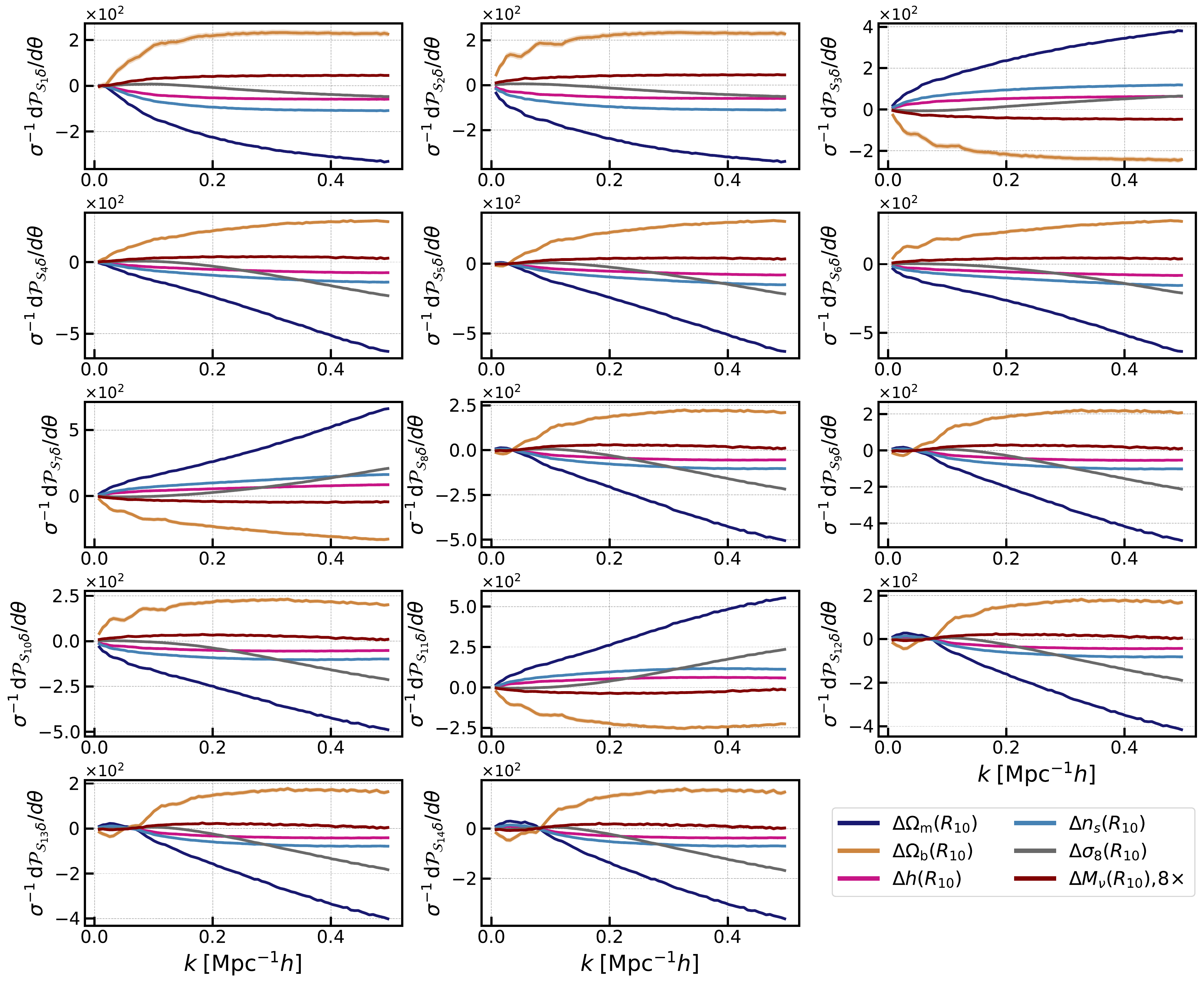}\vspace{-0.1in}
\caption{Same as figure \ref{fig:quijote_normed_response_smooth10_shotoff_6cosmo}, but for Molino galaxy catalogs. Note that we do not subtract shot noise here, which is responsible for the increasing trend in high $k$ when compared to figure \ref{fig:quijote_normed_response_smooth10_shotoff_6cosmo}.}
\label{fig:molino_normed_response_smooth10_shoton_6cosmo}
\vspace{0.2in}
\centering
\includegraphics[width=0.8\textwidth]{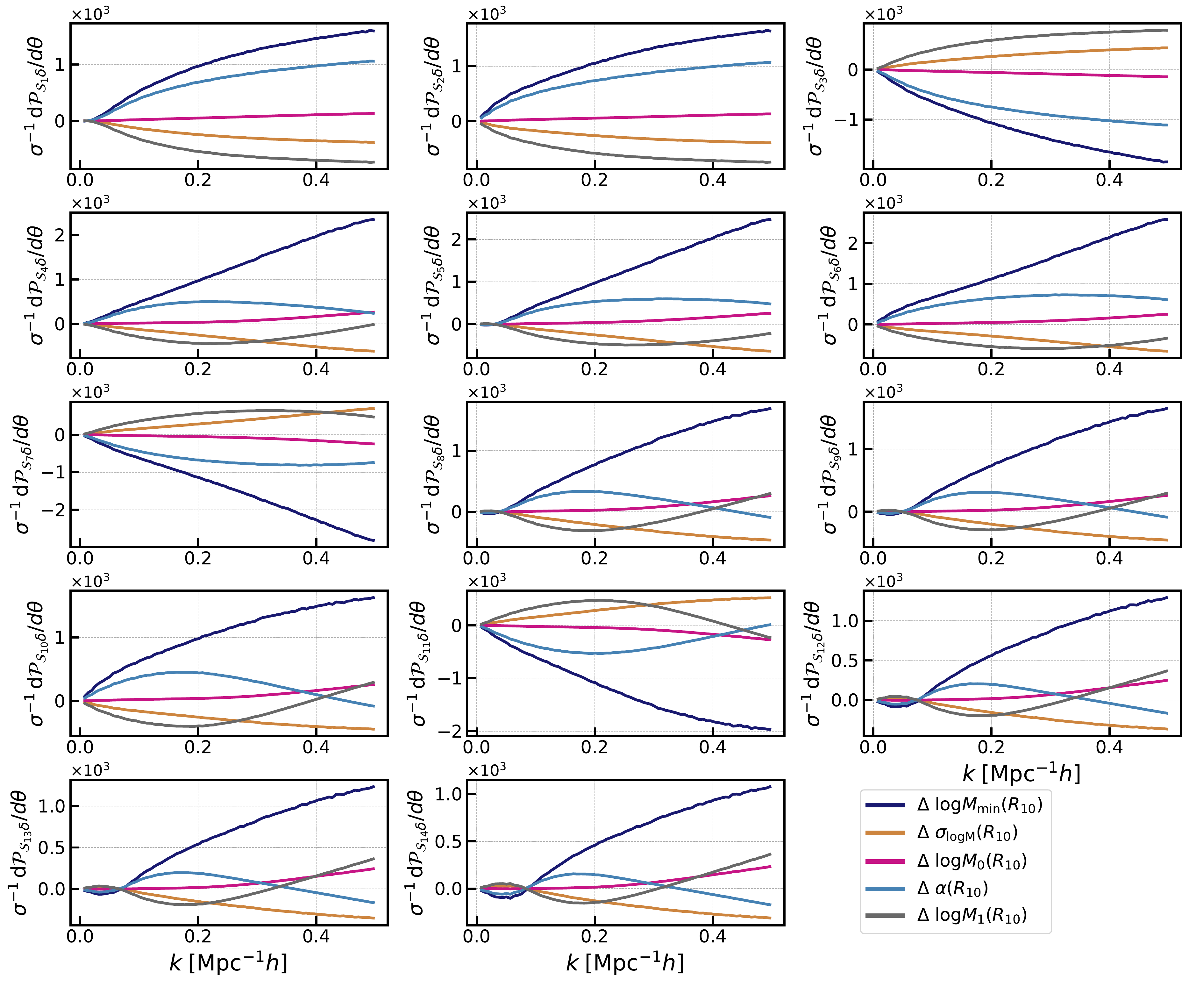} \vspace{-0.1in}
\caption{Similar to figure \ref{fig:molino_normed_response_smooth10_shoton_6cosmo}, but here shows responses of the 14 skew spectra of Molino galaxies to HOD parameters.}
\label{fig:molino_normed_response_smooth10_shoton_5hod}
\end{figure}

In figure \ref{fig:molino_normed_response_smooth10_shoton_6cosmo}, we show the response of Molino galaxies to variations of cosmological parameters, while in figure \ref{fig:molino_normed_response_smooth10_shoton_5hod}, we show the responses to the HOD parameters. As mentioned in the main text, for computational convenience, we do not subtract shot noise for Molino galaxies. Shot noise is responsible for the divergent behavior in responses with increasing $k$ for $\Omega_{\rm m}$ and $\sigma_8$ since these two paired simulations have the largest variation in the number density ($\sim\!10\%$ and $\sim\!7\%$, respectively, while the others are less than 2\%) and shot noise dominates over the small scales. One might expect that adding shot noise boosts the Fisher information and tighten constraints. However, this is not necessarily the case as the shot noise enhances the off-diagonal elements of the covariance, as will be demonstrated in Appendix~\ref{app:shot}. Moreover, we show in figure \ref{fig:2dcontour_ss_smooth10_shotonxoff_nuisMmin} that the impact of shot noise is marginal with Quijote halos as an example. For responses to HOD parameters, we find that the skew spectra are most sensitive to the logarithm of the minimum halo mass-cut, ${\rm log} M_{\rm min}$. The divergent behavior in the HOD parameters is again due to different number densities in the paired simulations, which results in an artificial response due to shot noise when approaching small scales. The responses to $\alpha$ and ${\rm log}M_{1}$ almost mirror each other and exhibit a turnover point starting from $\mathcal{P}_{\mathcal{S}_4\delta}$. Indeed, the contour plot in figure \ref{fig:molino_2dcontour_pk4shot1xssshot1_k0d25} shows that there is a strong correlation between these two parameters. We also see that the skew spectra can significantly improve the constraints for HOD parameters such as ${\rm log}M_{\rm min}$, $\alpha$, and ${\rm log}M_{1}$.

\subsection{Marginalized Constraints from Individual Skew Spectra}

\begin{figure}[htbp!]
    \centering
    \includegraphics[width=0.75\textwidth]{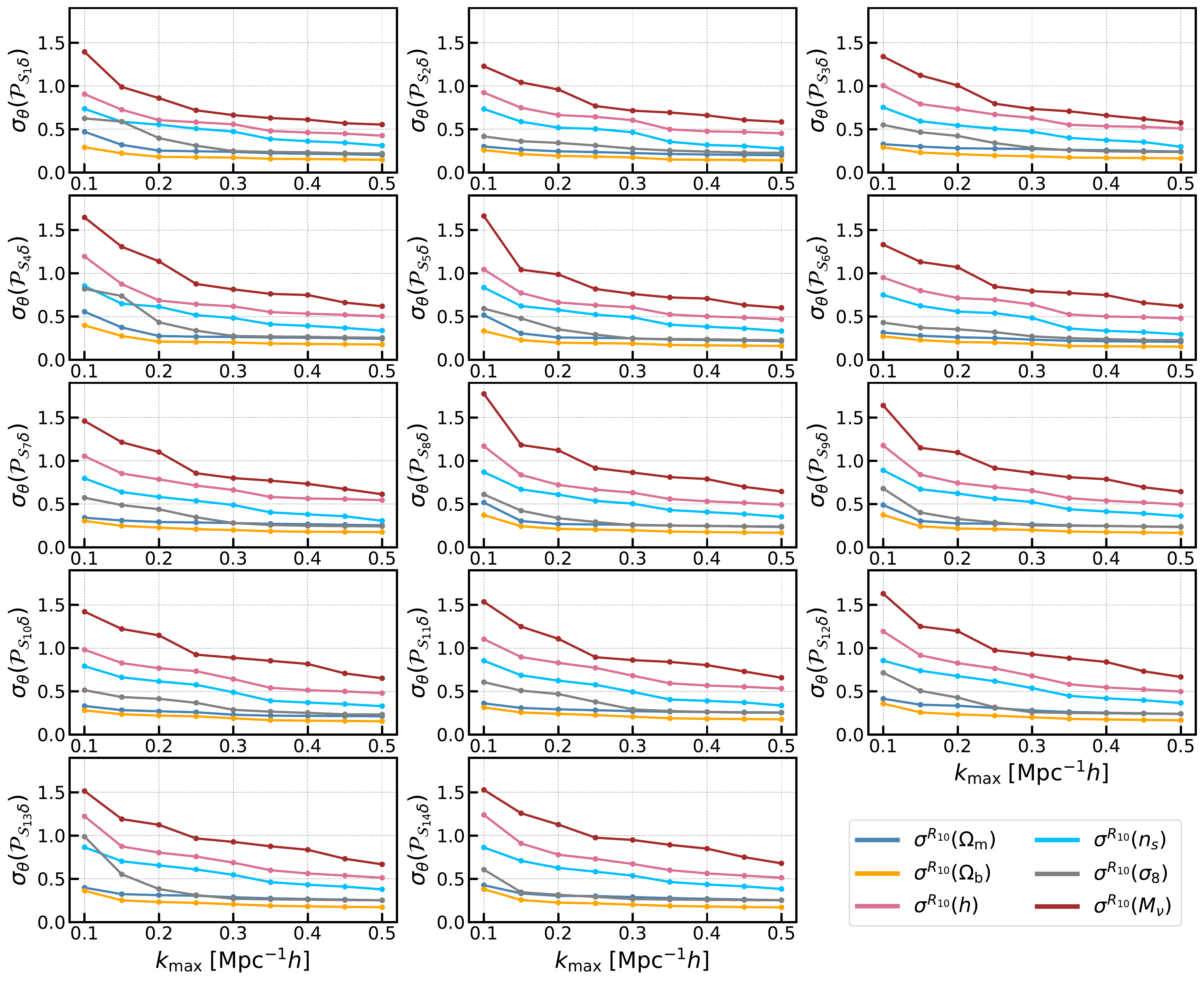}
    \caption{Marginalized 1-$\sigma$ constraints on cosmological parameters $\theta=\{\Omega_{\rm m}, \Omega_{\rm b}, h, n_s, \sigma_8, M_{\nu}\}$ from individual skew spectra as a function of $k_{\rm max}$. The smoothing scale is set to $R=10\, \mpch$. }
    \vspace{0.1in}
    \label{fig:1sigma_kmax_R10_ind}
    \end{figure}

Next, we explore the information content of individual skew spectra on cosmological parameters. In figure \ref{fig:1sigma_kmax_R10_ind}, we show the marginalized 1-$\sigma$ parameter constraints (different colors correspond to different cosmological parameters) from each of the 14 skew spectra as a function of $k_{\rm max}$. We have set the smoothing scale to be $R=10\, \mpch$. Overall, we see a similar trend for the $k_{\rm max}$-dependence of the constraints for all parameters and from all spectra; the parameter uncertainties decrease until reaching a plateau. 

\begin{figure}[t] 
    \centering
    \includegraphics[width=0.8\textwidth]{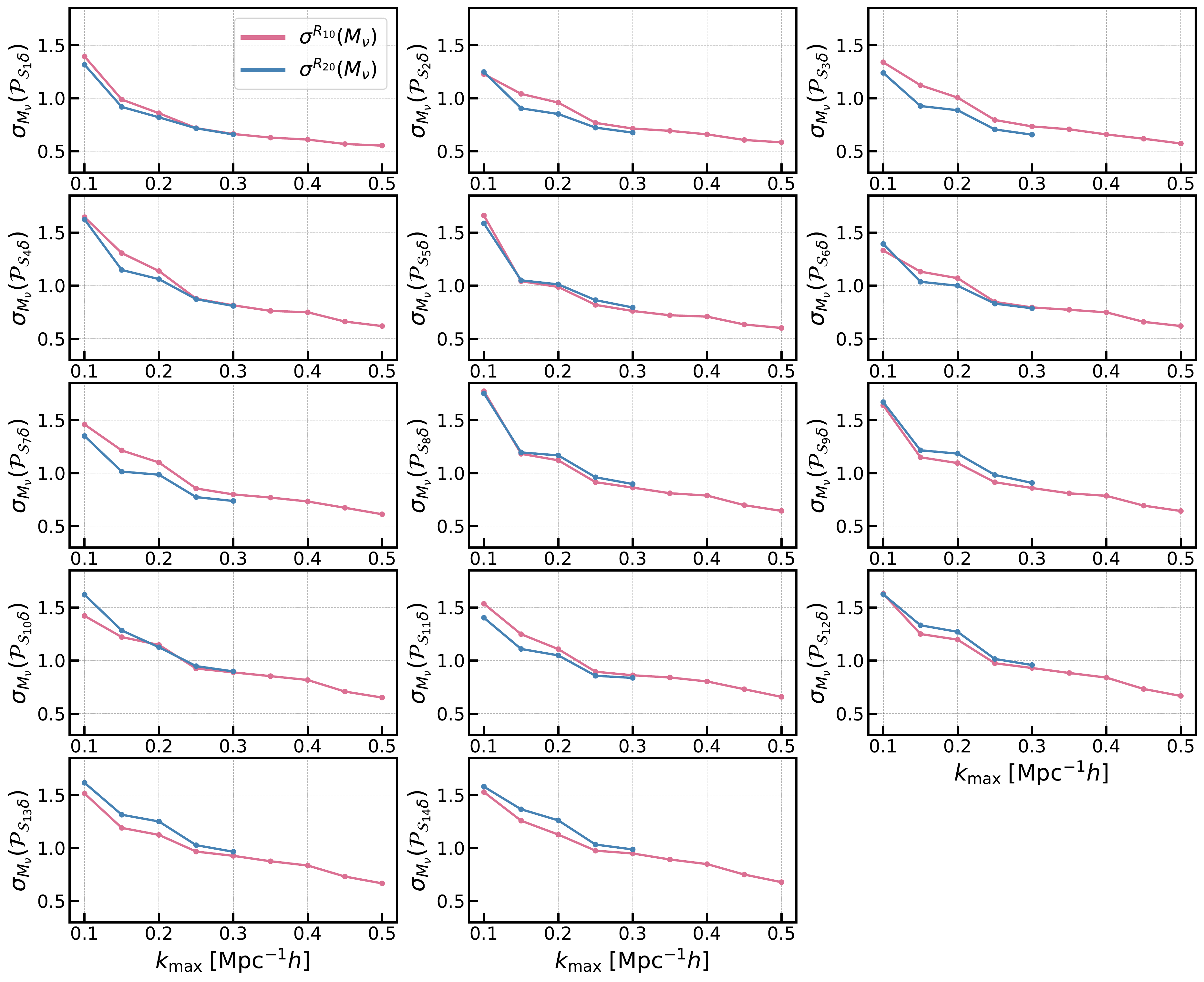}
    \caption{Same as figure \ref{fig:1sigma_kmax_ind_smoothing_omegam}, but for total mass of neutrinos, $M_\nu$.}
    \label{fig:1sigma_kmax_ind_smoothing_mnu}
\end{figure}

\begin{figure}
    \centering
    \includegraphics[width=0.75\textwidth]{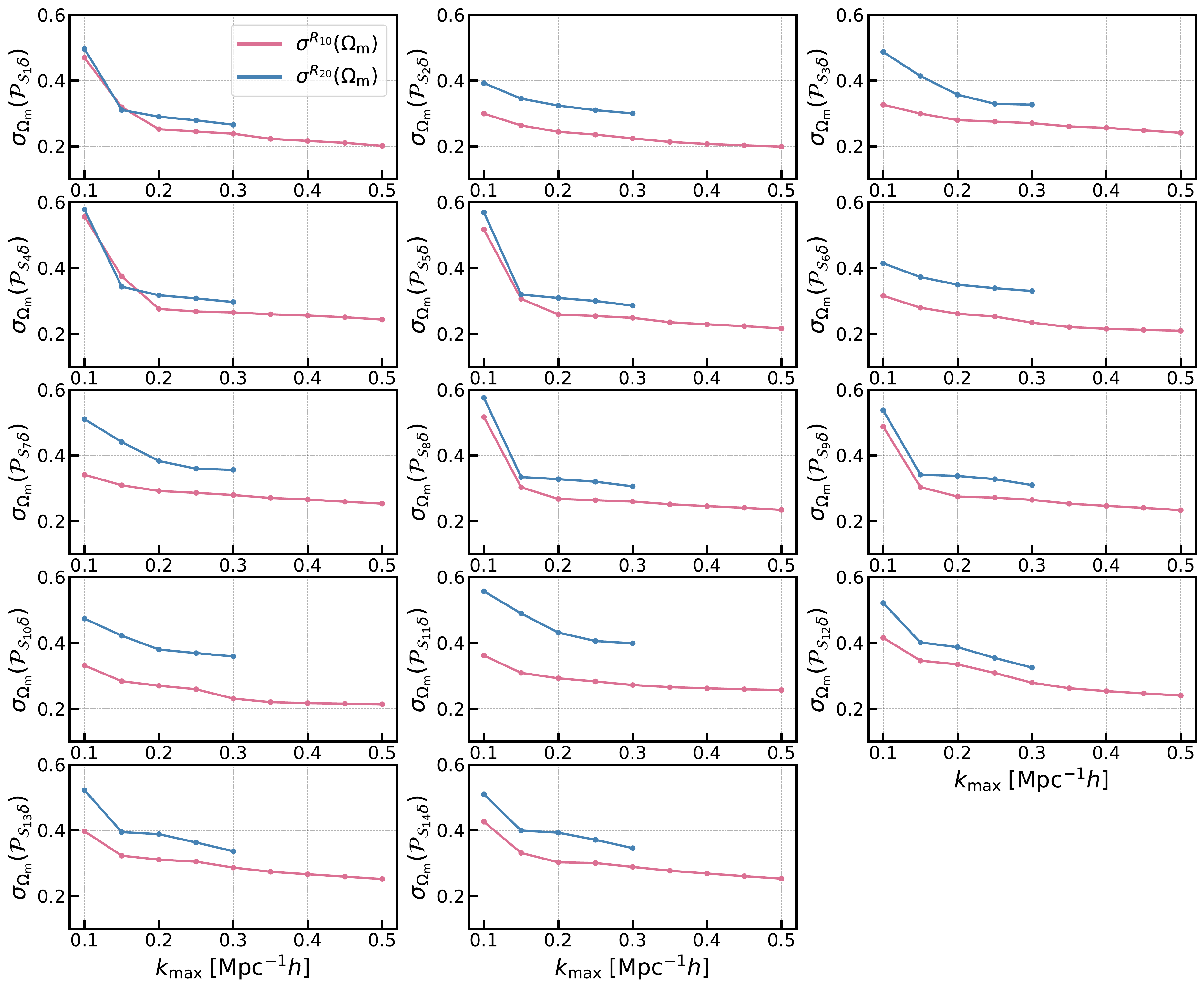}
    \caption{Marginalized 1-$\sigma$ constraints on total matter overdensity, $\Omega_m$, from individual skew spectra as a function of $k_{\rm max}$, assuming smoothing scales of $R=20\, \mpch$ (blue) and $R=10\, \mpch$ (magenta).}
    \label{fig:1sigma_kmax_ind_smoothing_omegam}\vspace{0.2in}
\end{figure}

To highlight the impact of the choice of smoothing scale on the obtained constraints for a given value of the small-scale cutoff in Figs. \ref{fig:1sigma_kmax_ind_smoothing_omegam} and \ref{fig:1sigma_kmax_ind_smoothing_mnu}, we show the marginalized 1-$\sigma$ constraints on $\Omega_m$ and $M_\nu$ from individual skew spectra with $R=20\, \mpch$ (blue) and $R=10\, \mpch$ (magenta). For other parameters, the trend is similar to that of $\Omega_m$, so we do not plot them here. For $\Omega_m$, whenever the skew spectra have similar shape and amplitude, the constraints for the two smoothing scales are very similar. For example, for $\cP_{\cS_1 \delta}$ and $\cP_{\cS_4 \delta}$ on scales $k_{\rm max}\lesssim 0.15$, the constraints are the same since the overall shape of these skew spectra fully overlap on very large scales. The constraints on $M_\nu$ are nearly independent of the choice of the smoothing scale on almost all scales and for all skew spectra, with the exception of the results on $k_{\rm max} < 0.1$, where the constraints from some of the skew spectra with smaller smoothing scale are tighter. 

\section{Stability Tests} \label{app:convergence}

To test the robustness of the numerical Fisher forecast, we investigate the stability of the constraints w.r.t. variation of the number of mocks used in measuring the covariance and the derivatives. Regarding the covariance, we follow Ref. \cite{Dodelson:2013uaa} to estimate the additional variance in the parameter constraints due to uncertainty in the covariance matrix estimation (resulting from an insufficient number of mocks). For the derivatives, we test the variation of the 1-$\sigma$ marginalized constraints as a function of the number of mocks used in the estimation of the derivatives, $N_{\rm deriv}$. In this case, we are able to quantify the amount by which the reported error bars are underestimated due to the lack of convergence of the measured derivatives and how many mocks are needed to achieve full convergence. We also investigate whether applying a smoothing using Gaussian processes (GP), as was proposed for the bispectrum in Ref. \cite{Coulton:2022rir}, provides better convergence.

\subsection{Additional variance due to covariance noise}
\label{app:cov_variance}

The sampling covariance matrix inferred from a limited number of mocks is contaminated by noise. It was shown in~\cite{Dodelson:2013uaa} that the covariance noise propagates into parameter constraints and adds an additional variance component, which up to the second order (s.o.) is given by
\begin{eqnarray}
    \left.\left\langle \theta_\alpha \theta_\beta\right\rangle\right|_{\text {s.o. }}=B F_{\alpha \beta}^{-1}\left(N_{\rm b}-N_{\rm p}\right).
\end{eqnarray}
Here, $N_{\rm b}$ is the data vector's size, and $N_{\rm p}$ is the number of free parameters. Throughout the paper, we have assumed that the data vector of skew spectra follows a multi-variate Gaussian distribution given by the central limit theorem; in this limit, the coefficient $B$ derived in Ref.~\cite{Taylor:2012kz} 
\begin{eqnarray}
    B=\frac{N_{\rm s}-N_{\rm b}-2}{\left(N_{\rm s}-N_{\rm b}-1\right)\left(N_{\rm s}-N_{\rm b}-4\right)}.
\end{eqnarray}
In the limit where the number of simulations $N_{\rm s}$ is much larger than the size of the data vector, the additional correction approaches one. 

\begin{figure}[htbp!]
\centering
\includegraphics[scale=0.38]{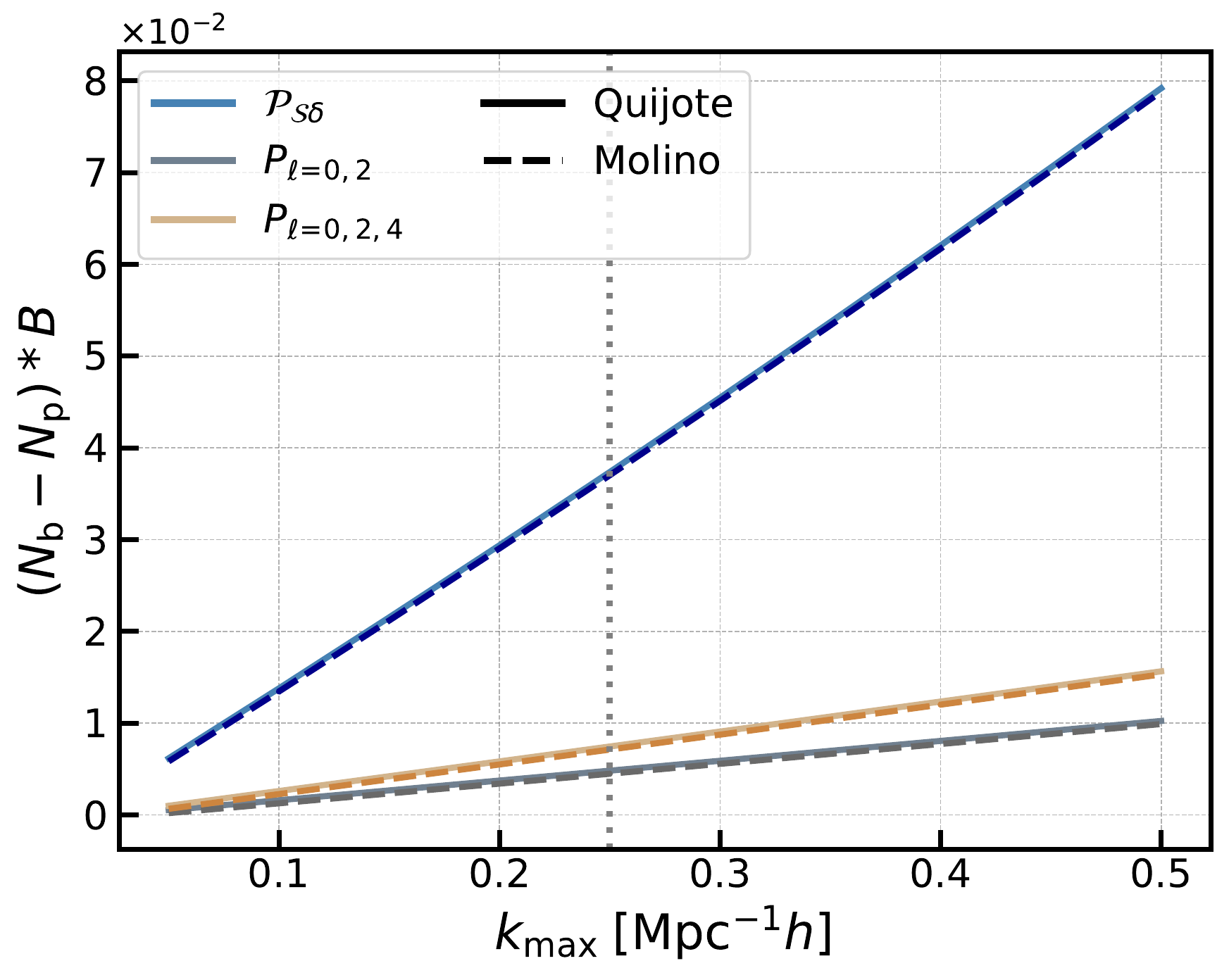} 
\caption{Correction to the parameter constraints due to the noise in the sampling covariance matrix.}
\label{fig:facB_correction_SSxPk}
\end{figure}

In figure~\ref{fig:facB_correction_SSxPk}, we show the correction to parameter constraints due to misestimation of the covariance of the skew spectra (blue) and power spectrum multipoles ($\ell=0,2$ in grey and $\ell=0,2,4$ in yellow) from Quijote (solid lines) and Molino (dashed lines) samples. At $k_{\rm max}=0.25 \,\hmpc$, using 15,000 mocks leads to an additional variance $\sim\! 3.6\%$ for skew spectra or $\sim\! 1.8\%$ in terms of 1-$\sigma$ constraints, while the additional correction is at the sub-percent level for the power spectra given the smaller data vector size. These corrections are significantly smaller than the uncertainties due to the derivative estimation for the current number of simulations, as will be discussed in the following section~\ref{app:unsmooth_deriv}. 

\subsection{Convergence of the Numerical Derivatives Estimates}
\label{app:unsmooth_deriv}

Numerical derivatives obtained from a finite number of simulations may suffer from statistical noise; this additional variance can lead to underestimation of the forecasted parameter uncertainties, rendering the commonly used convergence tests unreliable. Therefore, it is crucial to test the stability of constraints w.r.t. variation of the number of mocks.

To estimate the stability of the numerical derivatives, we calculate the 1-$\sigma$ constraint at a given number of mocks $N_{\rm deriv}$ by averaging over the three LoS dependencies. It is important to notice that the convergence rate is sensitive to the correlation between the error inferred from each individual realization. Mixing correlated data points (different LoS from the same realization) and uncorrelated data points (different realizations) can artificially enhance the convergence rate. Hence we average the three LoS for each realization as a conservative estimate of the convergence rate. By definition, the ratio between the 1-$\sigma$ constraints should approach the truth when $N_{\rm deriv}\rightarrow \infty$, we expect the curve to show an asymptotic behavior. We therefore fit the constraints for each parameter $\sigma_{\theta}(N_{\rm deriv})$ using an empirical function with an exponential form, $f(x) = a + b(1-e^{-c x})$, which takes three fitting parameters $\{a,b,c\}$, to fit the points up to $N_{\rm deriv}=500$. The fitting procedure calls the \textsc{Scipy} routine $\mathsf{scipy.optimize.curve\_fit}$.

\begin{figure}[htbp!]
\centering
    \begin{subfigure}{0.44\linewidth}
    \centering
        \includegraphics[width=0.75\textwidth]{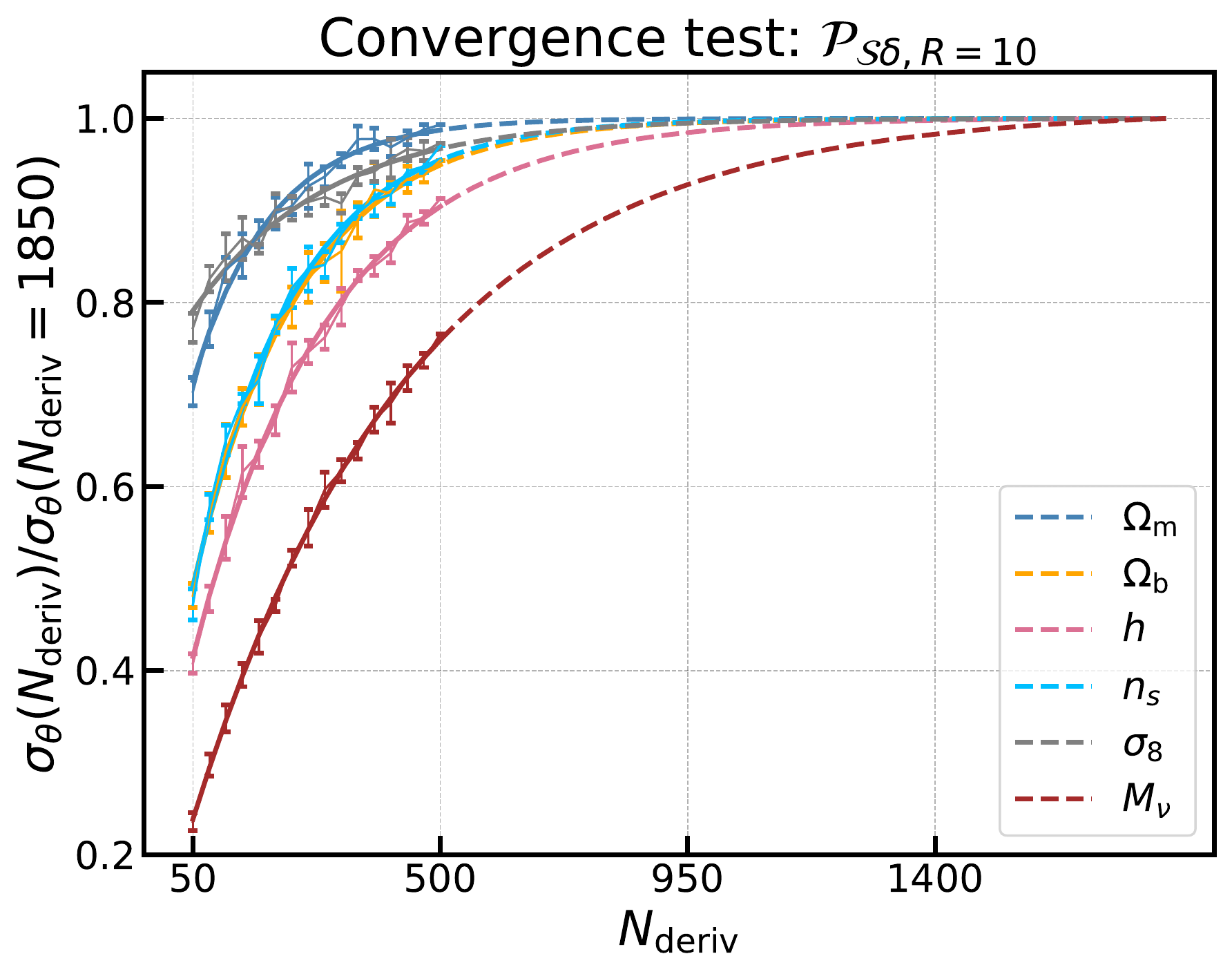}
        \includegraphics[width=0.75\textwidth]{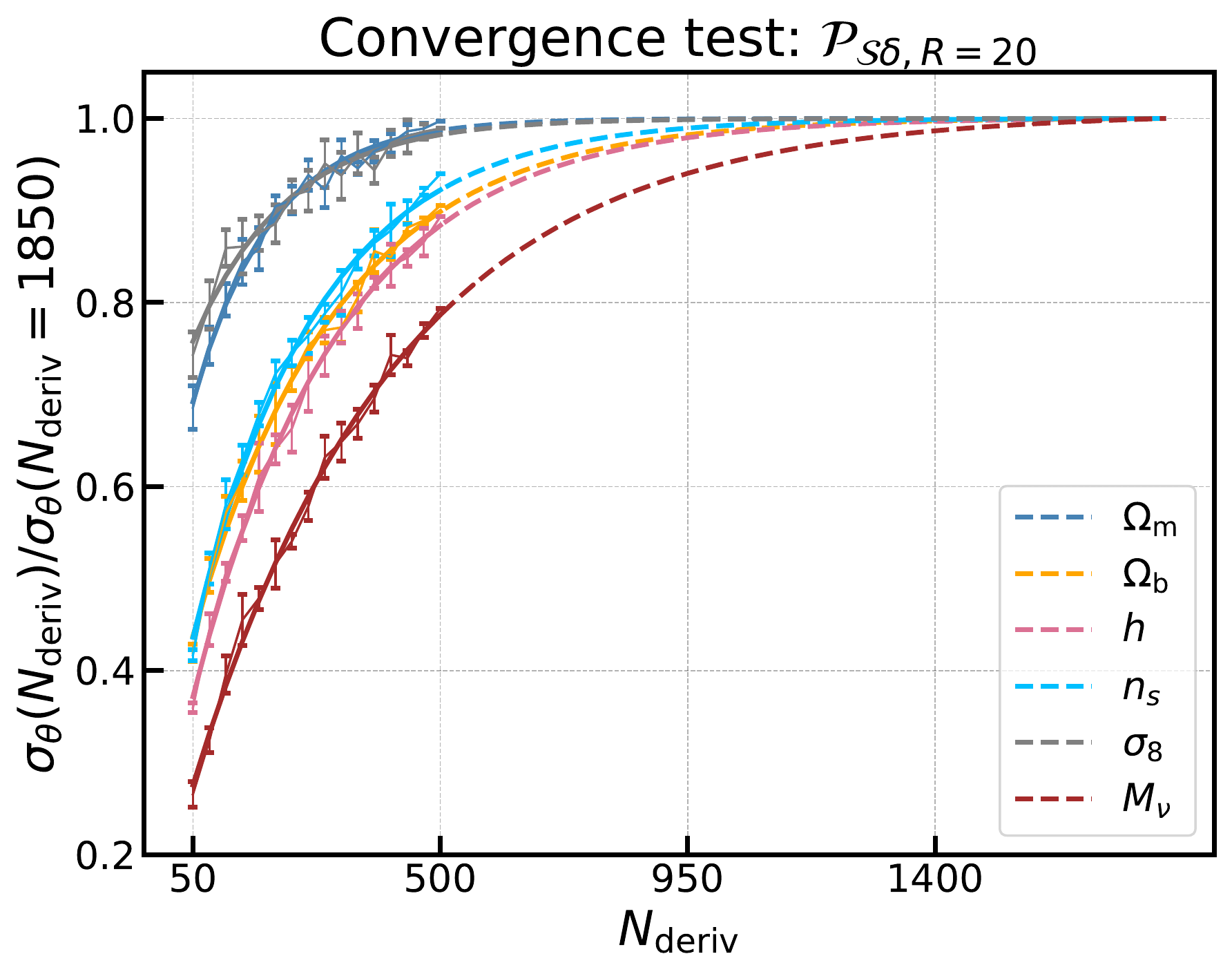}
        \includegraphics[width=0.75\textwidth]{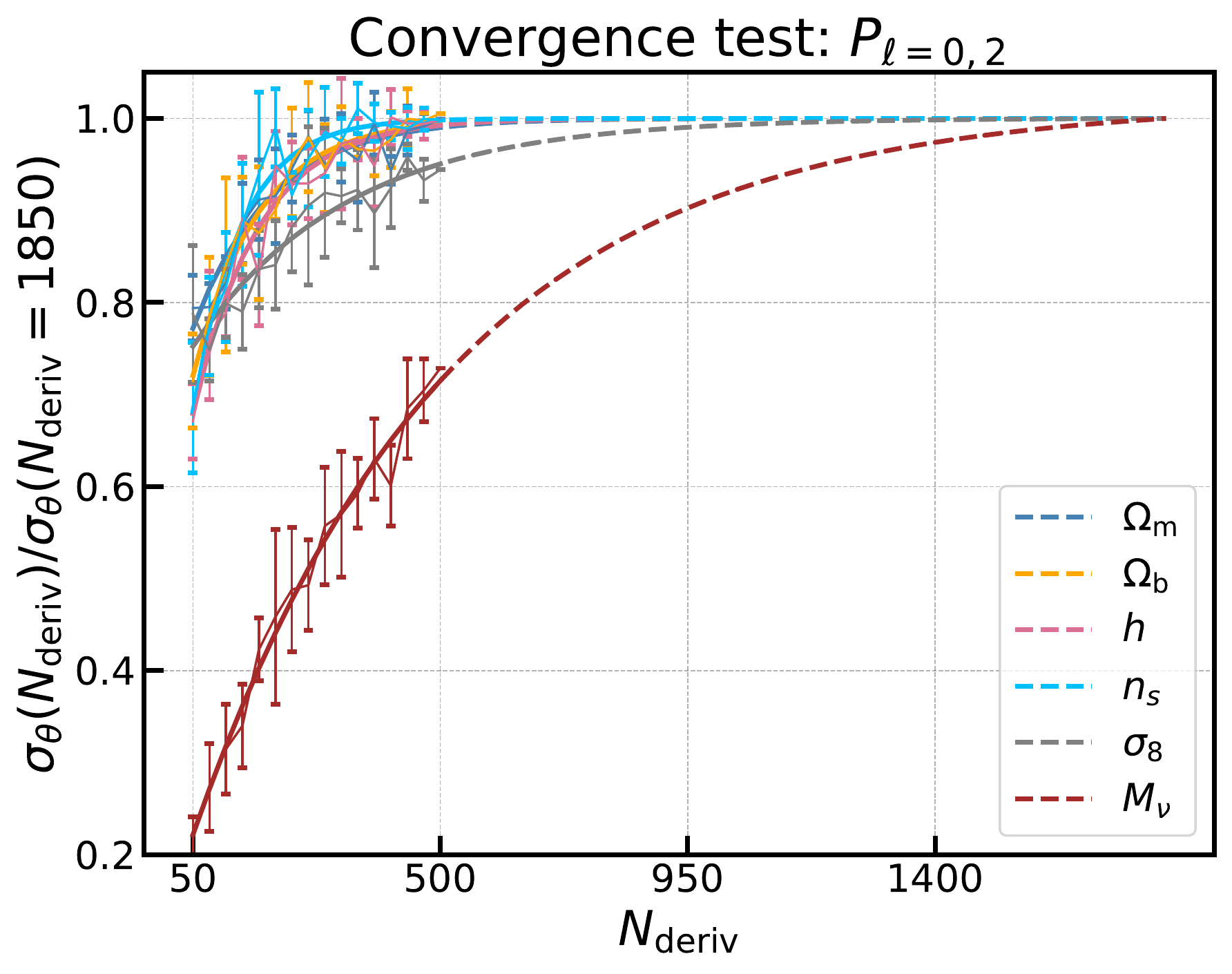}
        \caption{Quijote Halo catalogs}
    \end{subfigure}
    \begin{subfigure}{0.44\linewidth}
    \centering
        \includegraphics[width=0.75\textwidth]{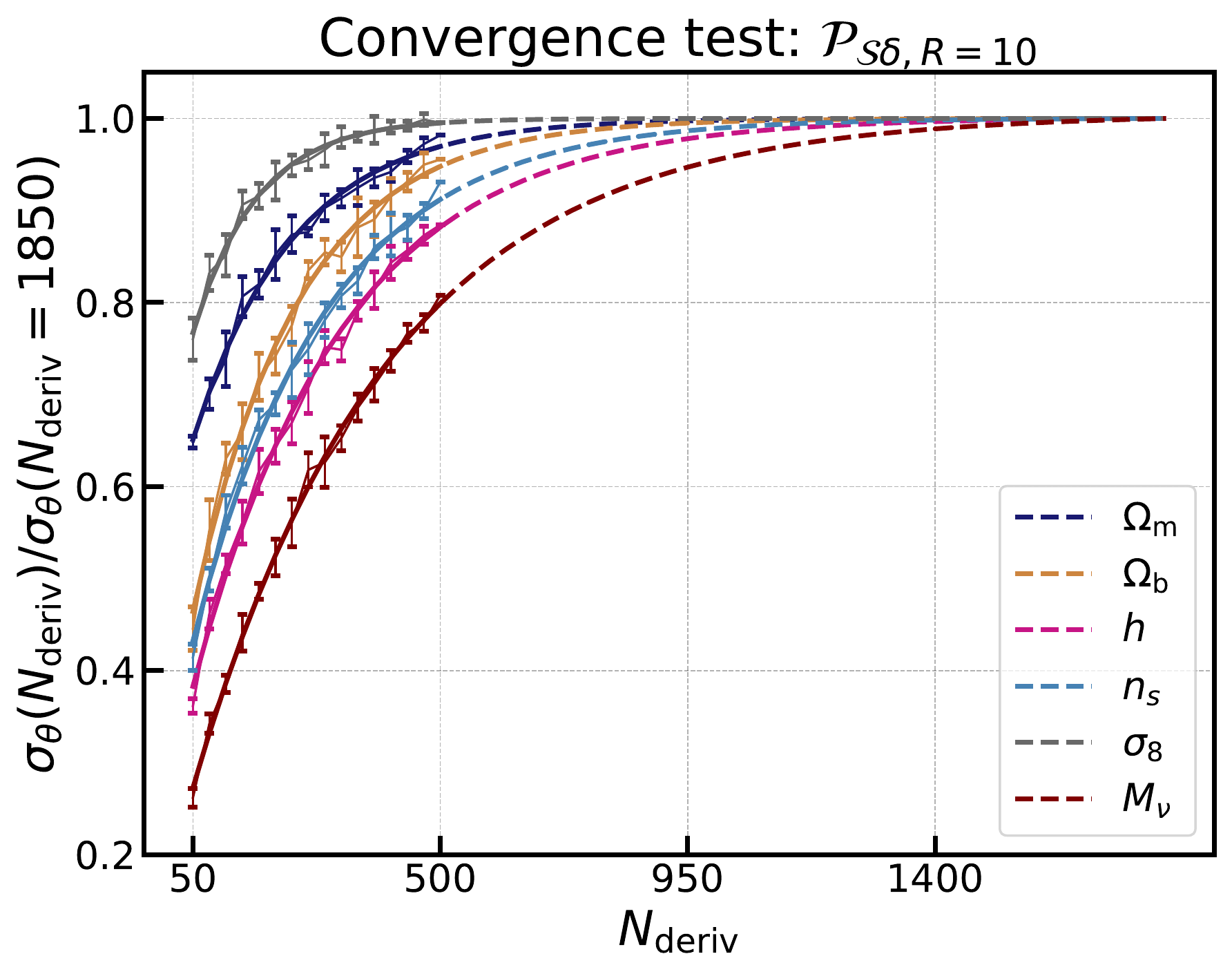}
        \includegraphics[width=0.75\textwidth]{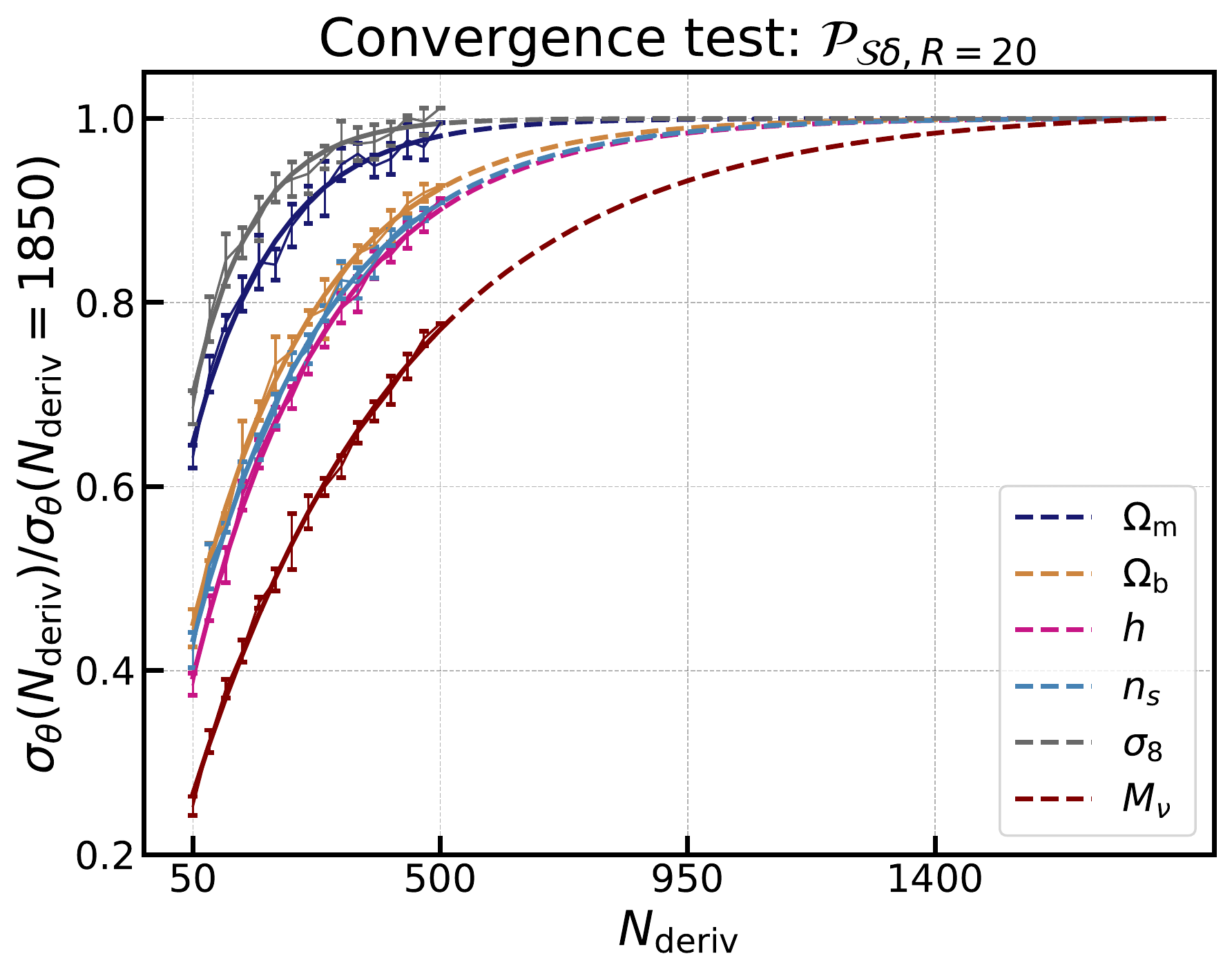}
        \includegraphics[width=0.75\textwidth]{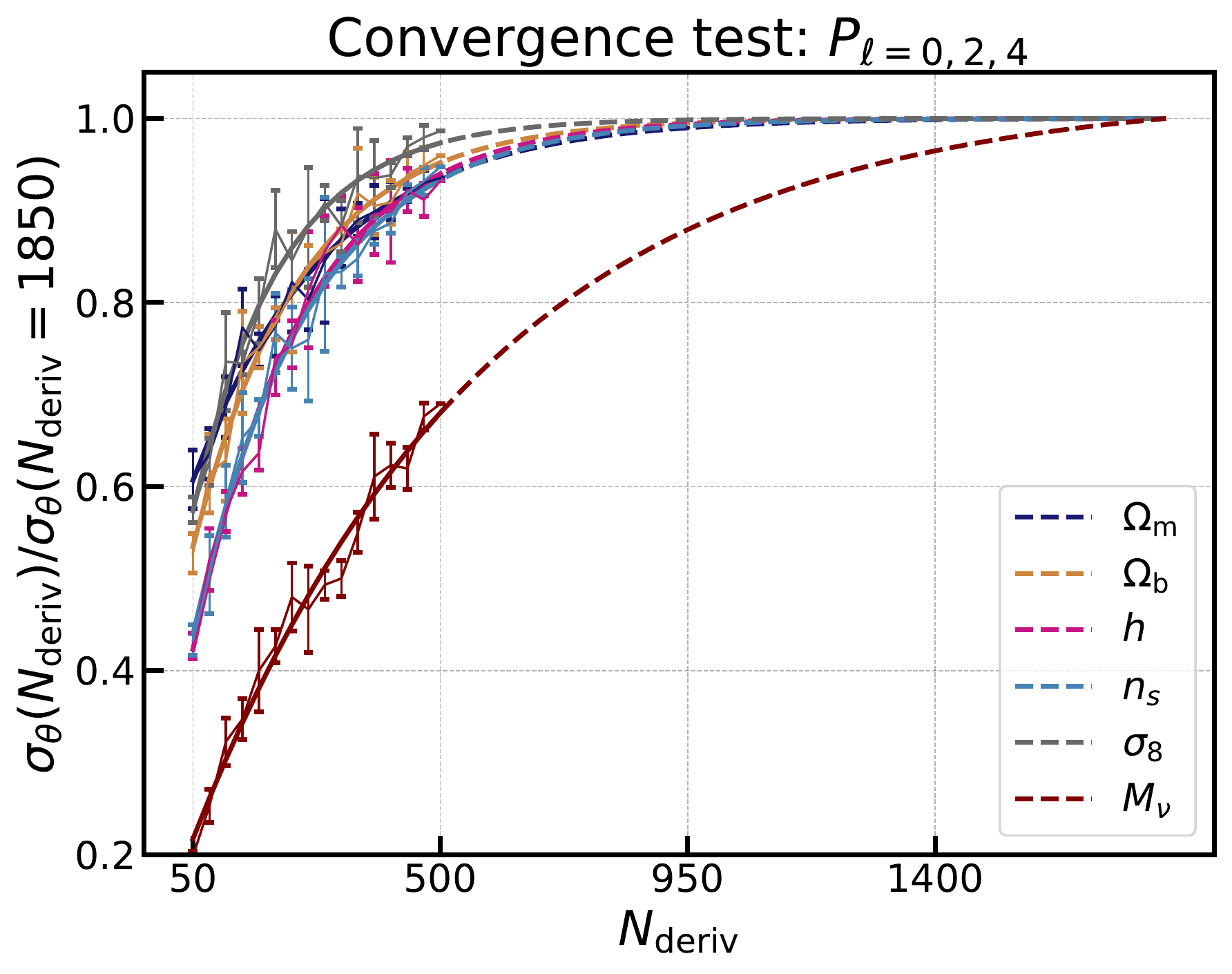}
        \caption{Molino Galaxy catalogs}
    \end{subfigure}
\caption{The convergence of marginalized 1-$\sigma$ parameter constraints from skew spectra, as a function of the number of mocks used for measuring the derivatives, $N_{\rm deriv}$. The left panels show the constraints from Quijote halos, while the ones on the right are from Molino galaxies. The dashed lines are the extrapolation of the convergence curve to $N_{\rm deriv}=1850$. The smoothing scale is set to $R=10\, \mpch$.}
\label{fig:quijote_molino_convergence_deriv_ss10_kmax0.25}
\end{figure}

In figure \ref{fig:quijote_molino_convergence_deriv_ss10_kmax0.25}, we show the the dependence of 1-$\sigma$ marginalized constraints on the number of mocks used in computing the derivatives, $N_{\rm deriv}$. We fit $\sigma_{\theta}(N_{\rm deriv})$ for the first 500 points and extrapolate the curve to $N_{\rm deriv}=1850$ (dashed curve). The y-axis is normalized to $\sigma(N_{\rm deriv}=1850)$. The panels on the left correspond to results for Quijote halos, and the ones on the right are from Molino galaxies. The first two rows show the convergence for skew spectra with $R=10\, \mpch$, and $R=20\, \mpch$, respectively. As a reference, we also show the convergence of the power spectrum multipoles, $P_{\ell=0,2}$ for Quijote and $P_{\ell=0,2,4}$ for Molino. The error bar in each plot is inferred by randomly drawing $N_{\rm deriv}$ from the LoS-averaged mocks repetitively by 5 times. Given the non-vanishing error bar, the sequence of drawing the realizations can also have an impact on the convergence behavior.

Overall, we find that for either Quijote or Molino, the uncertainty on the total mass of neutrinos is most under-predicted. The under-prediction ranges from $(20-30)\%$ for both power spectrum and skew spectra on both synthetic datasets. For the rest of the parameters, the power spectrum converges faster than the skew spectra in general. For either Quijote or Molino, the under-prediction by the skew spectra varies from $(1-15)\%$ and for power, spectrum ranges from sub-percent to $7\%$.

\subsection{Post-processing of the Numerical Derivatives}
\label{app:convergence_GP}

Using the procedure described in the previous subsection, we confirmed the stability of our results w.r.t. variation in $N_{\rm deriv}$. In order to make a connection with the recent analysis of Ref. \cite{Coulton:2022rir}, here we discuss whether smoothing the numerical derivatives provides an alternative way to assure the robustness of the forecasts for skew spectra. 

It was proposed in Ref. \cite{Coulton:2022rir} that smoothing the measured derivative, for example, by applying a Gaussian process (GP) \cite{books/lib/RasmussenW06} to obtain a non-parametric estimation of the underlying function, can mitigate the effect of noisy derivatives on forecasted parameter uncertainties. This issue was shown to be substantial for the bispectrum analysis. Even though unlike the bispectrum, the measured derivatives of the skew spectra are rather smooth, we investigate whether using the smoothed derivatives has the same effect on parameter constraints from skew spectra. Contrary to the findings of Ref. \cite{Coulton:2022rir}, we found that the GP smoothing artificiality reduces the parameter uncertainties. After discussing the basic procedure for GP fitting that we use, through analytic approximations, we illustrate why the situation is different for the skew spectra compared to the power spectrum and bispectrum.

\begin{figure}[t]
    \centering
    \includegraphics[width=.9\textwidth]{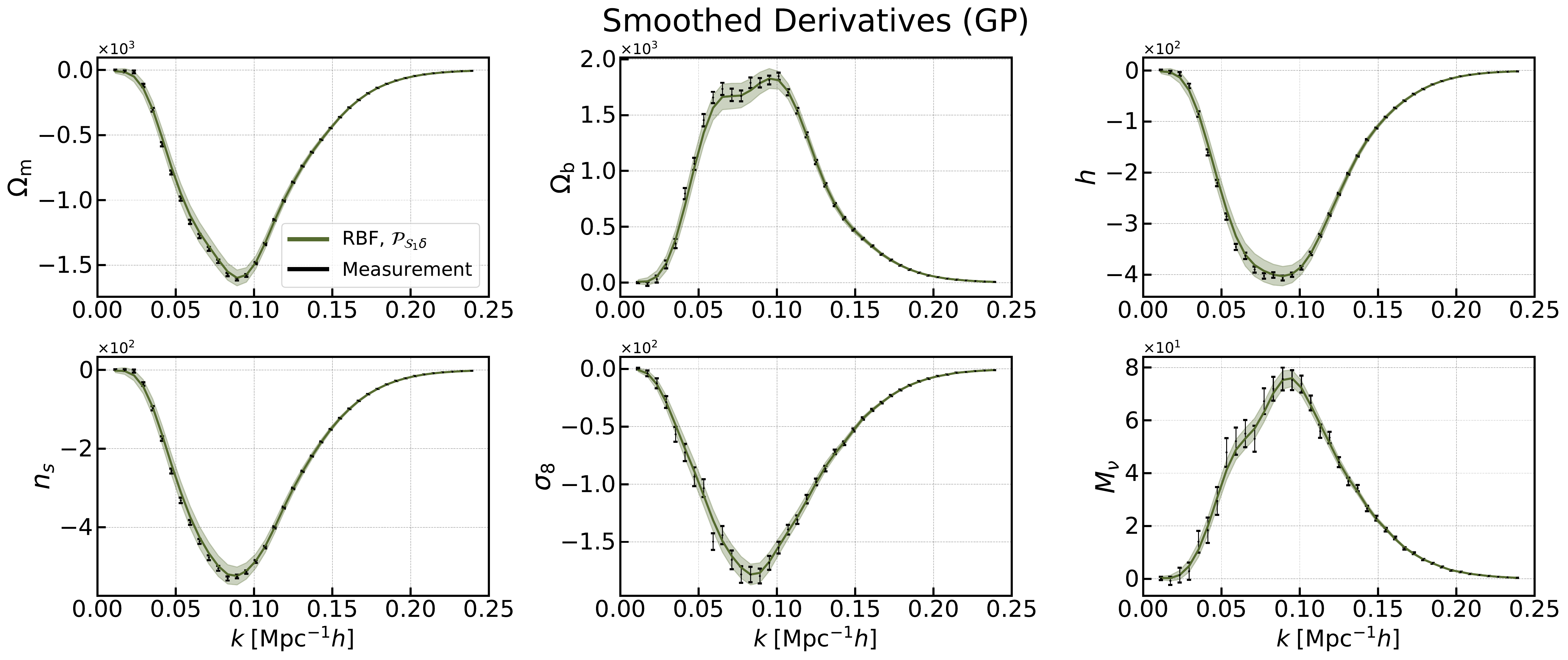}
    \caption{Fitted Gaussian processes to smooth the derivatives of the first skew spectrum $\mathcal{P}_{\mathcal{S}_1\delta}$. Here we display the smoothed derivatives (solid green), the expected error for GP (shaded green), and the original measurement (black) for the six cosmological parameters.}
    \label{fig:ss1_grp}\vspace{-0.1in}
\end{figure}

As in \cite{Coulton:2022rir}, we use the implementation of GP in \textsc{skickit-learn} and use an isotropic ``Radial-basis function" (RBF) to describe the correlation between the data points. In order to reduce the dynamical range of the data vector, we normalize both the data vector and its error by the mean of the covariance matrix. Due to the large difference in the amplitudes among the 14 skew spectra, we fit them individually. As an example, we show the smoothed derivatives of the first skew spectrum, $\cP_{\cS_1 \delta}$, and the fitted Gaussian Processes in figure \ref{fig:ss1_grp}. Here, we have fitted the data up to $k_{\rm max} = 0.25 \ \hmpc$. The shaded bands show the error for the GP fit. We see that the GP smoothed derivatives give a good description of the original data and remove the statistical fluctuations. The major difference for the data vector consisting of the 14 skew spectra is that the skew spectra have a large correlation between the 14 terms, whereas the covariance for either power spectrum multipoles or the bispectrum is almost diagonal. As we will see below, this difference is what is driving the reduction of parameter uncertainties in the case of smooth derivatives.

In the presence of noise, the Fisher matrix is modified as
\begin{align}
    F'_{\alpha\beta} &= \sum_{i,j = 1}^{N_d} \frac{\partial (\bar{d}_i+\delta d_i)}{\partial \theta_\alpha}\, {\mathbb C}^{-1}_{ij}\,  \frac{\partial (\bar{d}_j+\delta d_j)}{\partial\theta_\beta}\\ \nonumber
    &= \sum_{i,j = 1}^{N_d} \frac{\partial \bar{d}_i}{\partial \theta_\alpha}\, {\mathbb C}^{-1}_{ij}\,  \frac{\partial \bar{d}_j}{\partial\theta_\beta} + \left[\left(\frac{\partial \bar{d}_i}{\partial \theta_\alpha}\, {\mathbb C}^{-1}_{ij}\,  \frac{\partial \delta d_j}{\partial\theta_\beta} + i\leftrightarrow j\right)+ \frac{\partial \delta d_i}{\partial \theta_\alpha}\, {\mathbb C}^{-1}_{ij}\,  \frac{\partial \delta d_j}{\partial\theta_\beta}\right]\\ \nonumber
    &\equiv F_{\alpha\beta} + \huge[\ldots\huge],
\end{align}
where $\delta d_i$ denotes the statistical fluctuation in the data due to the limited number of mocks and scales as $1/\sqrt{N_{\rm mock}}$, and $\bar{d}_i$ denotes the true signal. In the last equation, apart from the Fisher matrix as the true response of the data vector to the change in parameters, there are two additional terms due to the noise in the square bracket $[\ldots]$. In the case of an almost diagonal covariance matrix, the first noise term in the square bracket averages to zero after the data bin summation for $i, j$. The second noise term is always positive
and enhances the Fisher matrix, which leads to artificially tighter constraints. However, this is not guaranteed in the case of skew spectra as there is a strong correlation across the skew spectra with both positive and negative signs. Also notice that the last term is proportional to the covariance $\av{\delta d_i \, \delta d_j} \sim \hat{\mathbb{C}}_{ij}$, which follows Wishart distribution $\hat{\mathbb{C}}_{ij} \sim W_p(\bar{\mathbb{C}}_{ij}/N_{\rm mock}, N_{\rm mock})$, while the variance of Wishart distributed sample is given by ${\rm Var}(\hat{\mathbb{C}}_{ij}) = (\bar{\mathbb{C}}^2_{ij}+\bar{\mathbb{C}}_{ii}\bar{\mathbb{C}}_{jj})/N_{\rm mock}$. In the limit of $N_{\rm mock} \rightarrow \infty$, the second term in the noise drops out, and we recover the true Fisher information matrix.

The goal of GP process is to remove the statistical fluctuation $\delta d_i$. However, we perform the fitting to each skew spectrum individually, and this step removes the correlation across the skew spectra. The new covariance between the data vector is thus modified to $\av{\tilde{\delta}d_i \, \tilde{\delta}d_j} \sim \tilde{\mathbb{C}}_{ij}$,
where $\tilde{\delta}d_i$ denotes the residual noise after the GP fitting and the covariance for the 14 skew spectra reduces to a subblock-diagonal structure $\tilde{{\mathbb C}} = {\hat{{\mathbb C}}}\, \delta^{\rm K}_{kl}$ for $\{k = {1,\ldots}, 14\}$ denoting the 14 skew spectra and $\delta^{\rm K}_{kl}$ being the Kronecker delta. Denoting the second term in the rectangular bracket as $\Delta F_{\alpha\beta}$ and using the Sherman-Morrison-Woodbury formula as in Eq.~\eqref{eqn:SM_expansion} to expand the noise-induced Fisher term, we have,
\begin{eqnarray}
    \Delta F_{\alpha\beta} &=& \partial \cP_{i,\alpha}\tilde{\mathbb{D}}^{-1} \partial \cP_{j,\beta} \delta_{ij}^{\rm K} - \boldsymbol{\lambda} \frac{(\delta \cP_{i,\alpha} \mathbb{D}^{-1} \delta \cP_{i}) (\delta \cP_{j} \tilde{\mathbb{D}}^{-1} \delta \cP_{j, \beta} )}{\mathbb{I}+\boldsymbol{\lambda} \delta \cP_i \tilde{\mathbb{D}}^{-1} \delta \cP_j}\\ \nonumber
    &\approx& \partial \cP_{i,\alpha}\tilde{\mathbb{D}}^{-1} \partial \cP_{j,\beta} \delta_{ij}^{\rm K} - \boldsymbol{\lambda} {(\delta d_{i,\alpha} \mathbb{D}^{-1} \delta d_{i}) (\delta \cP_{j} \tilde{\mathbb{D}}^{-1} \delta \cP_{j, \beta} )}
    \label{eqn:correction_fisher}
\end{eqnarray}
here we have generalized the notation and grouped the full covariance matrix into subblocks corresponding to the 14 skew spectra, with $i,j=\{1,\ldots,14\}$. $\tilde{\mathbb{D}}$ denotes the diagonal of subblock of the covariance matrix. Simplifying the partial derivative as $\delta \cP_{i,\alpha} \equiv \partial \delta \cP_i/\partial \theta_{\alpha}$. The correlation coefficients across the skew spectra is denoted by $\boldsymbol{\lambda}$ tensor and the GP process replaces $\boldsymbol{\lambda}\rightarrow \tilde{\boldsymbol{\lambda}}$ by further removing the sub-block of the covariance. Going from the first to the second line, we have used the fact that $\boldsymbol{\lambda}\ll \mathbb{I}$ and ignore higher order $\cO({\boldsymbol{\lambda}}^2)$. In the case of skew spectra, most of the coupling tensors $\boldsymbol{\lambda}$ are positive, and only those associated with $\cP_{\cS_{3,7,11}\delta}$ are negative. Thus removing off-diagonal subblocks overall help to improve the constraints.
While in the case of the power spectrum or bispectrum, different multipoles are much less correlated, and $\lambda$ is very close to zero. The second term in Eq.~\eqref{eqn:correction_fisher} does not contribute, and fitting each individual multipole is equivalent to fitting the whole data vector.

\section{Shot Noise Subtraction}\label{app:shot}

In our analysis of the Quijote halos, we used the shot-subtracted skew spectra, while for Molino galaxies, for computational convenience, we retained the shot noise in the estimator of the skew spectra. In this section, we first show the shape of the shot noise contribution to skew spectra and then discuss whether the parameter constraints are affected by the subtraction of shot noise. 

\begin{figure}[t]
    \centering
    \includegraphics[width=.75\textwidth]{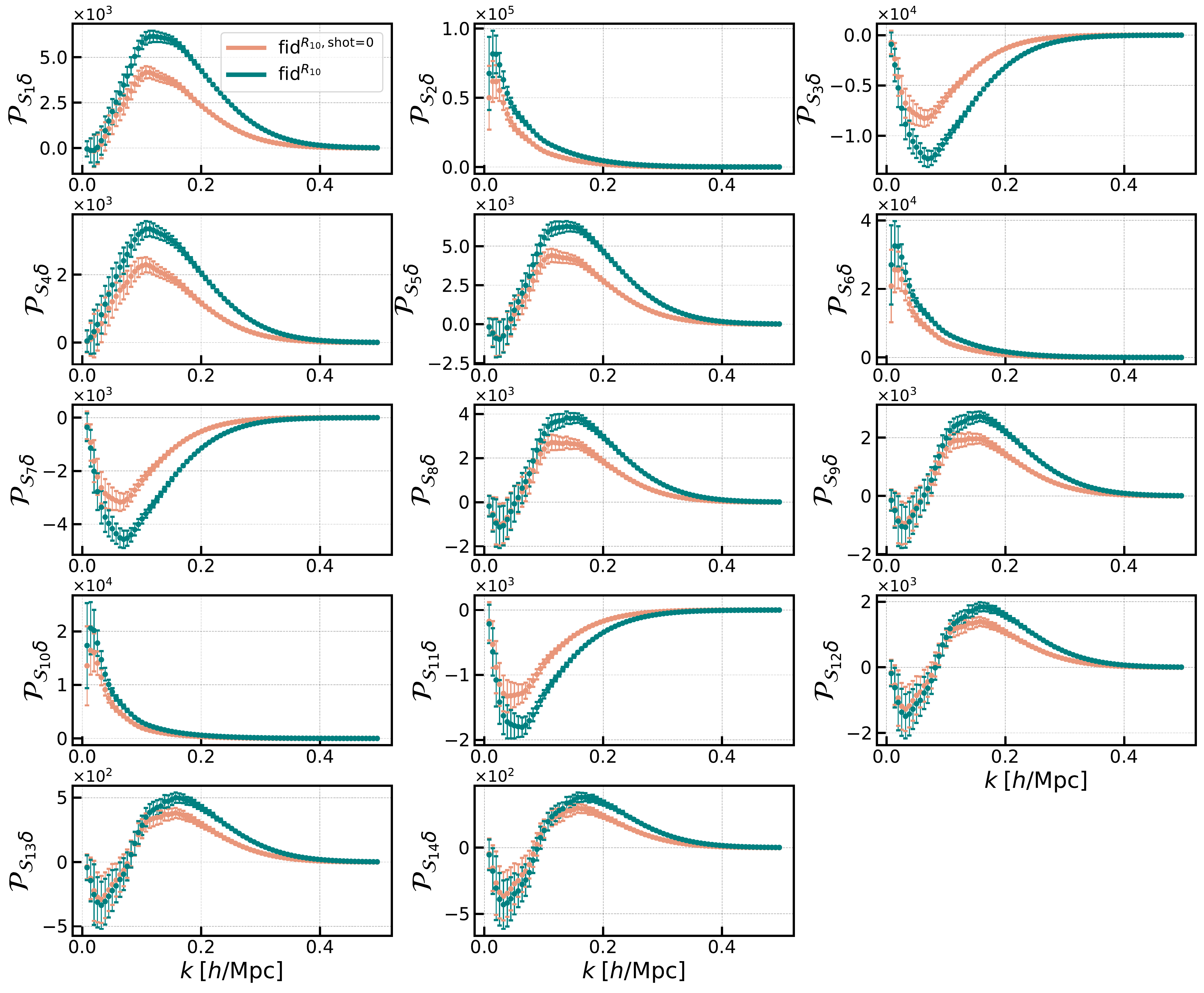}
    \caption{Impact of the shot noise on the skew spectra of Quijote halos. The green data points include the shot noise, while the orange points have the shot subtracted. The measurements are performed on Quijote halo catalogs at reference cosmology.}
    \label{fig:quijote_ss_smooth10_shot01_fid}
    \vspace{0.1in}
    \centering
    \includegraphics[width=.75\textwidth]{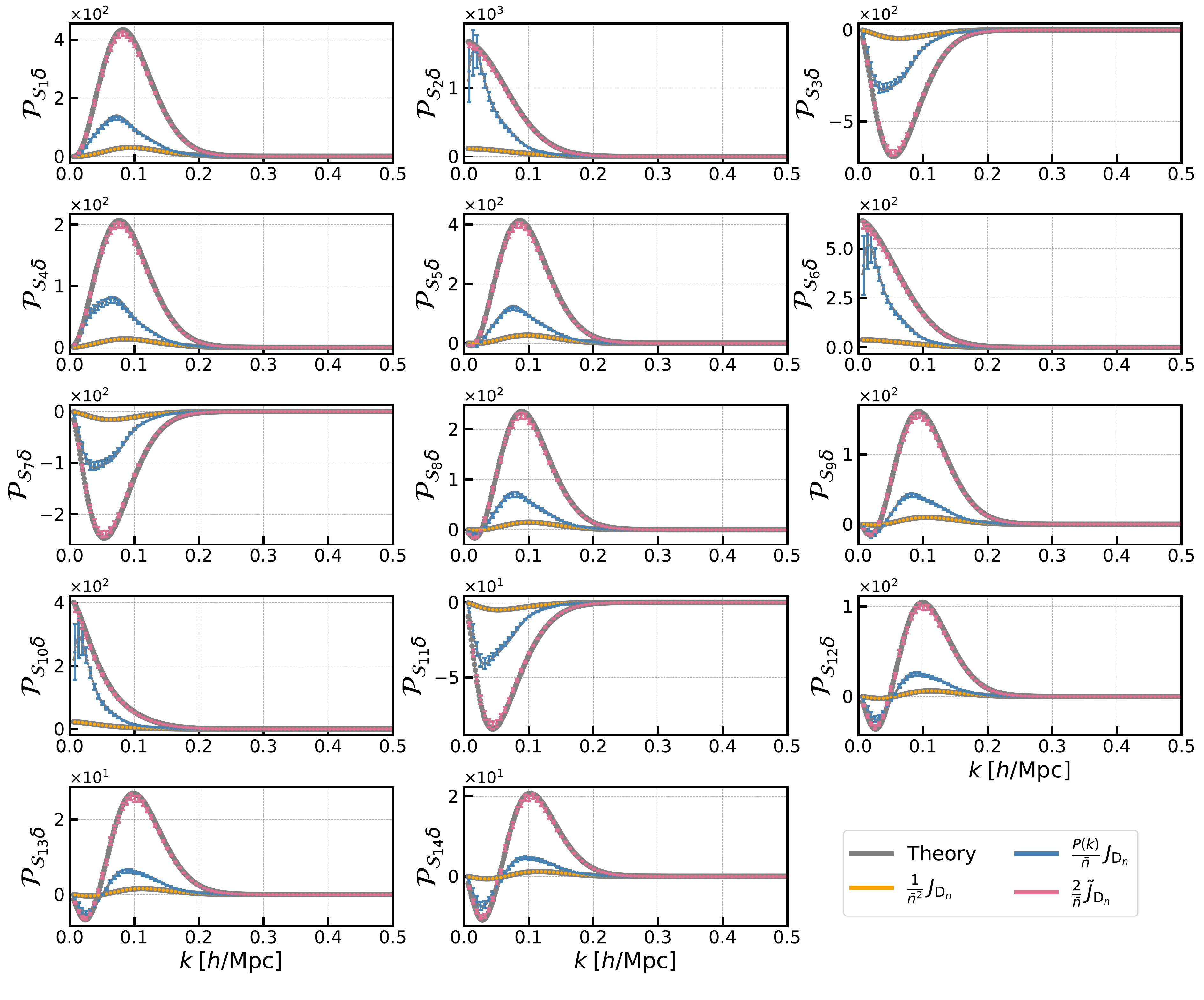}
    \caption{Measured shot noise contributions of skew spectra of Quijote halos (orange, blue, magenta) compared with the theoretical prediction (grey). The smoothing scale is set to $R=20\mpch$ since the perturbative theoretical model does not fare so well for the smaller smoothing scale. For demonstration, in this plot we show the mean of 100 mocks and with the error bar inferred from their standard deviation.}\vspace{-0.1in}
    \label{fig:quijote_ss_smooth20_shot123_fid}
\end{figure}
In figure \ref{fig:quijote_ss_smooth10_shot01_fid}, we show the measured skew spectra of Quijote halos, including shot noise (teal) and subtracting it (orange). The shot noise enhances the amplitude of the skew spectra but preserves their shape. This is due to the fact that the shape of the spectra is largely determined by the operators used in defining of the quadratic fields, $\cS_n$. In figure \ref{fig:quijote_ss_smooth10_shot01_fid}, we show the measurement of the three contributions to skew spectra shot noise, described in \S\ref{subsec:shot} and \S\ref{subsec:pipeline}, together with the theoretical predictions (grey). Here we input the linear bias, which is measured from the ratio between the Quijote halo power spectrum  and the matter power spectrum at large scale limit $b_1=\sqrt{P_{\rm halo}/P_{\rm mm}}=1.7$ for $k_{\rm max} = 0.1\, \hmpc$. In order to check the numerical shot noise implementation as discussed in \S\ref{subsec:shot}, we compare the measurement to the theory prediction by applying a smoothing scale $R=20 \,\mpch$.\footnote{It was shown in~\cite{Schmittfull:2020hoi} that perturbation theory-based prediction underestimates the power when compared to a smaller smoothing scale with $R=10\,\mpch$ since we aim for a sanity check for the numerical implementation we choose a scale that isolates the effects caused by theory uncertainties.} The error bar is derived from the standard deviation of the 100 simulations. Here we can see an excellent agreement between the theoretical prediction and the measurement.

\begin{figure}[t]
    \centering
    \includegraphics[width=0.8\textwidth]{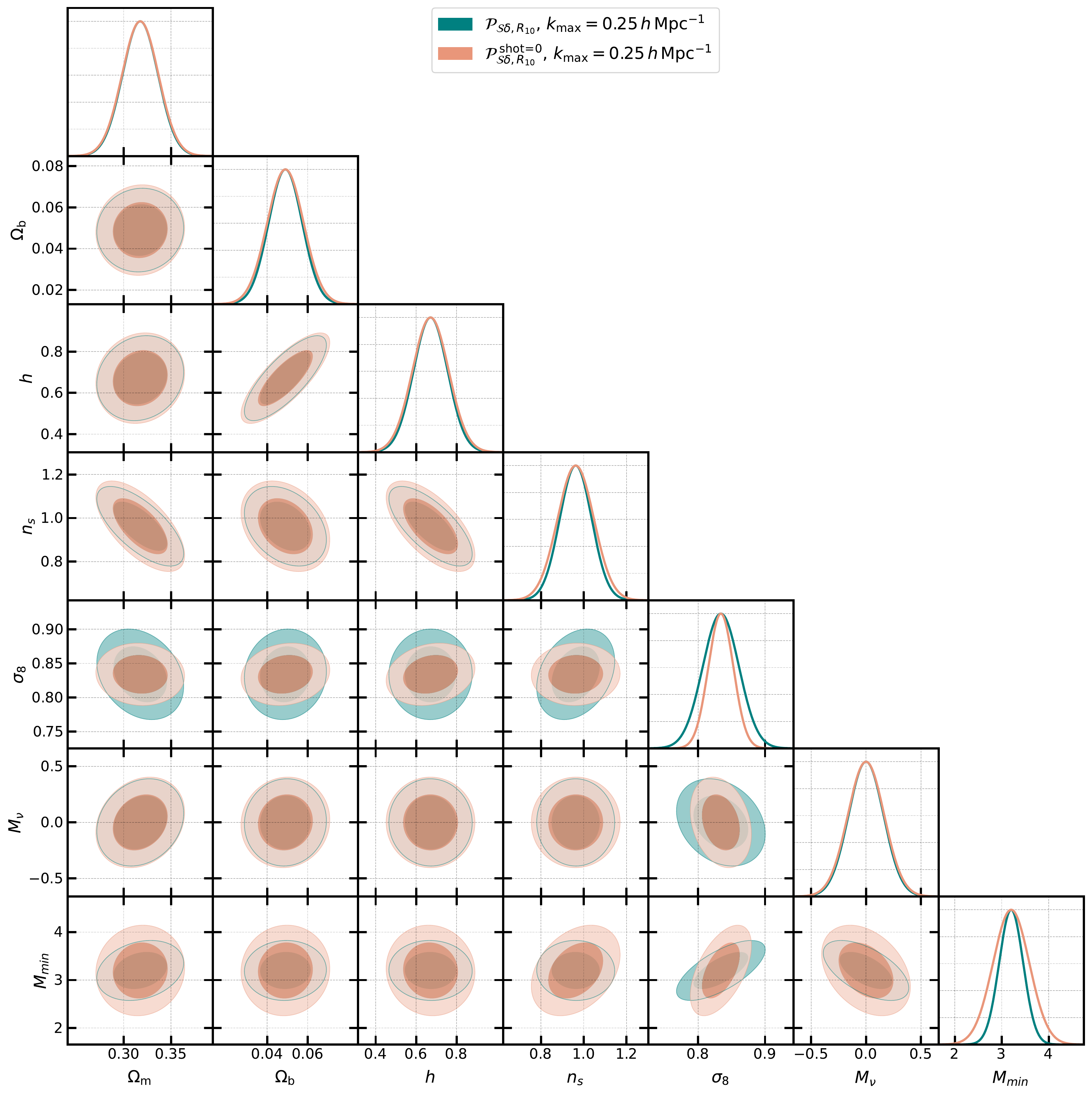}
    \caption{Marginalized 1- and 2-$\sigma$ parameter constraints with (orange) and without (teal) shot subtraction. Shot noise affects mostly $\sigma_8$ and $M_{\rm min}$.}
    \label{fig:2dcontour_ss_smooth10_shotonxoff_nuisMmin}
\end{figure}

In figure \ref{fig:2dcontour_ss_smooth10_shotonxoff_nuisMmin}, we show the 2D marginalized constraints from skew spectra of Quijote halos with (orange) and without (teal) shot noise subtraction. We see that the shot noise subtraction has a marginal impact on almost all parameters except for $\sigma_8$ and the nuisance parameter $M_{\rm min}$. Since $M_{\rm min}$ is inferred from the halo catalogs with two different mass cuts $M_{\rm cut}^+=3.3\times 10^{13}\, M_{\odot}$ and $M_{\rm cut}^-=3.1\times 10^{13}\, M_{\odot}$, each of them has different number density. Shot noise can lead to an artificial response to the bias constraints encoded in $M_{\rm min}$. Given that $\sigma_8$ is positively correlated with $M_{\rm min}$, this effect is also propagated into the $\sigma_8$ constraints.

\section{Comments on Molino Galaxy Sample}\label{app:Molino}

\begin{figure}[t]
\centering
\includegraphics[width=0.4\textwidth]{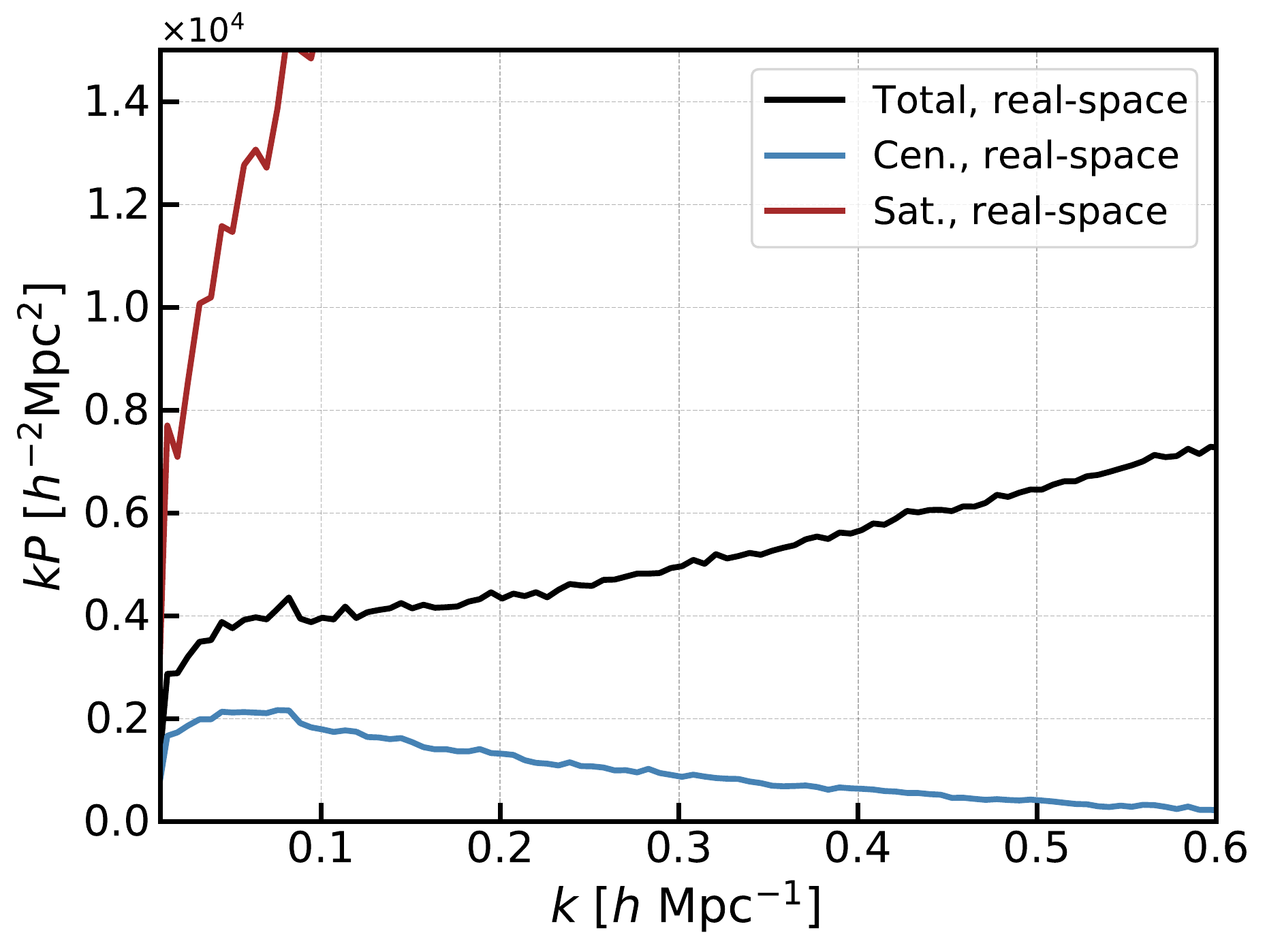}
\includegraphics[width=0.4\textwidth]{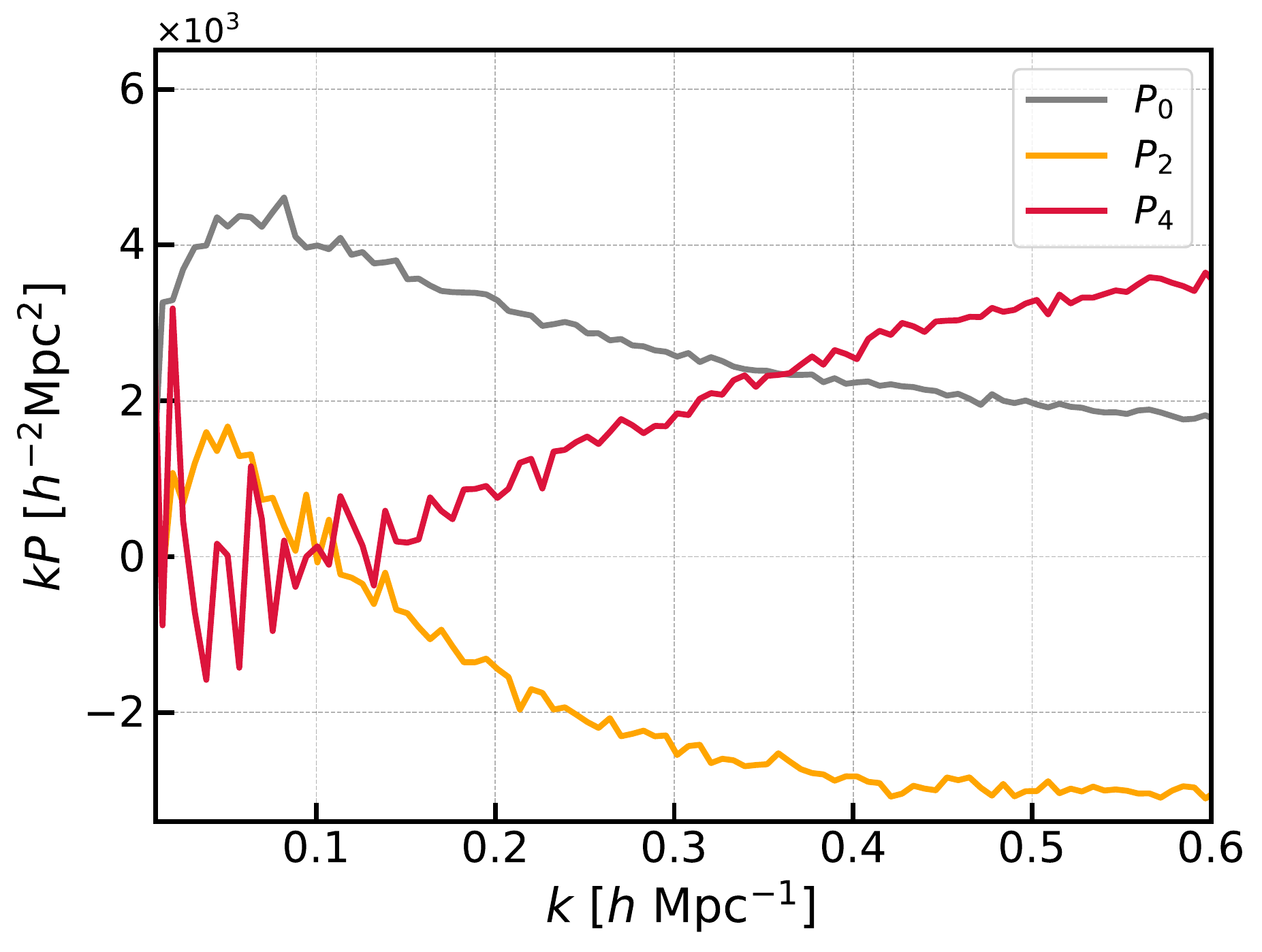}   
\caption{{\it Left:} Molino power spectrum in real-space split into the central galaxies (blue), satellite galaxies (red), and the total sample (black). {\it Right:} Molino power spectrum in redshift-space, with monopole (grey), quadrupole (orange), and hexadecapole (red).} 
\label{fig:pkz_nmul024_tot_molino}
\end{figure}

Molino galaxy catalogs are built on the low-resolution version of Quijote simulations, where low-mass halos can not be properly resolved. Ref.~\cite{Hahn:2020lou} adopted the best-fit HOD parameters for the SDSS sample with luminosity cuts $M_r<-21.5$ and $M_r<-22$, based on Ref. \cite{Zheng:2007zg}. However, these HOD parameters lead to a power excess in real space. This is shown in the left panel of figure \ref{fig:pkz_nmul024_tot_molino}, where in addition to the power spectrum of the full sample, the power spectrum of central (blue) and satellite (red) galaxies are also shown. The increasing power spectrum towards small scales indicates a very large one-halo term~\cite{Okumura:2016mrt}. One hypothesis for this power excess is the low resolution of Quijote simulation; HOD parameters tuned for higher resolution simulation may lead to excessive satellite galaxies residing in a single massive halo. However, more follow-up studies on the Molino sample will provide further insight into HOD parameter choice for a given resolution. Considering the redshift-space power spectrum, we show in the right panel of figure \ref{fig:pkz_nmul024_tot_molino}, the total galaxy monopole (grey), quadrupole (orange), and hexadecapole (red). The quadrupole drops to a negative value already at relatively large scales $k \sim 0.1 \ \hmpc$, and the hexadecapole becomes quickly positive also at a similar scale. This is likely due to the velocity component of the satellite galaxies towards small scales. In a realistic galaxy sample (such as the LRG sample from BOSS), the hexadecapole is usually very close to zero with a low SNR. In contrast, the hexadecapole has non-negligible information  for the case of Molino galaxies. 

\begin{figure}
\centering
\includegraphics[width=0.8\textwidth]{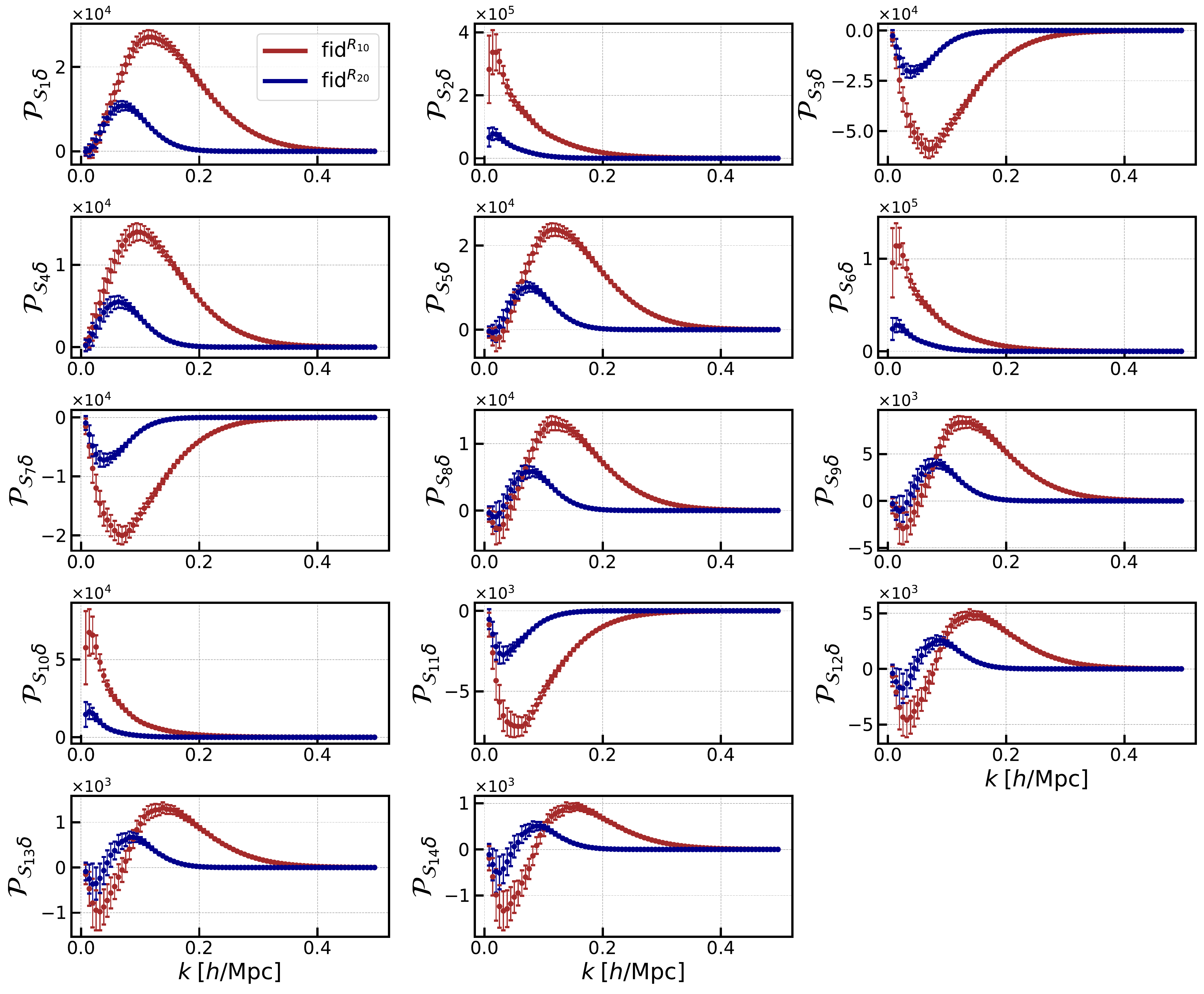}
    \caption{Measured skew spectra with fiducial cosmology on Molino galaxy catalogs, using the smoothing scales of  $R=10\, \mpch$  (red) and $R=20\, \mpch$ (blue). The error bars correspond to the average of over 15,000 simulations at reference cosmology.}
\label{fig:molino_ss_smooth10x20_shoton_fid} \vspace{-0.1in}
\end{figure}

When analyzing skew spectra using the Molino sample, due to the convolution in Eq.~\ref{eqn:quad_field} by integrating out the vector ${\bf q}$, different scales are mixed, and the extra power induced by satellite galaxies are spread over all the skew spectra. Therefore, when we compare the information gain, it is fairer to include higher-order multipoles. Interestingly, despite the high power induced by the satellite galaxies, the general shape of the 14 Molino skew spectra remains unchanged as shown in figure \ref{fig:molino_ss_smooth10x20_shoton_fid}, we can identify the three classes of the skew spectra as the Quijote case (see also figure \ref{fig:quijote_ss_smooth10x20_shoton_fid}). We also observe that when using a smaller smoothing scale of the density field ($R=10\,\mpch$, red), the overall amplitude gets amplified, and the peaks (troughs) shift towards smaller scales compared to the larger smoothing scale ($R=20\,\mpch$, blue).
\begin{figure}
    \centering
    \includegraphics[width=\textwidth]{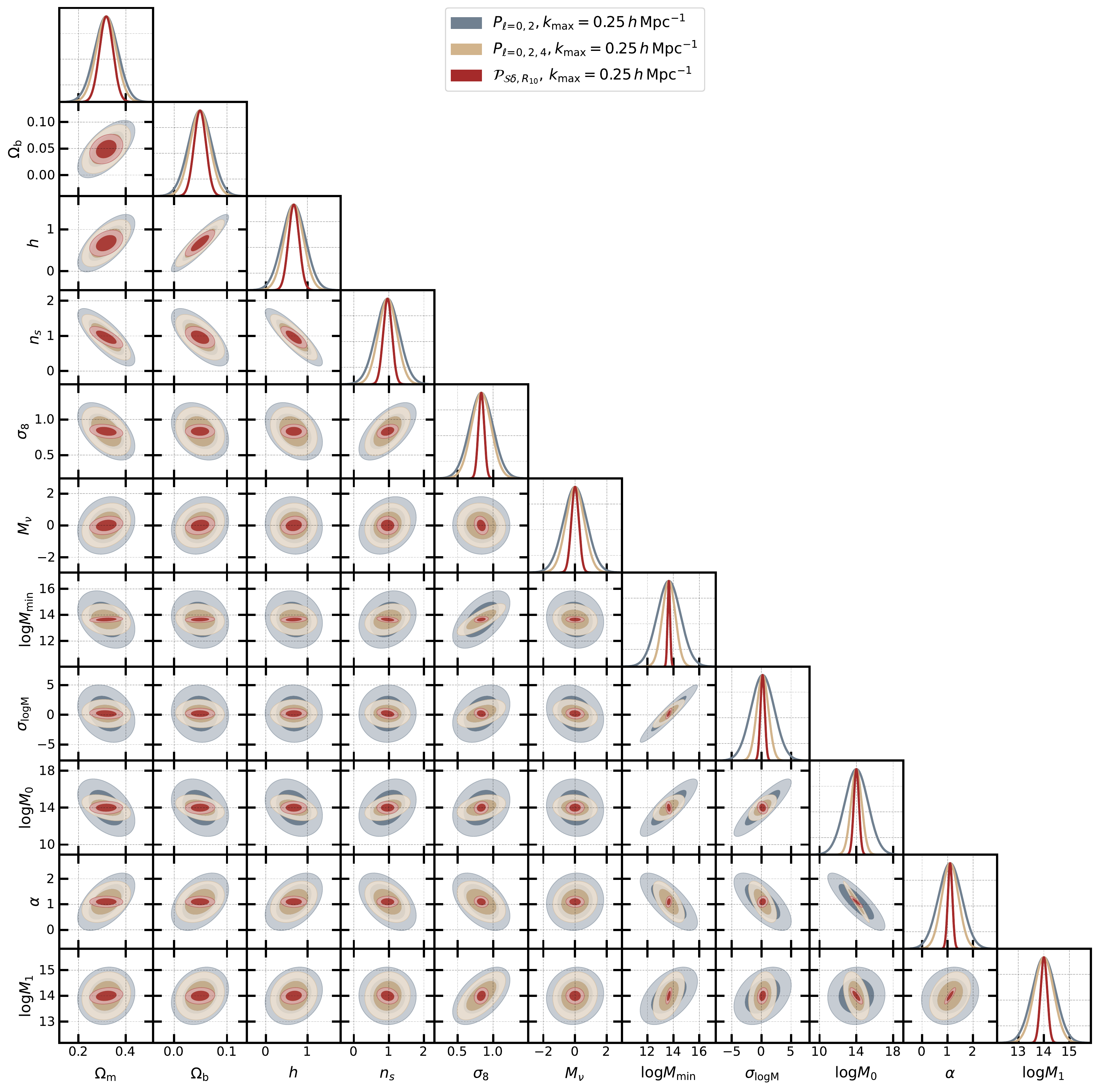}\vspace{-0.1in}
    \caption{Marginalized 1- and 2-$\sigma$ 2D parameter constraints from the power spectrum and the skew spectra measured on Molino galaxy catalogs. The grey contours include only the monopole and quadrupole of the power spectrum, while the blue ones also include the hexadecapole. For the skew spectra, we set the smoothing scale of $R=10\, \mpch$. The small-scale cutoff is set to $k_{\rm max} = 0.25\, \hmpc$.}
    \label{fig:molino_2dcontour_pk_shot0_ss_shot1_k0d25}\vspace{-0.1in}
\end{figure}
\noindent In figure \ref{fig:molino_2dcontour_pk_shot0_ss_shot1_k0d25}, we show the 2D marginalized constraints on cosmological and HOD parameters for the skew spectra and power spectrum multipoles measured on Molino galaxies. While adding hexadecapole further helps to improve constraints by $(17-31)\%$ for the 6 cosmological parameters, the power spectrum $P_{\ell=0,2}$ and $P_{\ell=0,2,4}$ share almost the same degeneracy directions for the 6 cosmological parameters. However, on the HOD parameters, the degeneracy directions can change when adding the hexadecapole information for power spectrum, {\it e.g.,} ${\rm log}M_0-{\rm log}M_1$ plane, which can be a consequence of the satellite galaxies. 

\newpage
\bibliographystyle{utphys}
\bibliography{skewspecs_info}

\end{document}